%% file: acl_latex.tex
\documentclass[11pt]{article}

\usepackage{acl}

\usepackage{times}
\usepackage{latexsym}
\usepackage[T1]{fontenc}
\usepackage[utf8]{inputenc}
\usepackage{microtype}
\usepackage{inconsolata}

\AtBeginDocument{%
  }

\usepackage{booktabs,tabularx}
\usepackage{graphicx}
\usepackage{multirow}
\usepackage{placeins}
\usepackage{amsmath,amssymb}
\newcommand{\ci}[1]{\ensuremath{^{\pm\text{#1}}}}%
\newcolumntype{L}{>{\raggedright\arraybackslash}X}
\newcolumntype{C}{>{\centering\arraybackslash}X}

\newcolumntype{Y}{>{\centering\arraybackslash}X}

\title{ABISS: Evaluating Text-to-SQL Systems Through Agent Interaction}
\author{{Giovanni Sullutrone\textsuperscript{1}, Luca Sala\textsuperscript{1}, Sania Aftar\textsuperscript{2}, Georgia Koutrika\textsuperscript{3,4}, Sonia Bergamaschi\textsuperscript{1}} \\
  {\textsuperscript{1}University of Modena and Reggio Emilia, Modena, Italy} \\
  {\textsuperscript{2}Independent Researcher} \\
  {\textsuperscript{3}Athena Research Center, Athens, Greece} \\
  {\textsuperscript{4}University of Nicosia, UNIC Athens, Department of Computer Science, Athens, Greece} \\
  {\small\texttt{\{giovanni.sullutrone, luca.sala, sania.aftar, sonia.bergamaschi\}@unimore.it, georgia@athenarc.gr}}}

\begin{document}
\maketitle

\begin{abstract}
    Real-world users frequently pose ambiguous or unanswerable questions that current Large Language Model (LLM)-based Text-to-SQL systems handle poorly. Three interconnected gaps hinder progress: incomplete taxonomies, {the lack of} scalable generation of realistic benchmark data, and evaluation limited to fixed, idealized conversations.
    We address these gaps through three contributions: (1) a unified taxonomy of 8 categories covering ambiguous and unanswerable questions; 
    (2) a multi-agent generation pipeline that generates Natural Language Questions (NLQs), grounds them in SQL, and applies a ten-stage validation pipeline with council-based checks; and (3) ABISS (Ambiguity Benchmark using Interaction-Simulated Sessions), a dynamic simulation environment where Text-to-SQL agents interact with style-aware simulated users across multi-turn dialogues.
    {Experiments with ten models, nine of them open-source, on BIRD- and Spider-based datasets reveal two bottlenecks. First, models often fail to recognize that a question is ambiguous, although they usually name its subcategory correctly once they do, and the correct label helps mostly on the questions whose type they miss. Second, models lose most when they answer without asking: once the simulated user supplies the missing information, their SQL is nearly as accurate as when it is given in advance.} We release our code and {a link to the datasets} on GitHub\footnote{{\def\UrlBreaks{\do\/\do\-}\url{https://github.com/giosullutrone/ABISS-Evaluating-Text-to-SQL-Systems-Through-Agent-Interaction}}}.
\end{abstract}

\section{Introduction}

Large Language Models (LLMs) have pushed Text-to-SQL\footnote{The task of converting a natural language question into {an} SQL query executable on a given database.} past 80\% execution accuracy on curated benchmarks such as {Spider~\cite{gao2024text}}. Such benchmarks are designed to contain only well-formed and answerable questions. In real-world settings, however, users also pose questions that are ambiguous or unanswerable given the available schema and domain knowledge~\cite{EvaluatingDataModel}.

\begin{figure}[b]
    \centering
    \includegraphics[width=0.9\linewidth]{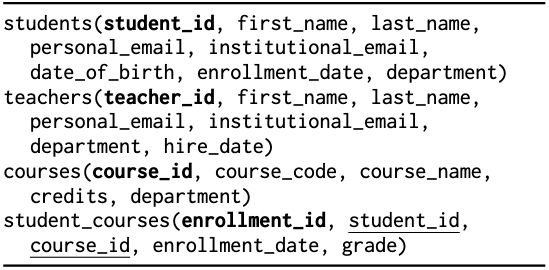}
    \caption{University database schema used in the running examples
throughout the paper.}
    \label{fig:university_schema}
\end{figure}

Consider the university schema in Figure~\ref{fig:university_schema}. The request \textit{``List recent courses''} would be ambiguous because \textit{recent} has no specified temporal threshold, whereas \textit{``List the administrative staff in the engineering department''} would be unanswerable because the schema contains no staff information. Beyond such failures arising from the question or schema, modern Text-to-SQL systems may receive external domain knowledge or policies, commonly provided as \textit{evidence}, to assist question interpretation~\cite{Li2023bird}. Knowledge augmentation introduces further failure modes when the required knowledge is missing, depends on user-specific information, or is internally inconsistent. A robust system should ask targeted clarifications for ambiguous questions and explain the missing information for unanswerable ones, rather than commit to an arbitrary SQL query. Supporting this behavior requires progress on three connected problems: defining the relevant failure modes, constructing data that instantiates them reliably, and evaluating systems in interactions where they must diagnose and resolve those failures.
\paragraph{Incomplete taxonomies.} Determining whether to generate SQL, request clarification, or provide feedback requires identifying why a question is problematic. Existing taxonomies, however, {rarely separate} failure modes introduced by knowledge augmentation~{\cite{Li2023bird}}: missing domain knowledge (\textit{``grades in alphabetic notation''} presupposes a numeric-to-letter mapping), missing user context (\textit{``professors in my department''} presupposes the user's affiliation), and inconsistent evidence (two incompatible definitions of \textit{student performance}).
\paragraph{Data scalability.} Once these distinctions are defined, a benchmark requires sufficiently broad examples that are verifiably ambiguous, admitting more than one valid interpretation, or operationally unanswerable under {a} validation procedure. Although Wizard-of-Oz annotation can produce realistic interactions, it {requires expensive work by SQL experts}~\cite{cosql}. Existing pipelines instead either target only {answerable} questions~\cite{OMNI-SQL}, {leave the questions to human annotators}~\cite{Ambrosia}, or rely on a single generative model that cannot account equally well for the distinctive requirements of every subcategory~\cite{practiq,benchmarking-ambiguity}.
\paragraph{Fixed interaction histories.} Even with sufficiently broad data, evaluation must test whether systems can diagnose and resolve problematic questions through interaction. Existing benchmarks provide a fixed conversation history~\cite{cosql,practiq}{, so their SQL evaluation tests only whether the system incorporates a supplied clarification}. This removes key parts of the interaction from evaluation: {acting on its own diagnosis}, formulating an appropriate clarification or explanation, and responding to user answers that may be incomplete {or only partly relevant}.

We address these challenges with three contributions. First, we introduce a \textbf{unified taxonomy for knowledge-augmented Text-to-SQL} that adds \textit{Missing User Knowledge} and \textit{Conflicting Knowledge} categories and separates schema-level from domain-level knowledge gaps (Section~\ref{sec:taxonomy}).
Second, we develop an \textbf{automated generation pipeline} that builds both answerable and problematic questions from any seed database and enforces their quality through ten validation stages, several of which use majority voting by a council of open-source models (Section~\ref{sec:generation_pipeline}).
Third, we present \textbf{ABISS (Ambiguity Benchmark using Interaction-Simulated Sessions)}, a dynamic environment that evaluates Text-to-SQL agents over multi-turn dialogues with simulated users. Agents must identify a question's subcategory, ask for clarification when it is ambiguous, and explain why it cannot be answered when it is unanswerable. Simulated-user responses are generated by an LLM council and preserve the linguistic style of the original question (Section~\ref{sec:abiss}).

{We evaluate nine open-source models ranging from 7B to 125B parameters and Claude Haiku~5.5 on ABISS-BIRD and ABISS-Spider, the datasets that our pipeline builds from BIRD and Spider databases (Section~\ref{sec:experiments}), and identify two main bottlenecks. The first is recognizing the question type. Models call about half of the ambiguous questions answerable, but usually name the subcategory correctly once they recognize the type. Supplying the correct subcategory raises execution accuracy on ambiguous questions from 53.7\% to 64.6\% and feedback accuracy from 82.4\% to 89.7\% on ABISS-BIRD, almost entirely on the questions whose type models miss. The second is asking. Compared with an oracle that states the intended meaning in advance, models lose accuracy on ambiguous questions mainly where they answer without a clarification. Once the simulated user gives a relevant reply, their SQL is on average within 3 points of this oracle over both datasets.}

\section{Related Work}
\label{sec:related_work}

\subsection{Text-to-SQL Systems and Interactive Approaches}

Contemporary Text-to-SQL systems leverage LLMs through prompt engineering~\cite{chang2023prompt,gao2024text,DINSQL}, task decomposition~\cite{li2023resdsql,li2024codes,DINSQL}, and agent-based refinement~\cite{wang2025macsql}, with strong results on Spider~\cite{yu2018spider} and BIRD~\cite{Li2023bird}. All of them optimize for well-formed questions under single-turn evaluation, whereas deployed users frequently ask ambiguous or unanswerable ones~\cite{KnowWhatIDontKnow,Floratou2024NL2SQLIA,EvaluatingDataModel}. Early interactive systems target this gap, detecting errors through question-query misalignment~\cite{li-etal-2020-mean,li2014constructing} or error detectors~\cite{yao2019model,gur2018dialsql} {and resolving them with} template-based clarifications. Templates, however, only cover recurrent error types and break down whenever resolution requires an open-ended response.

\subsection{Taxonomies for Problematic NLQs}

Existing taxonomies of problematic {natural language questions (NLQs)} fall into two groups. Some~\cite{Ambrosia,benchmarking-ambiguity,InteractiveT2SQLViaExpected} focus on specific forms of ambiguity, such as structural ambiguity, semantic mapping ambiguity, or lexical vagueness, and therefore do not cover unanswerability or failures involving external knowledge. Others~\cite{didaskgoodquestion,cosql,KnowWhatIDontKnow,Floratou2024NL2SQLIA,practiq} cover a broader range of problematic questions but use coarse labels that merge distinct phenomena, omit structural ambiguity, or do not represent vagueness and missing user information as separate categories. Moreover, none explicitly covers failures arising in knowledge-augmented systems, where a Text-to-SQL model receives external evidence in addition to the database schema~\cite{Li2023bird}. We address these limitations by introducing explicit \textit{Missing User Knowledge} and \textit{Conflicting Knowledge} categories and separating schema-level from domain-level knowledge gaps.

\subsection{Synthetic Data Generation}
\label{sec:synthetic_data_generation}

Datasets of problematic questions have been built through manual annotation~\cite{didaskgoodquestion,Ambrosia}, Wizard-of-Oz collection~\cite{cosql}, rule-based perturbations~\cite{KnowWhatIDontKnow}, and LLM-driven synthesis~\cite{benchmarking-ambiguity,practiq}. PractiQ~\cite{practiq} perturbs Spider SQLs and schemas, then expands each example into a fixed four-turn conversation with a single proprietary LLM, inheriting model bias, dependence on the source SQL structure, and pre-scripted rather than autonomous disambiguation; OmniSQL~\cite{OMNI-SQL} reaches {high} synthesis quality but only for {answerable} NLQs. We retain OmniSQL's complexity- and style-controlled generation while starting from any seed database, guiding each prompt with our taxonomy, and validating with councils of open-source LLMs, which limits the self-evaluation bias of single-model pipelines~\cite{LLMFavor}.

\subsection{Benchmarks for Problematic Questions}

Existing benchmarks evaluate along narrow dimensions, such as {classification}~\cite{didaskgoodquestion,KnowWhatIDontKnow,practiq} or SQL translation under fixed, oracle clarifications~\cite{cosql,practiq}. PractiQ covers {all three question types}, but its conversations are preconstructed. These settings therefore do not jointly test whether a system autonomously decides to request clarification, provides informative feedback on unanswerable questions, and refrains from unnecessary clarification on answerable ones.
BIRD-Interact~\cite{huo2025birdinteractreimaginingtexttosqlevaluation}, concurrent
with our work, offers a dynamic environment with hierarchical knowledge bases and a function-driven user simulator, but is ambiguity-only: it covers neither answerable
nor unanswerable cases under a unified taxonomy, omits {conflicting-knowledge} and {user-centered}
failures, and does not isolate category diagnosis from dialogue {behavior}. {BIRD-Interact and ABISS} are complementary; Appendix Table~\ref{tab:related_comparison} gives the full comparison.

\section{Taxonomy}
\label{sec:taxonomy}

\subsection{Definitions}
Building on {\citet{Floratou2024NL2SQLIA}}, let \textit{D} be a database, \textit{S} the set of all non-equivalent valid SQL queries\footnote{Two queries are non-equivalent if there exists a database state on which they produce different results~\cite{Floratou2024NL2SQLIA}.} formulable on \textit{D}, \textit{q} a natural language question, and \(K=\{k_1,\dots,k_n\}\) an external knowledge base whose elements are pieces of evidence that may assist in interpreting \textit{q}.

\textbf{Definition 1 (Mapping function).}
\label{def:mapping}
Define \(f(q,D,K)=\{s\in S \mid s\text{ correctly answers }q\text{ on }D\text{ given }K\}\).

\textbf{Definition 2 (Ambiguity).}\label{def:ambiguity}
\textit{q} is \textit{ambiguous} w.r.t.\ \textit{D} and \textit{K} if \(\lvert f(q,D,K)\rvert \ge 2\), i.e.\ it admits two or more non-equivalent valid queries.

\textbf{Definition 3 (Unanswerability).}\label{def:unanswerability}
\textit{q} is \textit{unanswerable} w.r.t.\ \textit{D} and \textit{K} if \(f(q,D,K)=\varnothing\).

\subsection{Taxonomy}
{Our} taxonomy focuses on the root cause of the \textit{ambiguity} and \textit{unanswerability} defined above. For each category, we specify whether it is ambiguous (Amb) or unanswerable (U). Some categories are further divided into subcategories denoted in \textit{italics}.

\paragraph{Structure Ambiguity (Amb).}
It occurs when the syntactic structure of an NLQ permits multiple valid interpretations of the relationships among its components, resulting in distinct formal query mappings.
It splits into \textit{Scope Ambiguity}, arising from unclear quantifiers (e.g.{,} each,
every, all) that admit a collective or a distributive reading, and \textit{Attachment
Ambiguity}, arising from uncertainty in how a modifier attaches within the
sentence~\cite{Ambrosia}{: a modifier that follows a conjunction of two nouns, each naming its own set of entities, can attach to the nearest noun only or to both, as in \textit{``professors and students in engineering''}}.

\paragraph{Semantic Mapping Ambiguity (Amb).}
It occurs when a term or reference in the NLQ can correspond to multiple schema elements, resulting in more than one plausible semantic grounding.
It splits into \textit{Lexical Overlap}, when two or more schema attributes share similar or
identical forms, and \textit{Entity Ambiguity}, when multiple plausible entities could satisfy
the same reference.

\paragraph{Lexical Vagueness (Amb).}
It arises when an NLQ contains terms whose meaning lacks a precise or objective boundary, leading to vague selection criteria during query generation (e.g., the term \textit{``recent''} in the previous example \textit{``List recent courses''}).

\paragraph{Missing Schema Elements (U).}
An NLQ is unanswerable due to Missing Schema Elements when the database schema lacks the tables, columns, or relationships required to express the information requested. It splits into \textit{Missing Entities or Attributes}, when key information is absent from the schema, and \textit{Missing Relationships}, when required entities exist but no linkage between them is represented.

\paragraph{Missing External Knowledge (U).}
An NLQ is unanswerable due to Missing External Knowledge when its interpretation depends on objective, domain-specific facts or policies not present in the database or knowledge base \textit{K}.

\paragraph{Missing User Knowledge (Amb).}
An NLQ has Missing User Knowledge when it contains a user-anchored expression such as
\textit{my}, \textit{our}, \textit{me}, or \textit{I}, but the required user-specific fact is
absent from \textit{D} and \textit{K}. For example, \textit{``What are my courses?''}
becomes answerable once the user's identity is known, at which point the SQL query can
filter the enrollments for that user. The same \textit{D} and \textit{K} admit one valid
query for each possible user identity, which is why we classify this category as
ambiguous.

\paragraph{Conflicting Knowledge (Amb).}
An NLQ is ambiguous due to Conflicting Knowledge if the knowledge base \textit{K} contains multiple, non-equivalent pieces of evidence for interpreting the same concept in the question.

\paragraph{Improper Question (U).}
An NLQ is Improper when it is unrelated to the domain of either the database \textit{D} or the knowledge base \textit{K}.

In total, the taxonomy comprises eight problematic categories, which yield eleven problematic subcategories. Together with the two Answerable subcategories, \textit{With Evidence} and \textit{Without Evidence}, these form the thirteen subcategories used in the generation pipeline (Section~\ref{sec:generation_pipeline}) and benchmark evaluation (Section~\ref{sec:abiss}).

\begin{figure*}[t]
    \centering
    \includegraphics[
        width=\textwidth,
        height=0.60\textheight,
        keepaspectratio
    ]{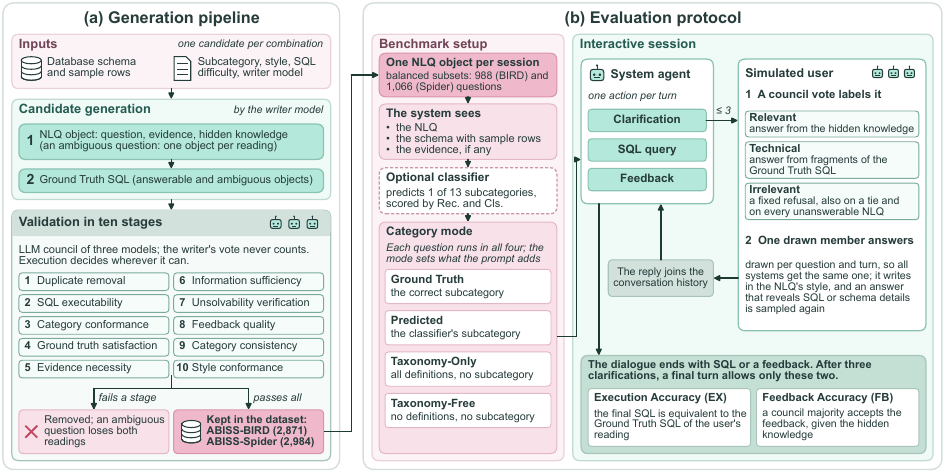}
    \caption{{Overview of ABISS. \textbf{(a) Generation pipeline}: the writer model generates the NLQ objects for each combination of inputs and, for a solvable object, its Ground Truth SQL. Ten validation stages, decided by duplicate checks, SQL execution, or a vote of the council without the writer, keep the questions of ABISS-BIRD and ABISS-Spider. \textbf{(b) Evaluation protocol}: each session takes one NLQ object, and the category usage mode determines which subcategory information the system receives. At every turn the system asks a clarification question, writes an SQL query, or gives feedback. The simulated user labels each clarification by council vote and answers through one drawn council member. An SQL query or feedback ends the dialogue.}}
    \label{fig:abiss-pipelines}
\end{figure*}

\section{Automated Generation Pipeline}
\label{sec:generation_pipeline}

Our generation pipeline leverages multiple open-source LLMs coordinated through category-aware prompting and staged validation. The pipeline operates in two phases: generation and validation. The generation phase produces candidate NLQs across multiple categories, styles, and difficulty levels using structured prompting {(Appendix~\ref{app:prompts})}. For each candidate, it records four fields: the \textit{NLQ}, the \textit{Evidence}, the \textit{Ground Truth SQL} (for answerable and ambiguous NLQs), and the \textit{Hidden Knowledge} (disambiguation information for ambiguous NLQs or actionable feedback for unanswerable ones). Together, these fields form the NLQ object used throughout the pipeline. The validation phase applies a cascade of automated checks specific to each category, employing multiple LLMs in a council-based approach with majority voting{, in which the model that wrote a question has no counted vote on it}.

\subsection{Category-Aware NLQ Object Generation}\label{sec:question_generation}
Generating realistic yet intentionally problematic NLQs requires explicit control over the target taxonomic category. Simply instructing an LLM to ``generate an ambiguous question'' is insufficient, as the model may produce questions that are merely difficult rather than exhibiting the intended ambiguity or knowledge gap. We therefore adopt a two-step generation process.

In the first step, we generate the linguistic portion of each NLQ object. For each target database, the prompt combines taxonomic grounding (e.g., {the} category definition), schema context with sample rows {(for BIRD, also the column descriptions distributed with the dataset, Appendix~\ref{app:setup})}, and controlled diversity over three question styles (formal, colloquial, interrogative) and four SQL difficulty levels following {\citet{OMNI-SQL}}. The model returns the NLQ with the {evidence}, where applicable, and the {hidden knowledge}.
This step produces one NLQ object for answerable and unanswerable categories, and two NLQ objects for ambiguous ones, since each ambiguous question encodes two distinct interpretations. {When a validation stage removes one interpretation, the other is removed as well.}

In the second step, we generate Ground Truth SQL for all solvable objects, i.e.\ answerable and ambiguous ones. A separate chain-of-thought prompt{, inspired by DIN-SQL~\cite{DINSQL} for its simplicity,} receives the schema, the question, the evidence, and the hidden knowledge, then performs schema linking {and SQL generation. The prompt asks for a query that returns only the values that the question requests. The validation stages in which council models write SQL reuse this prompt, so that their queries and the Ground Truth SQL follow the same conventions. Because the generated SQL does not always match the requested difficulty level, each solvable object stores the level measured from its Ground Truth SQL with keyword rules that approximate the criteria of the requested levels, next to the requested one (Appendix~\ref{app:dataset_composition}).}

\subsection{Multi-Stage Validation Pipeline}
\label{sec:validation}

A question {labeled} \textit{ambiguous} must demonstrably admit more than one valid interpretation rather than merely be hard, and an \textit{unanswerable} one must be operationally unanswerable. We therefore apply a cascade of ten validators ({Figure~\ref{fig:abiss-pipelines}a; }details{, and what each council member changes,} in Appendix~\ref{app:validation_detail}), each operating only on the questions that survived the previous stages, several of them deciding by majority vote across a council of open-source LLMs. These validation stages {are}: (1) {duplicate removal}, (2) SQL executability verification, (3) category conformance checking, (4) ground truth satisfaction, (5) evidence necessity checking, (6) {information sufficiency checking}, (7) unsolvability verification, (8) feedback quality checking, (9) category consistency checking, {and} (10) style conformance checking. {An ambiguous question thus passes only if the Ground Truth SQL of its two interpretations return results that our equivalence check tells apart (Stage~2) and if, given the hidden knowledge of each interpretation, a council model other than the writer reaches that interpretation's result (Stage~6).} {Stages~2 and~6 remove the most (Appendix Table~\ref{tab:validation_stages}): 27.9\% and 25.4\% of the questions that enter them for ABISS-BIRD, and 33.2\% and 19.6\% for ABISS-Spider. Most Stage~2 removals are Ground Truth SQL that return no rows, ambiguous questions whose two interpretations return the same result, and the partners of removed interpretations. Overall the cascade retains 36.3\% and 37.7\% of the candidate questions.}

\subsection{Dataset Generation and Analysis}
\label{sec:dataset_analysis}

For data generation, we employed three large open-source instruction-tuned language models from different families: GPT-OSS-120B~\cite{openai2025gptoss120bgptoss20bmodel}, Qwen3.5-122B-A10B-FP8~\cite{qwen3.5}, and NVIDIA-Nemotron-3-Super-120B-A12B-FP8~\cite{nvidia_nemotron_3_2025}{, hereafter GPT-OSS, Qwen3.5, and Nemotron}. We selected models from different families as diversity is essential for the council-based validation framework (Section~\ref{sec:validation}). Question generation proceeded across all 13 subcategories for both the development split of the BIRD dataset~\cite{Li2023bird}, comprising 11 databases spanning diverse domains, and 11 company-related databases {of the Spider release~\cite{yu2018spider}}. For each unique combination of generative model, subcategory, linguistic style, SQL difficulty, and database, we generated {one candidate} to be processed by the validation pipeline. {We set no minimum number of retained questions per combination. Every subcategory still covers all databases, styles, and requested difficulty levels in both datasets (Appendix~\ref{app:dataset_composition}).}

Using this setup, we produced two datasets: \textit{ABISS-BIRD}, containing {2{,}871} validated questions across 11 databases, and \textit{ABISS-Spider}, containing {2{,}984} validated questions across 11 databases, for a total of {5{,}855} questions{, where the two interpretations of an ambiguous question count as two questions}. {Ambiguous questions constitute the majority (53.8\% of ABISS-BIRD and 52.3\% of ABISS-Spider), followed by unanswerable (36.7\% and 38.4\%) and answerable ones (9.4\% and 9.2\%)}; {Appendix Figure~\ref{fig:category_comparison} counts each subcategory}.

Two experts with doctoral-level expertise {jointly} annotated {5 questions per subcategory and dataset (130 in total)} along different dimensions (Table~\ref{tab:human_eval} in Appendix~\ref{app:dataset_composition}). {Style conformance (100\%), realism (95.0\%), feedback accuracy (92.5\%), and category correctness (90.8\%)} are high, while {disambiguation (87.9\%) and SQL correctness (83.1\%)} remain at levels sufficient to support benchmark evaluation{, and both readings of an ambiguous question are natural in 84.3\% of cases}.

\section{ABISS: Interaction-Based Simulation}
\label{sec:abiss}

To address the evaluation limitations discussed in Section~\ref{sec:related_work}, we present ABISS {(Figure~\ref{fig:abiss-pipelines}b)}, a dynamic simulation environment in which Text-to-SQL system agents engage in multi-turn dialogues with an agent simulating the user. Each simulated user inherits the \textit{linguistic style} of the original question. ABISS supports both modular systems, i.e.\ with separate classification and dialogue components, and end-to-end integrated architectures.

\subsection{Evaluation Protocol}
\label{sec:evaluation_protocol}

For each NLQ object in the generated datasets, ABISS always conducts an interactive phase and may precede it with a classification phase.

\textbf{Classification Phase (Optional).} When a system exposes a classification component, it receives the NLQ, the complete database schema with sample data, the NLQ's evidence, if available, and the definitions and examples of our 13 subcategories (Section~\ref{sec:taxonomy}), and predicts one of them. This prediction allows ABISS to measure recognition and classification and, in the \textit{Predicted} mode described below, to examine how the diagnosis affects the subsequent dialogue. Systems that do not expose an intermediate classification skip this phase and enter the interaction directly.

\textbf{Category Usage Modes.} Before the interactive phase, the selected mode determines what subcategory information, if any, is supplied to the system. ABISS conducts a separate dialogue for each NLQ object under four modes: \textit{Ground Truth}, which provides the correct subcategory and isolates interaction from diagnostic errors; \textit{Predicted}, which provides the label produced by the optional classification phase and captures how misclassification propagates into the dialogue; \textit{Taxonomy-Only}, which provides the complete taxonomy definitions without assigning a subcategory; and \textit{Taxonomy-Free}, which provides neither an assigned subcategory nor the taxonomy definitions, leaving only the general policy of choosing among SQL, clarification, and feedback. The {last} two require no intermediate diagnosis and therefore support end-to-end systems that do not separate classification from interaction. {In every mode, the tested system's prompt (Appendix~\ref{app:prompts}) describes the three question types (answerable, ambiguous, and unanswerable) and the response expected for each. In the \textit{Ground Truth} and \textit{Predicted} modes, it also names the question's subcategory and type, framed as strong contextual evidence rather than as an infallible instruction.}

\textbf{Interactive Phase.} The conversation proceeds through iterative cycles until termination. At each cycle, the system agent produces one of three response types based on the NLQ and conversation history: an \textit{SQL Query}, a \textit{Clarification Question} to {resolve} ambiguities{,} or \textit{Feedback} for unanswerable NLQs. {Until the final turn described below, ABISS does not restrict the response type.}

When the system agent produces a clarification question, a simulated user response is generated before the next {system agent} turn. The {dialogue} ends when the system agent produces either an SQL query or unsolvability feedback. If neither is produced within a predefined turn limit, the system receives one final turn to produce one of the two without further clarifications.

\subsection{Simulated User Agent}
\label{sec:simulated_user}

When the system produces a clarification question during the interactive phase, a simulated user agent, using a council of models, generates an appropriate response {in} the same linguistic style {as} the original NLQ.
Each clarification question produced by the system agent is classified into one of three relevancy {labels} based on whether it addresses the NLQ's actual problematic nature: \textit{Relevant}, when it directly targets the core ambiguity or knowledge gap {on a point that the hidden knowledge settles, including a question about which columns to use when that choice resolves the ambiguity} (e.g.{,} asking \textit{What time period does ``recent'' refer to?} for the lexical vagueness question \textit{``List recent courses''}); \textit{Technical}, when it addresses a valid but secondary database concern unrelated to the primary issue (e.g.{,} asking about result ordering preference); or \textit{Irrelevant}, when it does not provide any informative contribution to the solution or attempts to extract the {Ground Truth} SQL.
The label is obtained by majority vote across the council, defaulting to Irrelevant under ties, and conditions the answer: Relevant answers are grounded in the hidden knowledge, Technical ones in structured fragments of the {Ground
Truth} SQL{, and the simulated user states no preference on a point that this information does not settle. One council member writes each Relevant or Technical answer, the same member for every tested system at a given question and turn. An Irrelevant label receives a fixed refusal sentence in the style of the NLQ}. Appendix~\ref{app:user_sim} details the procedure.

\subsection{Evaluation Metrics}
\label{sec:metrics}
Once a dialogue terminates, ABISS assesses performance for both diagnostic capabilities (whether the system correctly recognized and classified the problematic question) and interactive effectiveness (whether the dialogue ultimately produced a correct result). 
{We report \textbf{Recognition (Rec.)} and \textbf{Classification (Cls.)}, the shares of questions whose type and subcategory the classification phase predicts correctly; \textbf{Execution Accuracy (EX)}, the share of answerable and ambiguous questions whose final SQL is equivalent to the Ground Truth SQL of the user's interpretation (Appendix~\ref{app:equivalence}); and \textbf{Feedback Accuracy (FB)}, the share of unanswerable questions whose feedback is accepted by a majority of the council, which sees the hidden knowledge but not the subcategory. SQL on an unanswerable question fails EX, and feedback on a solvable one fails both EX and FB. On ambiguous questions, we call the simulated user's answer to a Relevant clarification a \textit{Relevant reply}. The \textbf{Answer Incorporation Rate (AIR)} is the share of first Relevant replies followed by a final answer, and \textbf{Turns to Relevant (TTR)} is the number of clarifications up to the first Relevant one. Metrics are computed per run (system, dataset, and mode) and averaged across systems as plain means. Appendix~\ref{app:metrics} gives the formal definitions.}

\section{Experiments}
\label{sec:experiments}

\subsection{Interaction Evaluation}
Each question {of the balanced subsets described below} was evaluated under all four \textit{category usage modes}. Conversations were limited to 3 clarification turns {for every category, since a deployed system does not know the category of a question in advance}.
{Nine} open-source instruction-tuned models from five families, spanning 7B to {125B} parameters, served as system agents; the same three large models used for data generation also served as the user simulation council. {Qwen3.8-Flash is held out: it is the only large open model outside the council. A tenth, Claude Haiku~5.5, is proprietary.} Appendix~\ref{app:setup} lists {the models, the decoding and compute settings, the limits of reproducibility, and the licenses}{, and Appendix~\ref{app:cost} the cost}. {Two control runs bound the value of interaction in the main run, which we call the \textit{dialogue}. In the \textit{floor}, run in every mode, systems must respond at once with SQL or feedback. In the \textit{oracle}, run in the \textit{Ground Truth} mode, each ambiguous question starts with a completed clarification that states the user's interpretation.}
{A manual audit of 220 simulated-user turns from both datasets found 94.6\% relevancy-label accuracy, weighted to all turns, and no answer leaking SQL or schema details (Appendix~\ref{app:user_sim}).}

The generated datasets are naturally imbalanced (see Section~\ref{sec:dataset_analysis}). To prevent this skew from inflating metrics, we {evaluate all systems} on subcategory-level balanced subsets. {Our draw of these subsets keeps the two interpretations of an ambiguous question together and takes questions from each database in turn.} This yields {988} questions for ABISS-BIRD and {1{,}066} questions for ABISS-Spider, corresponding to {76 and 82} questions for each subcategory, respectively. {Intervals are 95\% paired bootstrap intervals (Appendix~\ref{app:metrics}).}

\begin{table*}[t]
\centering
\fontsize{7.2}{9.4}\selectfont
\setlength{\tabcolsep}{2.4pt}

\begin{tabular}{@{}llrrrrrrrrrrrr@{}}
\toprule
& & \multicolumn{6}{c}{\textbf{ABISS-BIRD}} & \multicolumn{6}{c}{\textbf{ABISS-Spider}} \\
\cmidrule(lr){3-8} \cmidrule(lr){9-14}
& & & & \multicolumn{2}{c}{\textbf{EX}} & \multicolumn{2}{c}{\textbf{FB}} & & & \multicolumn{2}{c}{\textbf{EX}} & \multicolumn{2}{c}{\textbf{FB}} \\
\cmidrule(lr){5-6} \cmidrule(lr){7-8} \cmidrule(lr){11-12} \cmidrule(lr){13-14}
\textbf{Model} & \textbf{Size} & \textbf{Rec.} & \textbf{Cls.} & \textbf{Pred.} & \textbf{GT} & \textbf{Pred.} & \textbf{GT} & \textbf{Rec.} & \textbf{Cls.} & \textbf{Pred.} & \textbf{GT} & \textbf{Pred.} & \textbf{GT} \\
\midrule
Qwen3.5 & 122B & 72.6\ci{2.3} & 67.1\ci{2.3} & 63.6\ci{3.3} & 70.7\ci{3.6} & \textbf{88.3}\ci{3.6} & \textbf{94.1}\ci{2.8} & 74.5\ci{2.0} & 70.1\ci{2.1} & 73.2\ci{2.9} & 80.4\ci{3.0} & \textbf{89.1}\ci{3.3} & 92.7\ci{3.0} \\
GPT-OSS & 120B & 72.6\ci{2.2} & 67.4\ci{2.1} & \textbf{67.3}\ci{3.2} & \textbf{72.7}\ci{3.2} & \textbf{89.7}\ci{3.5} & 91.2\ci{2.9} & 73.6\ci{2.1} & 69.8\ci{2.2} & 75.6\ci{2.8} & 81.1\ci{2.9} & \textbf{92.5}\ci{3.0} & 93.4\ci{2.8} \\
Nemotron & 120B & 72.7\ci{2.2} & 66.6\ci{2.3} & \textbf{64.3}\ci{3.3} & \textbf{72.0}\ci{3.4} & \textbf{86.9}\ci{3.6} & 89.4\ci{3.4} & 73.2\ci{2.1} & 67.0\ci{2.2} & 74.0\ci{2.8} & 80.2\ci{3.0} & 88.3\ci{3.3} & 90.7\ci{3.0} \\
\midrule
Qwen3.8-Flash$^\dagger$ & 125B & \textbf{76.7}\ci{2.4} & \textbf{72.9}\ci{2.6} & 62.4\ci{3.5} & \textbf{74.8}\ci{3.6} & \textbf{87.3}\ci{3.7} & \textbf{96.4}\ci{2.5} & 78.0\ci{2.2} & 74.8\ci{2.4} & 73.6\ci{2.9} & \textbf{86.3}\ci{2.6} & 88.7\ci{3.5} & \textbf{96.4}\ci{2.3} \\
Qwen2.5-Coder-32B & 32B & 62.8\ci{2.1} & 54.7\ci{2.3} & 48.3\ci{3.6} & 60.5\ci{3.9} & 81.9\ci{4.3} & 90.0\ci{3.4} & 66.5\ci{2.1} & 57.3\ci{2.2} & 63.2\ci{3.4} & 75.5\ci{3.4} & 84.5\ci{3.7} & 91.9\ci{3.0} \\
Qwen2.5-32B & 32B & 65.8\ci{2.1} & 59.5\ci{2.1} & 49.9\ci{3.6} & 66.0\ci{3.9} & 80.2\ci{4.5} & \textbf{95.2}\ci{2.8} & 66.6\ci{2.3} & 61.4\ci{2.3} & 61.2\ci{3.4} & 77.5\ci{3.7} & 78.6\ci{4.6} & \textbf{92.7}\ci{3.2} \\
Gemma-3-27B & 27B & 56.7\ci{2.4} & 48.8\ci{2.3} & 41.3\ci{3.5} & 56.4\ci{4.1} & 73.7\ci{4.4} & \textbf{93.2}\ci{3.0} & 62.0\ci{2.5} & 52.7\ci{2.6} & 58.3\ci{3.3} & 72.9\ci{3.4} & 78.2\ci{4.3} & \textbf{94.6}\ci{2.5} \\
Mistral-Small-24B & 24B & 64.1\ci{2.5} & 58.7\ci{2.5} & 50.6\ci{3.7} & 59.6\ci{4.1} & 81.1\ci{4.4} & 88.0\ci{3.8} & 70.1\ci{2.4} & 63.5\ci{2.5} & 65.3\ci{3.5} & 75.8\ci{3.6} & 79.6\ci{4.3} & 89.3\ci{3.3} \\
Qwen2.5-7B & 7B & 51.6\ci{2.4} & 41.6\ci{2.7} & 30.5\ci{3.4} & 36.8\ci{3.8} & 66.9\ci{4.7} & 65.8\ci{4.7} & 56.1\ci{2.3} & 44.3\ci{2.6} & 49.5\ci{3.5} & 58.2\ci{3.8} & 70.6\ci{4.5} & 73.6\ci{4.4} \\
\midrule
Claude Haiku 5.5$^\ddagger$ & -- & \textbf{79.0}\ci{2.7} & \textbf{72.7}\ci{2.9} & \textbf{64.6}\ci{3.5} & \textbf{72.2}\ci{3.7} & \textbf{87.7}\ci{4.0} & 93.9\ci{2.8} & \textbf{83.3}\ci{2.3} & \textbf{77.3}\ci{2.5} & \textbf{79.2}\ci{3.0} & \textbf{84.3}\ci{2.8} & \textbf{90.6}\ci{3.3} & \textbf{95.0}\ci{2.5} \\
\bottomrule
\end{tabular}
\caption{{Overall results in the \textit{Predicted} (Pred.) and \textit{Ground Truth} (GT) modes; Rec.\ and Cls.\ come from the classification phase. Superscripts give the larger distance to the bounds of the 95\% interval; \textbf{bold} marks the best system and those whose paired difference from it has an unadjusted interval including zero. $^\dagger$Held-out system. $^\ddagger$Its size is not public.}}
\label{tab:overall_results}
\end{table*}

\subsection{Overall Results}
\label{sec:overall_results}

Table~\ref{tab:overall_results} reports the evaluation results under {the} \textit{Predicted} and \textit{Ground Truth} category {usage} modes{, Appendix~\ref{app:per_category} per subcategory, and Appendix~\ref{app:further} per dataset, system group, and question type}.

{\textit{The Question Type, Not the Subcategory, Limits Diagnosis}. A correct subcategory implies a correct type, so Cls.\ cannot exceed Rec.\ and in practice stays close to it: once a model recognizes the type of a question, it names the correct subcategory in about 90\% of cases, against 26\% for a uniform choice within the type (Appendix~\ref{app:further}). Diagnostic errors therefore lie mostly in the type: models call about half of the ambiguous questions answerable.}

{\textit{No Detected Advantage on Own Questions}. Claude Haiku~5.5 and Qwen3.8-Flash, which did not build the data, lead Rec.\ and Cls.\ (Claude Haiku~5.5 alone on ABISS-Spider), and council models show on average no detectable advantage on the questions they wrote (Appendix~\ref{app:score_checks}). Yet a pipeline run by one model can favor it: when Qwen3.5 alone builds an ABISS-BIRD set and simulates the user, its \textit{Ground Truth} EX rises while GPT-OSS's and Nemotron's fall, and a further run traces the gain to the data (Appendix~\ref{app:qwen_only}). A council keeps this gain far smaller (Appendix~\ref{app:score_checks}).}

\subsection{Impact of Category Knowledge}
\label{sec:category_impact}

{Table~\ref{tab:ablation} reports scores and dialogue statistics per mode, and Appendix~\ref{app:interaction_stats} the latter per model.}

\begin{table*}[t]
\centering
\fontsize{7.2}{9.4}\selectfont
\setlength{\tabcolsep}{3.2pt}

\begin{tabular}{@{}llrrrrrrrrrr@{}}
\toprule
& & \multicolumn{3}{c}{\textbf{Scores}} & \multicolumn{7}{c}{\textbf{Dialogue on ambiguous questions}} \\
\cmidrule(lr){3-5} \cmidrule(lr){6-12}
& \textbf{Mode} & \textbf{EX Ans.} & \textbf{EX Amb.} & \textbf{FB} & \textbf{Clar.} & \textbf{Rel.} & \textbf{Tech.} & \textbf{Irrel.} & \textbf{AIR} & \textbf{Reach} & \textbf{TTR} \\
\midrule
\multirow{4}{*}{BIRD} & GT & 69.7\ci{4.6} & 64.6\ci{3.1} & 89.7\ci{1.8} & 87.7\ci{2.0} & 90.7\ci{2.0} & 6.7\ci{1.5} & 2.7\ci{0.8} & 92.7\ci{1.6} & 83.9\ci{2.3} & 1.01\ci{0.01} \\
 & Pred. & 69.6\ci{4.6} & 53.7\ci{2.8} & 82.4\ci{2.6} & 56.2\ci{2.6} & 89.6\ci{2.5} & 6.7\ci{1.6} & 3.7\ci{1.4} & 92.2\ci{2.2} & 53.6\ci{2.5} & 1.01\ci{0.01} \\
 & Tax.-Only & 68.4\ci{4.8} & 55.6\ci{2.9} & 77.2\ci{2.5} & 62.1\ci{2.6} & 88.4\ci{2.3} & 8.0\ci{1.5} & 3.6\ci{1.5} & 92.4\ci{2.0} & 58.0\ci{2.7} & 1.01\ci{0.01} \\
 & Tax.-Free & 69.0\ci{4.9} & 54.4\ci{2.7} & 76.1\ci{2.6} & 59.9\ci{2.8} & 85.8\ci{2.3} & 9.6\ci{1.9} & 4.7\ci{1.1} & 92.8\ci{2.0} & 54.3\ci{3.0} & 1.01\ci{0.01} \\
\midrule
\multirow{4}{*}{Spider} & GT & 77.1\ci{4.7} & 78.9\ci{2.5} & 91.0\ci{1.6} & 91.8\ci{1.6} & 92.2\ci{1.2} & 4.3\ci{0.8} & 3.5\ci{0.9} & 93.9\ci{0.9} & 88.8\ci{1.9} & 1.01\ci{0.00} \\
 & Pred. & 77.9\ci{4.3} & 68.0\ci{2.3} & 84.1\ci{2.5} & 59.5\ci{2.5} & 92.1\ci{1.7} & 3.9\ci{1.2} & 4.0\ci{1.1} & 95.1\ci{1.1} & 57.1\ci{2.5} & 1.01\ci{0.01} \\
 & Tax.-Only & 76.6\ci{4.4} & 68.9\ci{2.3} & 79.7\ci{2.2} & 66.2\ci{2.4} & 89.2\ci{1.9} & 6.5\ci{1.5} & 4.3\ci{1.0} & 93.4\ci{1.2} & 61.9\ci{2.6} & 1.01\ci{0.01} \\
 & Tax.-Free & 77.7\ci{4.4} & 67.0\ci{2.4} & 78.0\ci{2.6} & 63.5\ci{2.5} & 88.2\ci{1.9} & 7.1\ci{1.3} & 4.8\ci{1.1} & 94.1\ci{1.1} & 58.2\ci{2.6} & 1.01\ci{0.01} \\
\bottomrule
\end{tabular}
\caption{{Effect of the \textit{category usage mode}, averaged across the ten systems. \textit{Clar.}, \textit{Reach}: conversations with a clarification, with a Relevant reply (\%); \textit{Rel.}, \textit{Tech.}, \textit{Irrel.}: relevancy labels (\%); \textit{AIR}: first Relevant replies followed by a final answer (\%); \textit{TTR}: clarifications up to the first Relevant one. Superscripts as in Table~\ref{tab:overall_results}.}}
\label{tab:ablation}
\end{table*}

{\textit{The Correct Subcategory Helps Where the Type Is Missed}. Compared with the \textit{Predicted} mode, the \textit{Ground Truth} mode leaves EX on answerable questions unchanged, but significantly raises EX on ambiguous questions for every model and FB for seven of the ten. The gain comes almost entirely from questions whose type the classifier missed. On the ambiguous questions that a model classifies as ambiguous, the two modes reach nearly the same EX. On those that it calls answerable (about half), it asks in only 15\% of \textit{Predicted} conversations, against 80\% with the correct subcategory, pooled over models. The FB gain comes from unanswerable questions classified as answerable or ambiguous, not from those given a wrong unanswerable subcategory. The ambiguous gains therefore concentrate on Attachment, Scope, and Entity Ambiguity and Lexical Overlap, the subcategories that models most often miss (Appendices~\ref{app:per_category} and~\ref{app:further}).}

{\textit{A Wrong Prediction Suppresses Asking}. Compared with the \textit{Taxonomy-Free} mode, a model's own prediction raises FB but not EX on ambiguous questions, because models follow it even when it is wrong. On the ambiguous questions that their classifier calls answerable, models that classify better tend to ask less (Spearman's $\rho=-0.67$ between this ask rate and Rec.\ over the ten models, Appendix~\ref{app:further}). The two best classifiers, Qwen3.8-Flash and Claude Haiku~5.5, ask in only 0.8\% and 3.0\% of these conversations, against 89.5\% and 59.6\% on the same questions with the correct subcategory.}

\subsection{Interaction Analysis}
\label{sec:interaction_analysis}

{\textit{Models Lose Most by Not Asking}. In the \textit{Ground Truth} mode, ambiguous EX rises from 20.3\% in the floor to 64.6\% in the dialogue and 72.4\% in the oracle on ABISS-BIRD, and from 24.8\% to 78.9\% and 84.6\% on ABISS-Spider. After a Relevant reply, EX lies on average only 1.8 to 2.8 points below the oracle on the same conversations, so most of the loss comes from conversations without a clarification, in every mode (Appendix~\ref{app:further}). Asking thus reorders the strongest models: over both datasets, Qwen3.8-Flash scores below GPT-OSS and Nemotron in the oracle but, asking in 96\% of conversations, above both in the dialogue.}

{\textit{Answers Without Asking Follow a Shared Default}. An answer without asking serves at most one of the two simulated users of a question unless it hedges by returning the columns of both readings (Appendices~\ref{app:score_checks} and~\ref{app:further}), and which one is not random: where at least five models lean to one reading, 94.0\% of them lean to the same one (Appendix~\ref{app:further}).}

{\textit{One Clarification Decides the Dialogue}. TTR is 1.01 in every mode: after a first Technical or Irrelevant clarification, a Relevant reply follows in only 7\% and 17\% of cases, respectively, and a quarter of the SQL written right after the reply to such a clarification takes the other interpretation.}

\section{Conclusion}
We addressed common gaps in LLM-based Text-to-SQL systems by introducing a unified taxonomy for ambiguous and unanswerable questions, a multi-agent generation pipeline, and the ABISS benchmark. Together they target a central limitation of curated datasets: they largely neglect ambiguous and unanswerable {questions}. {The taxonomy sorts these questions by root cause into 8 categories and 11 subcategories. The pipeline builds questions from any seed database and keeps those that pass ten validation stages, several decided by a council of three open models. It produced ABISS-BIRD and ABISS-Spider, with 2{,}871 and 2{,}984 questions, {on which the council models show on average no detectable advantage on their own questions, whereas a single-model pipeline can favor its own model}. ABISS evaluates a system in a dialogue with a style-aware simulated user who holds the missing information, against a floor that cannot ask and an oracle told the intended meaning in advance.}

{Our experiments with ten models identify two main bottlenecks. First, models often fail to recognize that a question is ambiguous, although they usually name its subcategory once they do, and the correct label helps mostly where they miss the type. Second, they lose most by answering without asking. Once the simulated user gives the missing information, their SQL is nearly as accurate as when it is given in advance, and a single clarification usually decides the dialogue. Benchmarks that supply a fixed clarification therefore cannot observe the main source of this loss, the decision to ask.} {Text-to-SQL research has mostly measured the translation of well-posed questions, whereas our results suggest that the larger gains lie in deciding whether a question can be answered as asked.}

\section*{Limitations}
A limitation concerns the operational definition of unanswerability. Stage~7 approximates unanswerability by asking a diverse council to generate a valid SQL {query} and discarding any question for which a correct query is found. Consequently, unanswerable questions in ABISS should be interpreted as questions that remain unsolved under this conservative validation procedure, rather than as questions proven formally unsatisfiable for all possible models or prompting strategies.

{A related limitation concerns the Ground Truth SQL. Human annotators judged 83.1\% of the sampled queries correct (Section~\ref{sec:dataset_analysis}), so absolute EX values include some noise from the references. Because every system answers the same questions, comparisons between systems and between modes are less affected. The same holds for FB. The graders accept some feedback that the annotators reject, so FB may be slightly too high (Appendix~\ref{app:score_checks}).}

A further limitation concerns the realism of the simulated user agent. The simulated user remains a controlled policy conditioned on the original NLQ, conversation history, and benchmark-specific hidden information, and therefore does not model user-side phenomena such as vague or incorrect clarifications, follow-up questions that redirect the conversation, or changes in intent across turns. Modeling less cooperative and more dynamic user behavior remains open.

Our current formulation also assigns each NLQ a single primary subcategory. As a result, a single gold subcategory may not capture the full difficulty of some questions.

Another limitation concerns the fixed choice of council models used throughout the benchmark. {Appendix Table~\ref{tab:roles} gives the role of each model, {Appendix~\ref{app:score_checks} checks the scores where these roles overlap,} and Appendix~\ref{app:qwen_only} compares the council with a single model, but} we did not study whether other council sizes or compositions would change dataset quality during generation or the measured performance of the evaluated systems in the interaction benchmark.

{Our system agents are LLMs prompted under a common protocol. Dedicated interactive systems such as AmbiSQL~\cite{ding2025ambisql} would need to be connected to our simulated user. We leave these adaptations and their evaluation to future work.}

A further limitation is that our evaluation includes {a single proprietary system, Claude Haiku~5.5, so we do not assess whether the same trends hold across proprietary systems}.

\section*{Acknowledgments}
{We acknowledge ISCRA for awarding this project access to the LEONARDO supercomputer, owned by the EuroHPC Joint Undertaking, hosted by CINECA (Italy). The authors acknowledge the HPE Private Cloud AI infrastructure provided through the ITSERR project and thank Carmelo Belfiore, Sandro Provenzano, Serena Russo (University of Palermo IT Services), and Francesco Lo Monaco (Department of Cultures and Societies, University of Palermo) for their support.}

{Claude Opus~5.5 was used to assist with proofreading, with testing and checking for bugs or other issues, and with verifying reported code issues. All final decisions, verifications, and manuscript contents were reviewed and approved by the authors.}

\bibliography{references}

\clearpage
\newpage
\appendix
\section{Comparison with Related {Benchmarks}}
\label{app:related_comparison}
Table~\ref{tab:related_comparison} compares ABISS with the closest interactive Text-to-SQL
{benchmarks}.

\begin{table*}[t]
\centering
\fontsize{8.4}{13}\selectfont
\setlength{\tabcolsep}{3 pt}

\renewcommand{\arraystretch}{1.45}
\begin{tabularx}{\textwidth}{@{}l l l C l C C l@{}}
\toprule

\textbf{{Benchmark}} &
\textbf{Dataset Size} &
\textbf{User Source} &
\textbf{Eval.\ Mode} &
\textbf{Task Scope} &
\textbf{Op.\ Scope} &
\textbf{Follow-up} &
\textbf{Taxonomy Scope} \\

\midrule

CoSQL$^\dagger$ &
3,007 dialogues &
Human &
Static &
Mixed dialogues &
Read-only &
Yes &
Dialogue acts \\

PractiQ &
$\sim$2,800 conv. &
Single-model LLM &
Static &
{Ans.+Amb.+Una.} &
Read-only &
No &
Coarse \\

BIRD-Interact &
600 tasks &
Function-driven sim. &
Dynamic &
Amb.-only &
CRUD &
Yes &
Ambiguity-only \\

\textbf{ABISS} &
{\textbf{5,855 questions}} &
\textbf{Council LLM} &
\textbf{Dynamic} &
\shortstack{\textbf{Ans.+Amb.}\\\textbf{+Una.}} &
\textbf{Read-only} &
\textbf{No} &
\textbf{Unified} \\

\bottomrule

\multicolumn{8}{@{}l}{%
\footnotesize
$^\dagger$CoSQL uses dialogue acts rather than a question-type taxonomy.
}

\end{tabularx}
\caption{Comparison of ABISS with related interactive Text-to-SQL
{benchmarks}.}
\label{tab:related_comparison}
\end{table*}

\section{{Models, Decoding, and Compute}}
\label{app:setup}

{This appendix lists the evaluated models and their roles, the inputs that all models share, the decoding of the open models, and the hardware, reproducibility, and licenses of the runs.}

{\textbf{Evaluated models.} As tested systems, we evaluate the large open models GPT-OSS-120B~\cite{openai2025gptoss120bgptoss20bmodel}, Qwen3.5-122B-A10B-FP8~\cite{qwen3.5}, NVIDIA-Nemotron-3-Super-120B-A12B-FP8~\cite{nvidia_nemotron_3_2025}, and Qwen3.8-Flash-Next-FP8~\cite{qwen38flash}. We also evaluate the smaller open models Qwen2.5-32B-Instruct~\cite{qwen2025qwen25technicalreport}, Qwen2.5-Coder-32B-Instruct~\cite{hui2024qwen25coder}, Gemma-3-27B-IT~\cite{gemmateam2025gemma3technicalreport}, Mistral-Small-3.2-24B-Instruct-2506~\cite{Mistral}, and Qwen2.5-7B-Instruct~\cite{qwen2025qwen25technicalreport}. All models follow the decoding protocol below, except Claude Haiku~5.5.

\textbf{Qwen3.8-Flash.} This model activates 6B of its 125B core parameters per token. It also stores 51B parameters of n-gram embedding tables and a 4B multi-token-prediction head. In {216} of its {8{,}216} conversations, or {2.6\%}, its reasoning or its answer reached the token budget of the decoding protocol below, mostly on hard questions. Compared with its other conversations of the same subcategory, these conversations lower its FB by between {0.3} points in the \textit{Ground Truth} mode and {3.5} points in the \textit{Taxonomy-Free} mode. They also lower its EX by {0.8} to {1.2} points.}

{\textbf{Claude Haiku~5.5.} The tenth system answers through Anthropic's Messages API. It receives the same prompts and faces the same simulated user and graders as the other systems. Each of its answers, however, is one request whose output the API constrains to the step's JSON format. The model therefore reasons only in its hidden reasoning, which we request at effort ``high''. The API takes no seed and no sampling values. In the dialogue and in the floor, 1.6\% to 2.8\% of its answers carry no reasoning, against 5\% to 6\% in the oracle. Anthropic's safety classifier refused six requests, all on ABISS-BIRD. In the dialogue and again in the floor, it refused the classification of one toxicology question and two responses. A refused classification counts as a wrong one and ends that question's \textit{Predicted} conversation, whereas a refused response is a failed turn.}

{\textbf{Roles of the models.} Table~\ref{tab:roles} gives the role of each model. The roles overlap only for the three council models, which are also tested systems. Each of them therefore takes part in labelling its own system's clarifications and in grading its own system's feedback. It also answers questions that its own model wrote. Appendix~\ref{app:score_checks} reports how often these overlaps decide an outcome and whether a council system scores higher on its own model's questions. Appendix~\ref{app:validation_detail} reports what each council member changes in validation.}

\begin{table}[t]
\centering
\small
\setlength{\tabcolsep}{3.5pt}
\begin{tabular}{@{}lccccc@{}}
\toprule
\textbf{Models} & \textbf{Write} & \textbf{Validate} & \textbf{User} & \textbf{Grade} & \textbf{Tested} \\
\midrule
Council (3) & $\bullet$ & $\bullet$ & $\bullet$ & $\bullet$ & $\bullet$ \\
Others (7) & \textendash & \textendash & \textendash & \textendash & $\bullet$ \\
\bottomrule
\end{tabular}
\caption{{Roles of the models. The council is GPT-OSS, Qwen3.5, and Nemotron, and the others are the seven other tested systems, Claude Haiku~5.5 among them. Validation votes leave out the writer of the question. The simulated user labels each clarification by a vote of the three council models and answers through one member drawn per turn. In the Qwen-only setting (Appendix~\ref{app:qwen_only}), Qwen3.5 alone writes, validates, and simulates the user, and the three council models still grade feedback.}}
\label{tab:roles}
\end{table}

{\textbf{Column descriptions.} For BIRD, every prompt that shows the schema includes each table's column descriptions as distributed with BIRD. These give each column's expanded name, a description, and a value description, and the prompt places them between the table definition and its sample rows. Spider provides no descriptions. We use the descriptions unedited, although some are inaccurate. In the \textit{financial} database, for example, dates described as YYMMDD are stored as YYYY-MM-DD, and loan amounts are described in US dollars although the bank is Czech.}

{\textbf{Decoding protocol.} Every open model answers in two steps. It first writes a free answer, reasoning included, with the sampling parameters recommended by its developers (Table~\ref{tab:sampling}). This answer has a budget of 8{,}192 tokens for every role, and we close reasoning that reaches the budget so that the model states its answer. The model then copies this answer into the required JSON format, continuing its own output with greedy decoding constrained to the format. An answer that already fits the format is used as it is. In the classification phase, the format admits only the subcategory labels. We constrain only the copy because \citet{tam-etal-2024-speak} find that format restrictions during generation can degrade reasoning, whereas constrained output helps classification tasks such as our labels. A council call makes up to three attempts when a copy fails or an answer is empty, and a call without a valid answer stops the run. Tested systems, by contrast, receive a single attempt per turn. Requests use a fixed seed, and sample $i$ of try $r$ uses the seed plus $rn+i$, where $n$ is the number of samples per request. Appendix~\ref{app:prompts} gives the prompt of every role.}

\begin{table}[t]
\centering
\small
\setlength{\tabcolsep}{4pt}
\begin{tabular}{@{}lcccl@{}}
\toprule
\textbf{Model} & \textbf{Temp.} & \textbf{Top-$p$} & \textbf{Top-$k$} & \textbf{Pen.} \\
\midrule
GPT-OSS & 1.0 & 1.0 & -- & -- \\
Qwen3.5 & 1.0 & 0.95 & 20 & pres.\ 1.5 \\
Nemotron & 1.0 & 0.95 & -- & -- \\
Qwen3.8-Flash & 1.0 & 0.95 & 20 & -- \\
Qwen2.5-32B & 0.7 & 0.8 & 20 & rep.\ 1.05 \\
Qwen2.5-Coder-32B & 0.7 & 0.8 & 20 & rep.\ 1.05 \\
Gemma-3-27B & 1.0 & 0.95 & 64 & -- \\
Mistral-Small-24B & 0.15 & -- & -- & -- \\
Qwen2.5-7B & 0.7 & 0.8 & 20 & rep.\ 1.05 \\
{Claude Haiku 5.5} & {--} & {--} & {--} & {--} \\
\bottomrule
\end{tabular}
\caption{{Sampling parameters of the free answer, as recommended by each model's developers (Pen.: presence or repetition penalty). Parameters that are not listed keep vLLM's defaults.} {Claude Haiku~5.5 takes no sampling values.}}
\label{tab:sampling}
\end{table}

{\textbf{Compute environment.} We ran all experiments with open models on a high-performance computing cluster. Data generation and the benchmark used nodes with two NVIDIA H200 NVL GPUs and vLLM~0.32.0, with tensor parallelism~2 and a context of 32{,}768 tokens. Each engine uses 85\% of each GPU's memory, except that every engine uses {75\%} in the runs of the five smaller tested systems, which run beside the council's engines. Qwen3.5 and Nemotron decode speculatively with their own multi-token-prediction heads. Between uses, we keep each council engine in host memory through vLLM's sleep mode instead of restarting it.}

{\textbf{Reproducibility.} Sampled runs cannot be reproduced exactly, because the outputs of vLLM depend on how requests are batched. Moreover, Nemotron's answers reproduce only in distribution even with fixed seeds. In addition, a system query whose execution time is close to the 3-second scoring limit may time out on a busy machine}{. A query that times out therefore runs once more, and the second outcome counts}{. Every run records its code version, library versions, model revisions, hardware, and the hashes of its question files, databases, and column-description files.}

{\textbf{Settings and licenses.}} We performed no task-performance hyperparameter search; generation, validation, interaction, and decoding settings were fixed before evaluation. We release our original code under the MIT License {and the generated datasets and result files under CC BY-SA 4.0}. The released artifact does not redistribute the BIRD or Spider databases, or model weights; these must be obtained under their respective original terms.

\FloatBarrier

\section{Validation Stage Details}
\label{app:validation_detail}

{Table~\ref{tab:validation_stages} summarizes the ten stages. Appendix~\ref{app:val_rules} gives the rules that all stages share, and Appendix~\ref{app:val_stages} describes each stage in order. Appendix~\ref{app:val_members} measures how much the decisions depend on a single council member, and Appendix~\ref{app:val_writers} tests whether writers favor their own questions.}

\begin{table*}[t]
\centering
\small
\begin{tabularx}{\textwidth}{@{}lccc>{\raggedright\arraybackslash\leavevmode}Xrrrr@{}}
\toprule
& & & & & \multicolumn{2}{c}{{\textbf{Of entering (\%)}}} & \multicolumn{2}{c}{{\textbf{Of checked (\%)}}} \\
\cmidrule(lr){6-7} \cmidrule(lr){8-9}
\textbf{Stage} & \textbf{Ans.} & \textbf{Amb.} & \textbf{Una.} & \textbf{Decided by} & \textbf{BIRD} & \textbf{Spider} & \textbf{BIRD} & \textbf{Spider} \\
\midrule
1 Duplicate removal & $\bullet$ & $\bullet$ & $\bullet$ & identical text, or identical value-masked SQL and hidden knowledge & 2.1 & 2.6 & 2.1 & 2.6 \\
2 SQL executability & $\bullet$ & $\bullet$ & \textendash & execution & 27.9 & 33.2 & 34.7 & 41.2 \\
3 Category conformance & $\bullet$ & $\bullet$ & $\bullet$ & vote & 7.4 & 7.4 & 7.4 & 7.4 \\
4 Ground truth satisfaction & $\bullet$ & $\bullet$ & \textendash & vote & 15.2 & 13.7 & 20.5 & 19.1 \\
5 Evidence necessity & $\bullet$$^{*}$ & \textendash & \textendash & execution, one witness & 2.5 & 2.2 & 46.0 & 47.9 \\
6 Information sufficiency & $\bullet$ & $\bullet$ & \textendash & execution, one witness & 25.4 & 19.6 & 36.9 & 29.5 \\
7 Unsolvability verification & \textendash & \textendash & $\bullet$ & execution and a vote on each attempt, one witness & 4.7 & 3.1 & 11.3 & 7.5 \\
8 Feedback quality & \textendash & \textendash & $\bullet$ & vote & 1.5 & 1.3 & 4.0 & 3.3 \\
9 Category consistency & \textendash & $\bullet$ & $\bullet$ & vote against each competitor & 2.3 & 2.1 & 2.5 & 2.3 \\
10 Style conformance & $\bullet$ & $\bullet$ & $\bullet$ & vote & 2.0 & 1.6 & 2.0 & 1.6 \\
\bottomrule
\end{tabularx}
\caption{{The ten validation stages. \textit{Vote} means a majority of the council members other than the writer, with a tie removing the question (at Stage~7, rejecting the attempt); \textit{one witness} means that a single such member decides.} {The last four columns give the share of the questions that the stage removes in the generation of each dataset, out of all questions entering the stage and out of those that it checks.} {\textit{Ans.}, \textit{Amb.}, and \textit{Una.} stand for answerable, ambiguous, and unanswerable questions; $\bullet$ marks the questions that the stage checks, and $^{*}$ only those with evidence.}}
\label{tab:validation_stages}
\end{table*}

\subsection{{Rules Shared by Several Stages}}\label{app:val_rules}

{\textbf{An ambiguous question passes or fails as a whole.} Its two interpretations are separate NLQ objects, but a stage that removes one also removes the other (Section~\ref{sec:question_generation}). Stages~3 and~9 judge the two interpretations together, in one prompt that shows both with their hidden knowledge and Ground Truth SQL. Stage~10 judges their shared question once, and Stage~1 removes a repeated ambiguous question as a whole.}

{\textbf{No model judges its own question.} We call the council member that wrote a question its \textit{writer}. In Stages~3 to~10, the writer still votes or writes its SQL attempt, but we only record its output and do not count it. The two other council members therefore make the decisions of these stages. A 1--1 tie between them removes the question in Stages~3, 4, 8, and~10 and makes it lose the comparison in Stage~9. In Stage~7, by contrast, a tie rejects the attempt under judgment and so does not remove the question. The simulated user of the benchmark instead counts every council member's vote (Appendix~\ref{app:user_sim}).}

{\textbf{Execution decides wherever it can.} Stages~5 to~7 ask each council model to write SQL with the prompt of Section~\ref{sec:question_generation}, which also produces the Ground Truth SQL. We count an attempt only if its SQL runs within 3 seconds, the limit that EX scoring sets for each query. The SQL must also have result columns and return at least one row with a non-\texttt{NULL} value. Finally, it must read at least one table of the database. A query that returns a constant message, such as \texttt{SELECT 'This question cannot be answered'}, thus answers nothing.}

{\textbf{What the judges see.} The judges of Stages~3, 4, 7, and~9 see the result of each SQL query with its row count, its column names, its first five rows, and the size of every table that it reads. They can thus notice, for instance, a \textit{``top 10\%''} that keeps only three rows. We truncate text values longer than 100 characters.}

\subsection{{The Ten Stages}}\label{app:val_stages}

{\textbf{Before Stage~1.}} {We remove generated objects whose question, hidden knowledge, or feedback merely copies a placeholder of the output template.}

{\textbf{Stage~1: duplicate removal.} Stage~1 removes a question that repeats a question already kept on the same database. A question repeats a kept one if it has the same text, or if it has the same value-masked Ground Truth SQL~\cite{OMNI-SQL} and the same hidden knowledge. Value masking replaces the strings and numbers of a query with a placeholder. We compare Improper Questions across databases, because they do not depend on the database. Only kept questions count, so a removed question never causes the removal of a later one.}

{\textbf{Stage~2: SQL executability.} Stage~2 keeps a solvable object only if its Ground Truth SQL runs within 1.5 seconds, half of the scoring limit, and returns at least one row with a non-\texttt{NULL} value. This last requirement matters because an aggregate over an empty selection returns a single \texttt{NULL} row, which any query with an empty selection would match. We also verify that the query returns at most 25 columns, since a system must reproduce every one of them. The released code takes this limit as an option (\texttt{-{}-max\_\allowbreak reference\_\allowbreak columns}), so that it can be raised for larger databases. The query also must not average or sum a column of words, which SQLite silently reads as zeros. It also must not read the current date or time, because that value changes from day to day. Moreover, the query must determine which rows it returns. Otherwise, a \texttt{LIMIT} that cuts through rows tied on every \texttt{ORDER BY} key would let a correct answer fail by chance. Rows that change when SQLite reads the tables in reverse order would do the same. Appendix~\ref{app:equivalence} lists the cases that this check does not cover. Finally, the equivalence check must tell apart the Ground Truth SQL of the two interpretations of an ambiguous question. Each of the two queries, used as an answer, must therefore fail against the other.}

{\textbf{Stage~3: category conformance.}} {Stage~3 asks whether a question conforms to its category, given the category's definition and examples.}

{\textbf{Stage~4: ground truth satisfaction.}} {Stage~4 asks whether each Ground Truth SQL answers its question, given the hidden knowledge for ambiguous interpretations.}

{\textbf{Stage~5: evidence necessity.} Stage~5 checks that an answerable question with evidence needs that evidence. Each council member writes SQL from the question without the evidence. The stage removes the question as soon as one model other than the writer still returns the Ground Truth result, under the equivalence check of Appendix~\ref{app:equivalence}.}

{\textbf{Stage~6: information sufficiency.} Stage~6 gives each council member the information that a system receives or must obtain during the benchmark. This information is the evidence and, for an ambiguous object, its hidden knowledge. The stage keeps a solvable object only if at least one model other than the writer reaches its Ground Truth result with this information. We thus verify that a clarification can lead a system to the intended answer. We also remove Ground Truth SQL that no model reaches, which is often erroneous. The writer used the same prompt to write the Ground Truth SQL. Stage~6 therefore samples with a different seed, so that the writer's recorded attempt does not merely reproduce its own Ground Truth SQL.}

{\textbf{Stage~7: unsolvability verification.} Stage~7 checks that an unanswerable question indeed cannot be answered. Each council member writes SQL from the question and its evidence, and the judges described below vote on each attempt. The stage removes the question as soon as they accept one counted attempt. The judges of Stage~7 see each attempt as a neutral candidate, without the explanation of why the question should be unanswerable. They apply the criteria of Stage~4 with two additions. First, a query may depend on a value or definition that the database, the question, and its evidence do not provide, such as an assumed threshold. Such a query does not answer the question. Second, rejecting a query that only approximates the request is preferable to accepting it.}

{\textbf{Stage~8: feedback quality.}} {Stage~8 asks whether the feedback of an unanswerable question correctly, specifically, and actionably explains why it cannot be answered.}

{\textbf{Stage~9: category consistency.} Stage~9 checks that an ambiguous or unanswerable question fits its own category better than any competing category. A competing category has the same type as the question's category and the same form. Two categories have the same form when their questions agree on whether they carry evidence, hidden knowledge, Ground Truth SQL, and paired interpretations. The stage keeps the question only if its own category wins the comparison against every competitor. Conflicting Knowledge is the only ambiguous category whose questions carry evidence, so this rule leaves it without competitors. Because judges tend to favor the option shown first, we draw the order of the two categories in Stage~9 from the content of each comparison. Moreover, a vote for the competing category counts only if the judge repeats it with the order swapped.}

{\textbf{Stage~10: style conformance.}} {Stage~10 gives each judge the description of the required style and asks whether the question follows it.}

\subsection{{How Much a Single Member Decides}}\label{app:val_members}

{Table~\ref{tab:council_disagreement} shows how much the council's decisions depend on a single member. A \textit{unit} is what a stage judges at once, which is an ambiguous question as a whole at Stages~3, 9, and~10 and a single question or interpretation otherwise. The council decides each unit by the rules of Appendices~\ref{app:val_rules} and~\ref{app:val_stages}. \textit{Removed} counts the units that the council removed, without the partner interpretations that Table~\ref{tab:validation_stages} also counts. \textit{Changed} counts the decisions that the rule of each stage, applied to all three votes or SQL attempts, the writer's included, would have reversed. The other columns compare the council with each member M deciding alone on the units that M did not write, that is, with the rule of each stage applied to M's vote or SQL attempt only. At Stage~7, the counted judges still vote on M's attempt. \textit{Removals kept} is the share of the council's removals that M alone would have kept, and \textit{Divergence} is the share of units on which M alone would have decided otherwise. Stage~9 passes its 268 Conflicting Knowledge questions without votes, because they have no competitor.}

{Divergence is at most 2.8\% at Stages~3 and~7 to~10 and reaches higher values at Stage~4. It is highest at Stages~5 and~6, where a member's own SQL is tested against the Ground Truth result. Many of the council's removals, however, rest on a single member, which means that the two counted members decided differently on them. At Stage~4, for instance, 647 of the 921 removed units rest on a single member. The strictest member changes with the stage: GPT-OSS alone would have kept the largest share of the council's removals at Stage~3 and the smallest at Stage~4, and Qwen3.5 the reverse.}

\begin{table*}[t]
\centering
\footnotesize
\setlength{\tabcolsep}{2.4pt}
\begin{tabular}{@{}lrrrrrrrrr@{}}
\toprule
& & & & \multicolumn{3}{c}{{\textbf{Removals kept (\%)}}} & \multicolumn{3}{c}{{\textbf{Divergence (\%)}}} \\
\cmidrule(lr){5-7} \cmidrule(lr){8-10}
\textbf{Stage} & \textbf{Units} & \textbf{Removed} & \textbf{Changed} & \textbf{GPT-OSS} & \textbf{Qwen3.5} & \textbf{Nemotron} & \textbf{GPT-OSS} & \textbf{Qwen3.5} & \textbf{Nemotron} \\
\midrule
3 Category conformance & 7{,}444 & 582 & 208 & 37.1 & 15.9 & 29.8 & 2.8 & 1.3 & 2.2 \\
4 Ground truth satisfaction & 7{,}252 & 921 & 547 & 10.6 & 75.8 & 37.9 & 1.7 & 7.0 & 4.7 \\
5 Evidence necessity & 427 & 200 & 29 & 15.4 & 22.7 & 23.3 & 7.3 & 11.7 & 9.6 \\
6 Information sufficiency$^{\dagger}$ & 5{,}613 & 1{,}285 & 510 & 8.3 & 14.5 & 13.5 & 6.7 & 11.1 & 10.1 \\
7 Unsolvability verification & 2{,}698 & 253 & 137 & 21.3 & 27.0 & 32.2 & 2.2 & 2.7 & 2.6 \\
8 Feedback quality & 2{,}445 & 89 & 40 & 31.9 & 45.1 & 31.7 & 0.9 & 2.0 & 1.1 \\
9 Category consistency & 3{,}950 & 124 & 50 & 11.1 & 49.2 & 37.6 & 0.4 & 1.3 & 1.4 \\
10 Style conformance & 4{,}377 & 76 & 66 & 76.5 & 52.9 & 6.0 & 1.4 & 0.9 & 0.1 \\
\bottomrule
\end{tabular}

\vspace{6pt}
\setlength{\tabcolsep}{3pt}
\begin{tabular}{@{}lrrrrrrrrr@{}}
\toprule
& \multicolumn{3}{c}{{\textbf{Writer GPT-OSS}}} & \multicolumn{3}{c}{{\textbf{Writer Qwen3.5}}} & \multicolumn{3}{c}{{\textbf{Writer Nemotron}}} \\
\cmidrule(lr){2-4} \cmidrule(lr){5-7} \cmidrule(lr){8-10}
\textbf{Stage} & \textbf{Own} & \textbf{Qwen3.5} & \textbf{Nemotron} & \textbf{Own} & \textbf{GPT-OSS} & \textbf{Nemotron} & \textbf{Own} & \textbf{GPT-OSS} & \textbf{Qwen3.5} \\
\midrule
3 & 96.2 & +1.1 & \textminus{}0.1 & 93.5 & \textminus{}0.3 & +0.9 & 94.3 & +0.9 & \textminus{}0.3 \\
 &  & {\scriptsize [\textminus{}0.1, +2.3]} & {\scriptsize [\textminus{}1.1, +1.0]} &  & {\scriptsize [\textminus{}1.4, +0.8]} & {\scriptsize [\textminus{}0.2, +2.1]} &  & {\scriptsize [\textminus{}0.1, +2.0]} & {\scriptsize [\textminus{}1.5, +1.0]} \\[1pt]
4 & 95.4 & +7.3 & +4.9 & 95.4 & +1.8 & +4.2 & 95.3 & +2.9 & +0.5 \\
 &  & {\scriptsize [+5.7, +8.9]} & {\scriptsize [+3.3, +6.5]} &  & {\scriptsize [\textminus{}0.2, +3.8]} & {\scriptsize [+2.8, +5.6]} &  & {\scriptsize [+1.0, +4.8]} & {\scriptsize [\textminus{}0.7, +1.7]} \\[1pt]
5 & 53.9 & \textminus{}8.2 & \textminus{}5.1 & 69.5 & \textminus{}3.4 & \textminus{}6.6 & 50.3 & \textminus{}5.4 & \textminus{}2.2 \\
 &  & {\scriptsize [\textminus{}19.0, +2.1]} & {\scriptsize [\textminus{}14.5, +4.2]} &  & {\scriptsize [\textminus{}13.7, +6.6]} & {\scriptsize [\textminus{}16.1, +3.1]} &  & {\scriptsize [\textminus{}14.9, +4.4]} & {\scriptsize [\textminus{}12.1, +7.6]} \\[1pt]
6 & 75.9 & +9.0 & +8.2 & 75.3 & +9.2 & +10.1 & 78.1 & +7.3 & +6.5 \\
 &  & {\scriptsize [+6.1, +12.1]} & {\scriptsize [+5.1, +11.2]} &  & {\scriptsize [+6.2, +12.3]} & {\scriptsize [+7.2, +13.0]} &  & {\scriptsize [+4.2, +10.3]} & {\scriptsize [+3.6, +9.3]} \\[1pt]
7 & 93.4 & +0.7 & +0.1 & 95.8 & +0.9 & +1.5 & 91.4 & +0.3 & \textminus{}0.3 \\
 &  & {\scriptsize [\textminus{}1.6, +2.9]} & {\scriptsize [\textminus{}1.9, +2.2]} &  & {\scriptsize [\textminus{}1.2, +3.0]} & {\scriptsize [\textminus{}0.3, +3.3]} &  & {\scriptsize [\textminus{}1.9, +2.5]} & {\scriptsize [\textminus{}2.4, +1.8]} \\[1pt]
8 & 95.9 & \textminus{}1.1 & \textminus{}0.9 & 99.2 & \textminus{}0.2 & \textminus{}0.4 & 97.1 & +0.4 & +0.5 \\
 &  & {\scriptsize [\textminus{}2.8, +0.6]} & {\scriptsize [\textminus{}2.5, +0.7]} &  & {\scriptsize [\textminus{}1.6, +1.1]} & {\scriptsize [\textminus{}1.9, +1.2]} &  & {\scriptsize [\textminus{}1.0, +1.7]} & {\scriptsize [\textminus{}1.2, +2.2]} \\[1pt]
9 & 96.8 & \textminus{}0.8 & +0.4 & 98.5 & +1.7 & +0.5 & 98.8 & \textminus{}0.5 & +0.7 \\
 &  & {\scriptsize [\textminus{}1.8, +0.3]} & {\scriptsize [\textminus{}0.8, +1.6]} &  & {\scriptsize [+0.5, +2.9]} & {\scriptsize [\textminus{}0.5, +1.6]} &  & {\scriptsize [\textminus{}1.6, +0.7]} & {\scriptsize [\textminus{}0.3, +1.6]} \\[1pt]
10 & 99.9 & \textminus{}0.6 & \textminus{}0.1 & 100.0 & +0.8 & +0.3 & 96.9 & \textminus{}0.8 & \textminus{}0.3 \\
 &  & {\scriptsize [\textminus{}1.2, +0.1]} & {\scriptsize [\textminus{}1.0, +0.8]} &  & {\scriptsize [+0.1, +1.4]} & {\scriptsize [\textminus{}0.6, +1.2]} &  & {\scriptsize [\textminus{}1.9, +0.4]} & {\scriptsize [\textminus{}1.5, +0.9]} \\[1pt]
\bottomrule
\end{tabular}
\caption{{What each council member changes in Stages~3 to~10, both datasets together. Top: what each member M, deciding alone on the units that it did not write, would have changed, with the columns defined in the text. Bottom: each writer's keep rate on its own units (\textit{Own}) and, in points, that rate minus the keep rate of each other member on the same units, adjusted by subtracting the same two members' difference on the units of the third member, with a 95\% paired bootstrap interval (10{,}000 rounds). $^{\dagger}$At Stage~6, the share of the council's keeps that M alone would have removed.}}
\label{tab:council_disagreement}
\end{table*}

\subsection{{Whether Writers Favor Their Own Questions}}\label{app:val_writers}

{Because we record the writer's uncounted votes and SQL attempts (Appendix~\ref{app:val_rules}), we can test whether writers favor their own questions. The lower part of Table~\ref{tab:council_disagreement} compares each writer's keep rate on its own units with each other member's rate on the same units. A raw difference would also reflect how lenient the two members are in general, so the table subtracts their difference on the units of the third member, which neither of them wrote. A positive value thus means that the writer keeps its own units more often than the two members' general leniency explains. Writers do so at Stage~4, where members judge a Ground Truth SQL, and at Stage~6, where a member's own SQL must reach the Ground Truth result. At Stage~4, four of the six 95\% intervals lie above zero, and at Stage~6 all six do. Elsewhere, only two of the 30 intervals at Stages~3 and~7 to~10 exclude zero, and none at Stage~5. Counting the writer's vote would have changed 1{,}587 of the 33{,}938 decisions of Stages~3 to~10 (column \textit{Changed}). The writers' recorded votes thus support excluding the writer's vote.}

\section{Dataset Composition and Human Evaluation}
\label{app:dataset_composition}
\begin{figure}[htbp]
\centering
\includegraphics[width=\columnwidth]{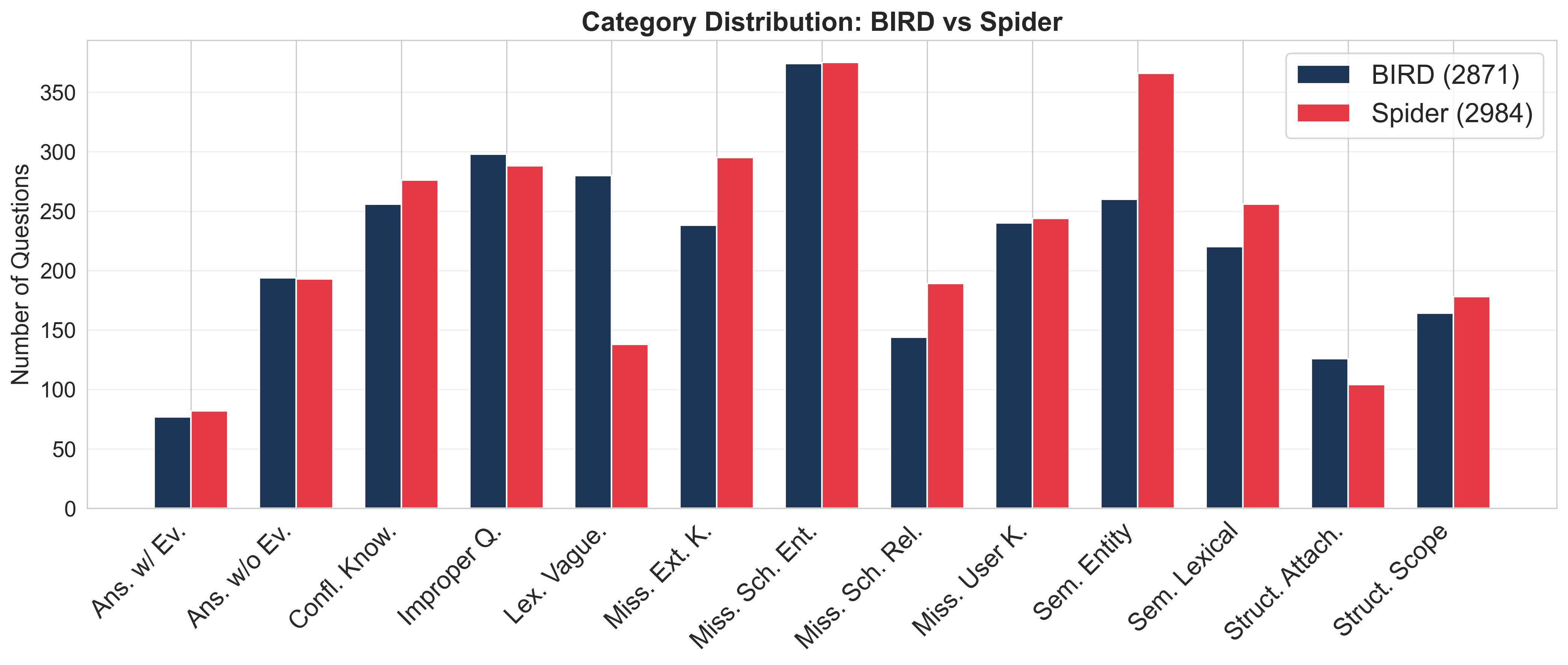}
\caption{{Subcategory} distribution of generated questions for {the} ABISS-BIRD and ABISS-Spider datasets after the validation pipeline. {Each interpretation of an ambiguous question counts as one question.}}
\label{fig:category_comparison}
\end{figure}

Figure~\ref{fig:category_comparison} reports the subcategory-level distribution of the
generated questions after validation, and Table~\ref{tab:human_eval} the results
of the human evaluation described in Section~\ref{sec:dataset_analysis}.

{\textbf{Composition and coverage.} Of the 11 databases of ABISS-Spider, nine come from Spider's training split and two from its test split. If we count each ambiguous pair once rather than each interpretation as in Section~\ref{sec:dataset_analysis}, ABISS-BIRD holds 2{,}098 distinct questions and ABISS-Spider 2{,}203. Unanswerable questions, not ambiguous ones, then form the majority, with 50.2\% of ABISS-BIRD and 52.1\% of ABISS-Spider. Ambiguous questions follow with 36.8\% and 35.5\%, and answerable ones with 12.9\% and 12.5\%. Counting each interpretation as one question instead, a database contributes between 211 and 332 questions to ABISS-BIRD and between 192 and 319 to ABISS-Spider. Of the 1{,}716 combinations of subcategory, database, style, and requested difficulty, 1{,}175 retain at least one question in ABISS-BIRD and 1{,}191 in ABISS-Spider, and every subcategory still covers all 11 databases, all three styles, and all four requested difficulty levels. Each writer model contributes about a third of the questions, between 31.5\% and 34.4\% of each dataset. The balanced subsets of Section~\ref{sec:experiments} stay close to this split on ABISS-BIRD, with 35.6\%, 31.7\%, and 32.7\% for GPT-OSS, Qwen3.5, and Nemotron. On ABISS-Spider, by contrast, Nemotron's share falls to 26.8\%, and those of GPT-OSS and Qwen3.5 rise to 35.8\% and 37.3\%.}

{\textbf{Difficulty.} We check whether the requested difficulty level carries through validation and into the Ground Truth SQL (Table~\ref{tab:difficulty}). First, validation removes harder requests more often. On both datasets, about half of the candidates requested as simple pass, against about a quarter of those requested as highly complex. Second, among the evaluated solvable questions, which are the solvable questions of the balanced subsets, the level measured from the Ground Truth SQL matches the requested one for {55.0\%} on ABISS-BIRD and {58.1\%} on ABISS-Spider. Where the two levels differ, the SQL is more often simpler than requested than harder, with a lower level for {25.9\%} and {25.7\%} and a higher one for {19.2\%} and {16.1\%}. Third, most evaluated solvable questions are simple or moderate by their SQL, although the candidates were requested equally often at each level.}

\begin{table}[htbp]
\centering
\small
\begin{tabular}{@{}lrrrr@{}}
\toprule
& \multicolumn{2}{c}{{Retained (\%)}} & \multicolumn{2}{c}{{Measured (\%)}} \\
\cmidrule(lr){2-3}\cmidrule(l){4-5}
Level & BIRD & Spider & BIRD & Spider \\
\midrule
Simple & 49.8 & 53.3 & 39.0 & 45.3 \\
Moderate & 38.0 & 36.6 & 45.2 & 40.1 \\
Complex & 30.4 & 33.4 & 7.2 & 5.7 \\
Highly complex & 26.9 & 27.4 & 8.6 & 8.9 \\
\bottomrule
\end{tabular}
\caption{{Difficulty of the generated questions. \textit{Retained} is the share of the candidate questions requested at each level that pass validation (1{,}980 candidates per level and dataset, each interpretation of an ambiguous candidate counted as one). \textit{Measured} is the distribution of the level that keyword rules assign to the Ground Truth SQL, over the solvable questions of the balanced subsets.}}
\label{tab:difficulty}
\end{table}

{\textbf{Annotation instructions.} The two annotators jointly inspected each NLQ together with the database schema, any accompanying evidence, and the assigned subcategory and linguistic style. They also inspected either the Ground Truth SQL and its execution result or, for an unanswerable question, the feedback provided. They first classified the NLQ as answerable, ambiguous, or unanswerable. They then marked each applicable quality dimension as correct or incorrect. They judged whether the NLQ was plausible and natural for the database domain, whether it conformed to the assigned style, and whether the assigned subcategory correctly described the question. For an ambiguous question, they also judged whether the disambiguation selected a plausible interpretation without changing the question's core intent, whether both readings were natural, and whether its ambiguity matched the assigned category. For an answerable or ambiguous question, they judged whether the SQL semantically answered the resolved question. For an unanswerable question, they judged whether the feedback accurately and actionably explained why the question could not be answered, and whether it was truly unanswerable.}

\begin{table}[htbp]
\centering
\footnotesize
\setlength{\tabcolsep}{3.5pt}
\renewcommand{\arraystretch}{1.2}
\begin{tabular}{@{}lrrrr@{}}
\toprule
\textbf{Quality Dimension} & \textbf{BIRD} & \textbf{Spider} & \textbf{Both} & $n$ \\
\midrule
Question Realism & 93.3 & 96.7 & 95.0 & 60 \\
Style Conformance & 100.0 & 100.0 & 100.0 & 65 \\
Category Correctness & 93.8 & 87.7 & 90.8 & 65 \\
{Assigned Type Recognized} & 93.8 & 87.7 & 90.8 & 65 \\
{Ambiguity Matches Category} & 94.3 & 97.1 & 95.7 & 35 \\
{Truly Unanswerable} & 95.0 & 100.0 & 97.5 & 20 \\
Disambiguation Correctness & 88.6 & 87.1 & 87.9 & 70 \\
{Both Readings Natural} & 91.4 & 77.1 & 84.3 & 35 \\
SQL Correctness & 73.8 & 92.5 & 83.1 & 80 \\
Feedback Accuracy & 85.0 & 100.0 & 92.5 & 20 \\
\bottomrule
\end{tabular}
\caption{{Human evaluation on 130 questions, 5 per subcategory and dataset, in \% of the applicable judgments; $n$ is their number in each dataset. Disambiguation Correctness, Both Readings Natural, and Ambiguity Matches Category apply only to ambiguous questions; SQL Correctness to answerable and ambiguous questions; Feedback Accuracy and Truly Unanswerable to unanswerable questions; and Question Realism excludes the Improper Question category. Assigned Type Recognized: the type that the annotators read from the text equals the assigned one.}}
\label{tab:human_eval}
\end{table}

\FloatBarrier
\section{Simulated User Agent}
\label{app:user_sim}

{\textbf{Two-stage user simulation.} The simulated user of Section~\ref{sec:simulated_user} answers each clarification in two stages. It first assigns a relevancy label (Relevant, Technical, or Irrelevant), and then gives the answer that this label calls for. In the first stage, each council member receives the NLQ and the conversation history. It also receives the hidden knowledge of an ambiguous question and the structured SQL fragments extracted from the Ground Truth SQL of a solvable (answerable or ambiguous) one. Each member then predicts the relevancy label and, for a Technical label, the fragments that the clarification concerns. The question type limits the labels that a member may choose. Answerable questions allow only Technical or Irrelevant. On them, a question about which columns or tables to use counts as Irrelevant, since the NLQ already states what it asks for. Ambiguous questions allow all three labels. Clarifications on unanswerable questions bypass the council's vote and receive the label Irrelevant. Every council member votes, including the model of the tested system when it belongs to the council. We keep its votes and only flag them, since excluding them would give each tested council member a different simulated user. The generation pipeline, by contrast, excludes the writer of a question (Appendix~\ref{app:validation_detail}).}

{In the second stage, a Technical label whose voters select no usable fragment becomes Irrelevant, so it receives the fixed refusal sentence of Section~\ref{sec:simulated_user}, like every Irrelevant label. Under a Relevant or Technical label, the first council member in an order drawn with a seed from the question and the turn writes the answer. The seed makes this order the same for every tested system. A Technical answer draws on the union of the SQL fragments that the Technical voters selected. The prompt asks for a statement, never a question. We check each answer for leakage, i.e.\ for SQL keywords written in capitals, \texttt{table.column} references, and table or column identifiers of the database that contain an underscore or an internal capital. A leaking answer is sampled again, and after three leaking attempts the next member in the drawn order answers. If every member leaks, we use the attempt with the fewest leaks.}

{\textbf{Manual audit.} To assess the quality of the simulated user, we manually audited {220} simulated-user turns from the benchmark runs on both datasets. We drew the turns from {51} strata, with at least one turn per stratum. Each stratum is one combination of dataset, tested system inside or outside the council, label, unanimous or split vote, and question type. We then weighted the results back to all {31{,}891} turns on answerable and ambiguous questions of the {nine} open systems' runs. Two annotators jointly judged the label of each turn and, on the 112 Relevant and Technical replies, the five other dimensions of Table~\ref{tab:user_audit}. For each turn, they inspected the NLQ, the conversation history, the simulated user's label and answer, and the available grounding context. They judged whether the label was correct, and whether the answer was grounded in the permitted information and stated everything that this information settled. They also judged whether the answer avoided revealing SQL or unnecessary schema details, and whether its style was natural and consistent with the requested style.}

{The relevancy label is the weakest dimension of Table~\ref{tab:user_audit}, and its errors concentrate where the council's vote is split. Split votes occur on {13.6\%} of all turns but account for {4.0} of the {5.4} error points. Only {70.2\%} of them give a correct label, against {98.4\%} of the unanimous ones. Irrelevant labels are the least reliable, with {62.9\%} correct, and account for {3.1} points although they occur on {8.3\%} of the turns. Technical labels occur on {7.6\%} and account for {1.2} points, with {83.8\%} correct. The {0.7\%} of turns whose Technical label became Irrelevant for lack of a usable fragment account for {0.2} points, with {66.7\%} correct. Relevant labels, whose answers draw on the hidden knowledge, are {99.0\%} correct. The simulated user thus rarely gives the hidden knowledge to a clarification that does not target it. Most of its label errors are instead refusals that withhold a deserved answer.}

\begin{table}[ht]
\centering
\small

\setlength{\tabcolsep}{16 pt}

\renewcommand{\arraystretch}{1.3}
\begin{tabular}{@{}lc@{}}
\toprule
\textbf{Dimension} & \textbf{Accuracy} \\
\midrule
Relevancy-label correctness & {94.6\%} \\
Answer-grounding correctness & {97.9\%} \\
{Answer completeness} & {99.0\%} \\
Absence of leakage & {100.0\%} \\
{Natural style} & {99.8\%} \\
{Consistency with the requested style} & {100.0\%} \\
\bottomrule
\end{tabular}
\caption{Manual audit of the simulated user agent {on 220 sampled turns of both datasets, weighted to all 31{,}891 turns on answerable and ambiguous questions of the nine open systems. Grounding, completeness, leakage, and style are judged on the 112 Relevant and Technical replies.}}
\label{tab:user_audit}
\end{table}

\FloatBarrier

\section{Relaxed SQL Equivalence Criteria}
\label{app:equivalence}

{We use the equivalence check in Stages~2, 5, and~6 of the validation pipeline and in the Execution Accuracy (EX) metric (Section~\ref{sec:metrics}). Following the relaxed semantic equivalence of \citet{Floratou2024NL2SQLIA}, it compares the results of two queries rather than their text. Two results are equivalent when (1) they have the same number of rows, (2) every column of the Ground Truth result is matched by content to a distinct column of the other result, and (3) their rows agree as multisets, i.e.\ the same rows occur equally often, in any order. Because of the second condition, column names and column order do not matter, and additional columns are accepted. If the Ground Truth SQL has a top-level \texttt{ORDER BY}, the order also counts, but only where the \texttt{ORDER BY} keys tell rows apart. Rows that tie on every key form a block. The blocks must appear in the order of the Ground Truth, but the rows within a block may appear in any order. For example, on the schema of Figure~\ref{fig:university_schema}, under \texttt{ORDER BY grade DESC}, two enrollments with the same grade may swap places, but one with a lower grade may not come first. When the keys cannot be read from the query (a compound query, \texttt{SELECT *}, \texttt{DISTINCT}, or \texttt{OFFSET}), we read them from the result if each key names one of its columns, and otherwise every row forms its own block, so the whole order must match. Floating-point values match after rounding to ten significant digits, but never more coarsely than whole units. A text that spells a plain number equals that number, so the text \texttt{'42'} equals the integer 42. Integers, other texts, and \texttt{NULL} must be equal.}

{Stage~2 checks that a Ground Truth SQL determines which rows it returns (Appendix~\ref{app:validation_detail}), but this check does not cover every case. It misses ties inside the \texttt{ORDER BY} of a window function and a bare column next to a \texttt{GROUP BY} or next to a tied \texttt{MIN} or \texttt{MAX}. It also misses a \texttt{LIMIT} inside a subquery or in a query whose \texttt{ORDER BY} keys cannot be read, as listed above. SQLite resolves these cases in a fixed but unspecified way, so a correct answer may occasionally fail on them.}

\FloatBarrier
\section{{Metric Definitions and Intervals}}
\label{app:metrics}

{We report Recognition (Rec.), Classification (Cls.), Execution Accuracy (EX), Feedback Accuracy (FB), Answer Incorporation Rate (AIR), and Turns to Relevant (TTR), each computed per run, i.e.\ for one system, dataset, and category usage mode, over the run's set $Q$ of questions. For $q\in Q$, let $t(q)\in\{\mathrm{ans},\mathrm{amb},\mathrm{un}\}$ and $c(q)$ be its type and subcategory, $\hat t(q)$ and $\hat c(q)$ their predictions in the classification phase, $a(q)\in\{\mathrm{sql},\mathrm{fb}\}$ the system's final response, $\hat s(q)$ its SQL, and $s(q)$ the Ground Truth SQL of the interpretation held by the simulated user. With $Q_{\mathrm{EX}}=\{q: t(q)\neq\mathrm{un} \lor a(q)=\mathrm{sql}\}$ and $Q_{\mathrm{FB}}=\{q: t(q)=\mathrm{un} \lor a(q)=\mathrm{fb}\}$,
\begin{align*}
\mathrm{Rec.} &= \textstyle\sum_{q\in Q} [\hat t(q)=t(q)]\,/\,|Q|,\\
\mathrm{Cls.} &= \textstyle\sum_{q\in Q} [\hat c(q)=c(q)]\,/\,|Q|,\\
\mathrm{EX} &= \textstyle\sum_{q\in Q_{\mathrm{EX}}} e(q)\,/\,|Q_{\mathrm{EX}}|,\\
\mathrm{FB} &= \textstyle\sum_{q\in Q_{\mathrm{FB}}} f(q)\,/\,|Q_{\mathrm{FB}}|,
\end{align*}
where $[\cdot]$ is 1 when its condition holds and 0 otherwise, $e(q)=[t(q)\neq\mathrm{un} \land \hat s(q)\equiv s(q)]$, and $f(q)=[t(q)=\mathrm{un} \land a(q)=\mathrm{fb} \land g(q)]$. Here $\equiv$ is the equivalence of Appendix~\ref{app:equivalence}, a missing $\hat s(q)$ is never equivalent, and $g(q)$ holds when a majority of the LLM council, whose members see the question{, its hidden knowledge, and the database schema} but not its subcategory, accepts the feedback. {A grader accepts feedback that names a problem that the hidden knowledge states and that by itself makes the question unanswerable. It rejects feedback that contradicts the hidden knowledge, states something false about the database, or is too vague to be useful.} Rec.\ and Cls.\ are computed in the \textit{Predicted} mode. SQL written for an unanswerable question counts as a failure in EX, and feedback given on an answerable or ambiguous question counts as a failure in both EX and FB. The denominators of EX and FB therefore vary across systems and modes. {The EX Ans.\ and EX Amb.\ columns of Table~\ref{tab:ablation} cover only answerable and only ambiguous questions, respectively.}}

{For an ambiguous conversation $q$, let $r(q)$ be the position, counting clarifications from~1, of the first clarification labeled Relevant, $R$ the ambiguous conversations that contain one, and $A\subseteq R$ those in which its reply is followed by another system turn $a'(q)$. Then
\begin{align*}
\mathrm{AIR} &= \textstyle\sum_{q\in A} [a'(q)\in\{\mathrm{sql},\mathrm{fb}\}]\,/\,|A|,\\
\mathrm{TTR} &= \textstyle\sum_{q\in R} r(q)\,/\,|R|,
\end{align*}
and TTR is reported with its reach rate, $|R|$ over the number of ambiguous conversations. Averages over systems are plain means of the unrounded per-run values, omitting runs whose denominator is empty.}

{\textbf{Intervals.} Running the benchmark several times would be costly, so we estimate the variation of each score from a single run by resampling its questions, similar to what the OpenAI Evals framework does~\cite{openai2023evals}. We draw 10{,}000 bootstrap samples of the questions~\cite{efron1993introduction}, with replacement within each dataset and subcategory. The two interpretations of an ambiguous question and all modes of a question are drawn together, so that differences between systems and between modes are paired, as in the paired bootstrap common in NLP~\cite{koehn-2004-statistical,dror-etal-2018-hitchhikers}. An interval is the middle 95\% of the drawn values, and we recompute averages over systems in each draw. We call one system or mode better than another only when the difference is significant after Holm's correction for multiple comparisons~\cite{holm1979simple}. The correction runs over the comparisons that one statement makes, for example the ten systems of a contrast between two modes, so that the chance of any false difference within the statement is at most 5\%.}

\FloatBarrier
\section{{Token Use and Cost}}
\label{app:cost}

{\textbf{Measurement and prices.} We report what it costs to generate the datasets and to evaluate one system. We take token counts and times from the logs of every model call in our runs. They include the tokens spent on candidates that a later stage removed. To turn tokens into dollars, we use the most common provider price on OpenRouter per million input and output tokens. For Nemotron, whose listings all differ, we use the middle of the listed prices. The input and output prices are thus \$0.15 and \$0.60 for GPT-OSS, \$0.26 and \$2.08 for Qwen3.5, and \$0.08 and \$0.45 for Nemotron. We call them the council models' provider prices.}

\begin{table*}[!htb]
\centering
\small
\setlength{\tabcolsep}{5pt}
\begin{tabular}{@{}lrrrrrr@{}}
\toprule
& & & & \multicolumn{2}{c}{\textbf{Per retained question}} & \\
\cmidrule(lr){5-6}
\textbf{Dataset} & \textbf{Questions} & \textbf{Input (cached)} & \textbf{Output} & \textbf{Input} & \textbf{Output} & \textbf{GPU-s} \\
\midrule
ABISS-BIRD & 2{,}871 & 455.1M (179.8M) & 84.3M & 158{,}527 & 29{,}353 & 20.7 \\
ABISS-Spider & 2{,}984 & 213.6M (57.8M) & 79.4M & 71{,}584 & 26{,}599 & 14.5 \\
Both & 5{,}855 & 668.7M (237.6M) & 163.6M & 114{,}217 & 27{,}949 & 17.6 \\
\midrule
ABISS-BIRD (Qwen-only) & 3{,}046 & 180.8M (61.9M) & 68.6M & 59{,}349 & 22{,}506 & 14.6 \\
\bottomrule
\end{tabular}
\caption{{Tokens and time of the generation. Input tokens in parentheses were served from vLLM's prefix cache. Time is GPU-seconds per retained question on two H200 GPUs. The last row is the Qwen-only generation of Appendix~\ref{app:qwen_only}.}}
\label{tab:tokens}
\end{table*}

{\textbf{Generation.} Table~\ref{tab:tokens} gives the input and output tokens and the time of the generation. Per retained question, ABISS-BIRD needs about twice the input of ABISS-Spider but a similar output, because its schemas, with column descriptions and sample values, are longer. At the council models' provider prices, the generation of both datasets costs \$350 through API providers, or \$0.060 per retained question. Qwen3.5 accounts for \$268 of this total, because it writes 100.1M of the output tokens, more than half, at the highest output price. On our own hardware, the generation took 28.6 H200 GPU-hours. At Together AI's on-demand price of \$5.99 per H200 GPU-hour, this time costs \$171, or \$0.029 per retained question. We also apply the list prices of frontier models, which charge cached input at the cache-hit price. The same token counts would then cost \$0.022 per question at Claude Haiku~5.5 or GPT-6 Luna prices. They would cost \$0.43 per question at Claude Sonnet~5.5 or GPT-6.1 Sol prices and \$0.86 at Claude Opus~5.5 prices. The Qwen-only generation of Appendix~\ref{app:qwen_only} costs \$0.062 per retained question through providers, slightly less than the \$0.069 of the three-model generation of ABISS-BIRD. It is cheaper because it needs fewer tokens per retained question, above all fewer input tokens, although Qwen3.5 writes all its output at the highest price.}

{\textbf{Evaluation.} Evaluating one system in the dialogue, the main run of the benchmark, takes 8{,}216 conversations on both datasets in all four modes. ABISS's own side, which runs the simulated user and the graders, uses 36.0M input and 17.0M output tokens per evaluated system, or 4{,}380 and 2{,}063 per conversation. At the council models' provider prices, these tokens cost \$35.5. They would cost \$11 at Claude Haiku~5.5 or GPT-6 Luna prices, \$219 at Claude Sonnet~5.5 or GPT-6.1 Sol prices, and \$438 at Claude Opus~5.5 prices. On the other side, an open tested system uses 66M to 97M input tokens in the dialogue, mostly the repeated prefix of the multi-turn dialogue. It writes 1.6M to 16.2M output tokens, the most for the reasoning models. Claude Haiku~5.5 as a tested system cost \$26.62 through Anthropic's API, for 202.5M input and 12.7M output tokens. This total covers the dialogue and the two control runs, the floor and the oracle, on both datasets. The dialogue alone cost \$13.82, for 105.2M input and 6.6M output tokens. On our hardware, the dialogue runs took about 5.8 GPU-hours per evaluated system, including the system's own inference.}

{\textbf{Caveats.} The providers of the council models list no cache price, so we charge cached input at the full price. Listed prices also differ between providers. At the lowest and the highest listed input and output prices, the generation would cost \$302 to \$558, and ABISS's side of one evaluation \$32.9 to \$55.7. The costs at frontier models' prices are extrapolations, because another model would use another tokenizer and produce reasoning of another length. For the Qwen-only generation, we count the prompt of each three-sample request once, because providers that return several samples per request bill it once. We read every price on 15 November 2026.}

\FloatBarrier
\section{Per-Subcategory Analysis}
\label{app:per_category}

\begin{figure*}[t]
\centering
\includegraphics[width=\textwidth]{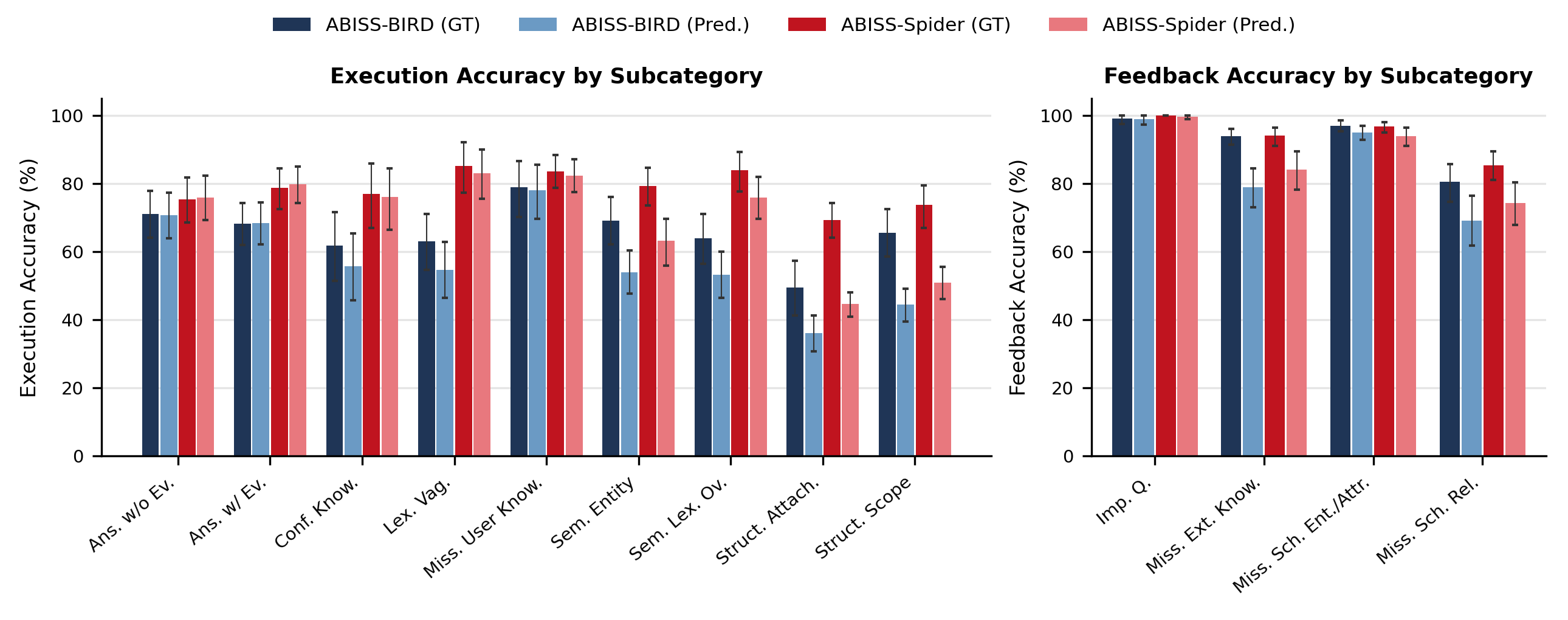}
\caption{Per-subcategory performance under the \textit{Ground Truth} (GT) and \textit{Predicted} modes, averaged across the evaluated models. Left: Execution Accuracy (EX) for answerable and ambiguous subcategories. Right: Feedback Accuracy (FB) for unanswerable subcategories. {Error bars give 95\% intervals.}}
\label{fig:per_category}
\end{figure*}

Figure~\ref{fig:per_category} compares per-subcategory performance under the \textit{Ground Truth} and \textit{Predicted} modes, averaged across the models.

{{\textit{Answerable questions}.} On both datasets, models reach similar EX on answerable questions with and without evidence. In nearly all questions with evidence, a second, unrelated piece of evidence accompanies the relevant one. Using supplied evidence is therefore not a bottleneck for current models.}

{\textit{Ambiguous questions}. In the \textit{Predicted} mode, models recognize Attachment and Scope Ambiguity, the two subcategories of Structure Ambiguity, as ambiguous in a quarter of the conversations or fewer. These two subcategories also have the lowest \textit{Predicted} EX among the ambiguous ones, and the correct label raises their EX by 13 to 25 points on both datasets. On average over the two datasets, Structure Ambiguity is therefore the hardest ambiguity to spot. Entity Ambiguity and Lexical Overlap are the next hardest to spot. Across these four subcategories, the four large systems (GPT-OSS, Qwen3.5, Nemotron, and Qwen3.8-Flash) recognize the ambiguity in 14\% to 51\% of \textit{Predicted} conversations. The other five (the open systems of 7B to 32B parameters) recognize it in 5\% to 20\%. On average over all ten models, Claude Haiku~5.5 included, models recognize these four subcategories as ambiguous in only 13\% to 38\% of \textit{Predicted} conversations. By contrast, models usually recognize Conflicting Knowledge, Lexical Vagueness, and Missing User Knowledge, and these three subcategories gain less from the label. The subcategories also show where the two datasets differ. In the \textit{Ground Truth} mode, ABISS-Spider's advantage over ABISS-BIRD on ambiguous questions (Appendix~\ref{app:further}) lies mostly in Lexical Vagueness, Lexical Overlap, and Attachment Ambiguity. On Lexical Overlap, this advantage partly comes from hedged answers, which return the columns of both interpretations (Appendix~\ref{app:score_checks}).}

{{\textit{Unanswerable questions}.} Among the unanswerable subcategories, Missing Relationships has the lowest classification rate and the lowest \textit{Predicted} FB. The correct label raises its FB by 3 to 4 points for the four large systems and by 18 to 19 points for the other five. Even in the \textit{Ground Truth} mode, however, it stays the lowest, because 17\% of its conversations on ABISS-BIRD and 11\% on ABISS-Spider still end in SQL. This SQL mostly joins the two entities through an existing path in place of the missing relationship.}

{{\textit{Questions without a label}.} In the \textit{Taxonomy-Free} mode, about a third of the Missing External Knowledge and Missing Relationships questions of the two datasets together end in SQL. Over all modes, 64\% of the SQL answers to Missing External Knowledge questions use a number, in a comparison or a \texttt{LIMIT}, that appears neither in the question nor in its evidence, such as a reputation of at least 1{,}000 for a \textit{``gold badge''}. Moreover, 78\% of these answers return rows, so the user receives a plausible answer with no sign that the question cannot be answered. The taxonomy text alone already helps here: on Missing External Knowledge, FB rises from 62.4\% in the \textit{Taxonomy-Free} mode to 74.1\% in the \textit{Taxonomy-Only} mode on ABISS-BIRD, and from 74.8\% to 83.5\% on ABISS-Spider. The correct label then raises it to 93.9\% and 94.1\%, respectively.}

\FloatBarrier

\section{{Further Results}}
\label{app:further}

\subsection{{Diagnosis}}

{\textbf{Naming the subcategory.}} {We check whether models that recognize the type of a question also name its subcategory. They do so in 90.0\% of cases on ABISS-BIRD and 90.2\% on ABISS-Spider, whereas a uniform choice within the type would reach 26\% to 27\%. The share stays high for each type. Pooled over models, it is 94\% to 95\% for ambiguous questions on each dataset, against 14\% for a uniform choice. It is 83\% to 84\% for unanswerable questions, against 25\%, and 96\% to 97\% for answerable questions, against 50\%.}

{\textbf{Recognizing the type.}} {We compare recognition (Rec.) across the three question types. Models recognize 98\% to 99\% of the answerable questions and 85\% to 89\% of the unanswerable ones. On the ambiguous ones, however, the four large systems (GPT-OSS, Qwen3.5, Nemotron, and Qwen3.8-Flash) recognize only 58\% on ABISS-BIRD and 59\% on ABISS-Spider. The other five (the open systems of 7B to 32B parameters) recognize 36\% and 42\%. Because 53.8\% of each evaluated set is ambiguous, Rec.\ largely measures ambiguity detection. For the same reason, a constant \textit{ambiguous} prediction would reach a Rec.\ of 53.8\%, close to the Rec.\ of Qwen2.5-7B in Table~\ref{tab:overall_results}.}

{\textbf{Classifying and acting.}} {We compare what a model's classifier says about a question with what the model does first in the dialogue. On ambiguous questions, the other five recognize ambiguity more often by acting than by classifying. In the \textit{Taxonomy-Free} mode, they open with a clarification on 64\% of the ambiguous questions, far more than their classifiers recognize. For the four large systems, by contrast, the two shares are close, with 60\% of conversations opening with a clarification. On unanswerable questions, the reverse holds on average over models. The classifiers recognize most of these questions, but the dialogue opens with feedback on only 70\%. The classifier is thus ahead of the dialogue on unanswerable questions, but not on ambiguous ones. This is why a predicted label raises FB but not ambiguous EX (Section~\ref{sec:category_impact}). For the other five, the label raises FB over both label-free modes (\textit{Taxonomy-Only} and \textit{Taxonomy-Free}). For the four large systems, it raises FB over the \textit{Taxonomy-Free} mode.}

{\textbf{Asking where the classifier errs.}} {We examine what models do on the ambiguous questions that their own classifier calls answerable. In these \textit{Predicted} conversations, the ask rate of individual models ranges from 0.8\% to 28.4\%. On average, the four large systems ask in 10\% of them, against 15\% to 20\% without a label. The other five ask in 18\%, against 45\% to 47\%. The wrong label thus suppresses asking, most of all for the other five. Models that classify better also tend to ask less in these conversations. Over the nine open systems alone, the Spearman correlation between this \textit{Predicted} ask rate and a model's Rec.\ is $\rho=-0.57$, against $-0.67$ over all ten models.}

{\textbf{Classification and score.}} {We ask whether the models that classify better also reach better \textit{Predicted} scores. GPT-OSS classifies less accurately than Qwen3.8-Flash and Claude Haiku~5.5. Yet it is among the best models in three of the four \textit{Predicted} EX and FB columns of Table~\ref{tab:overall_results}, and the best open system in the fourth. Moreover, averaged over the two datasets, its EX changes least between the \textit{Predicted} and \textit{Ground Truth} modes. A better classifier thus does not always yield a better \textit{Predicted} score.}

{\textbf{The held-out system.}} {We check whether the held-out Qwen3.8-Flash, the only large open model outside the council, behaves like the other models. Section~\ref{sec:category_impact} already shows that a wrong predicted label suppresses its asking, as it does for the other models. In the \textit{Ground Truth} mode, its EX of 83.9\% after a Relevant reply also matches the 84.3\% of its oracle. It thus follows the same patterns as the other models.}

\subsection{{Datasets and Systems}}

{\textbf{Datasets.}} {In the \textit{Ground Truth} mode, ABISS-Spider is easier than ABISS-BIRD on ambiguous questions for every model, by 9 to 24 EX points. The difference is 14 to 24 points for the open systems of 7B to 32B parameters and 9 to 12 points for the others. Part of this gap may come from the Ground Truth SQL, which the annotators judged correct for 92.5\% of the sampled queries on ABISS-Spider and 73.8\% on ABISS-BIRD (Table~\ref{tab:human_eval}). Feedback accuracy, by contrast, does not differ significantly between the datasets for any model in the \textit{Ground Truth} and \textit{Predicted} modes.}

{\textbf{The four large systems and the other five.}} {We compare the two groups of systems on EX and on feedback. As a group, the four large systems score above the other five on EX on both datasets. Part of this gap comes from SQL that does not run. Errors and time-outs end 0.5\% to 1.5\% of the solvable conversations of the four large systems, against 4.3\% to 15.7\% of those of the other five. On feedback with the correct label, however, smaller models such as Qwen2.5-32B and Gemma-3-27B are among the best, whereas GPT-OSS and Nemotron are not.}

{\textbf{Difficulty of ambiguous questions.}} {We check whether the difficulty of the SQL affects the comparison between ambiguous and answerable EX. In the evaluated sets, ambiguous interpretations have simpler SQL than answerable questions, with 88\% against 73\% simple or moderate. We therefore reweight ambiguous EX to the difficulty of the answerable questions. In the \textit{Ground Truth} mode on ABISS-BIRD, ambiguous EX then lies 9.4 points below answerable EX, against 5.2 points unweighted. Table~\ref{tab:ablation} thus understates the cost of ambiguity.}

{\textbf{Code specialization.}} {We compare Qwen2.5-Coder-32B with Qwen2.5-32B on both datasets, on Rec., Cls., and on answerable EX, ambiguous EX, and FB in each mode. The two models differ significantly in only 3 of these 28 comparisons. The Coder model is better once, on ambiguous ABISS-Spider questions in the \textit{Taxonomy-Free} mode, where it asks more often. {We thus detect no gain from code specialization.}}

\subsection{{Labels}}

{We examine what a label does when the model's own prediction is correct, and what the taxonomy text does without a label. On the ambiguous questions that a model itself classifies as ambiguous, EX is 77.1\% in the \textit{Predicted} mode and 77.9\% in the \textit{Ground Truth} mode, pooled over models and both datasets. On these questions, the predicted label is thus nearly as useful as the correct one. A correct prediction also helps over the label-free modes. On the unanswerable questions that it recognizes, a model writes SQL in 10.5\% of \textit{Taxonomy-Free} conversations, but in 2.0\% when it is shown its own prediction. The taxonomy text alone, which the \textit{Taxonomy-Only} mode adds to the \textit{Taxonomy-Free} mode, changes little on average. Individual models, however, react to it in opposite directions. On the ambiguous questions of both datasets together, it raises the ask rate of some models by up to 18 points and lowers that of others by up to 15 points.}

{We also ask whether models follow the correct label when their own classifier disagrees with it. They do not always. On the unanswerable questions that their own classifier calls answerable or ambiguous, GPT-OSS and Nemotron still write SQL in 50\% and 41\% of \textit{Ground Truth} conversations, respectively. Qwen2.5-32B, by contrast, does so in 3\%. Across models, the gain in accepted feedback on unanswerable questions from the correct label tracks the share of misclassified questions times the share on which the label is followed. The correlation is $r=0.95$.}

\subsection{{Dialogue}}

{\textbf{The floor.}} {We check what systems do on ambiguous questions in the floor, where they must respond at once with SQL or feedback. In the floor's \textit{Ground Truth} mode, the four large systems give feedback on 58\% to 60\% of the ambiguous questions, and the other five on 30\% to 32\%. The floor's EX therefore also reflects how often a system declines to answer.}

{\textbf{The gap to the oracle.}} {We split the gap between the dialogue and the oracle by what happens in each conversation. In every mode, most of it comes from conversations without a clarification. In the \textit{Ground Truth} mode, these conversations account for 4.0 of the 6.7 points of the gap. In the other modes, models ask less often (\textit{Clar.}\ in Table~\ref{tab:ablation}). There, the same conversations account for 13 to 15 of the 16 to 18 points of the gap to the same oracle. Conversations whose clarifications never obtain a Relevant reply account for only 1 to 2 points. After a Relevant reply, the other five lose less than the four large systems, except in the \textit{Taxonomy-Free} mode. Their EX lies 1.0 to 2.9 points below the oracle, against 2.3 to 3.7 points, depending on the mode. In the \textit{Ground Truth} mode, GPT-OSS, Nemotron, and Qwen3.5 ask less often than the held-out system (\textit{Clar.}\ in Tables~\ref{tab:interaction_stats_bird} and~\ref{tab:interaction_stats_spider}). On each dataset, their conversations without a clarification account for 4.0 to 8.6 of the 8.7 to 12.0 points between their dialogue and their own oracle. This loss explains why the held-out system overtakes GPT-OSS and Nemotron in the dialogue (Section~\ref{sec:interaction_analysis}). Over both datasets, Claude Haiku~5.5 follows the same pattern, with 5.5 of its 8.1 points between the dialogue and the oracle from conversations without a clarification.}

{\textbf{Clarifications not labeled Relevant.}} {We check what a first clarification asks about when it is not labeled Relevant. On Attachment, Scope, and Entity Ambiguity and Lexical Overlap questions, we count the first clarifications that use a word distinguishing the hidden knowledge of the two interpretations. A Relevant first clarification uses such a word in 65\% to 95\% of cases, whereas a non-Relevant one does so in 15\% to 37\%. A first clarification that is not labeled Relevant thus usually asks about a point other than the one that separates the two interpretations.}

{\textbf{Models with fewer Relevant replies.}} {We examine the models whose first clarification least often obtains a Relevant reply. Qwen2.5-7B, Mistral-Small-24B, Qwen2.5-Coder-32B, and Gemma-3-27B have the lowest shares of such replies. The same four models also ask again most often after a Relevant reply and ask most often on answerable questions. Yet in the \textit{Ground Truth} mode, Qwen2.5-7B, Mistral-Small-24B, and Gemma-3-27B are about as accurate after a Relevant reply as with the oracle. Their lower scores therefore come from the SQL itself rather than from using the reply. Repeated asking rarely reaches the limit of three clarifications, which binds in 1.7\% of ambiguous conversations. About half of these conversations are Qwen2.5-7B's.}

{\textbf{A shared default reading.}} {We ask whether models that answer without a clarification choose between the two interpretations of an ambiguous question at random. We call these two interpretations, or readings, a \textit{pair}. We say that a model \textit{leans} to one reading of a pair when at least three of its SQL answers without a clarification take either of the two readings, and most of these answers take the same one. We count answers from both the floor and the dialogue. On the 259 pairs where at least five models lean, 94.0\% of these models lean to the same reading on average, against 64.0\% if each model chose its reading at random. Models also ask less where this default is strong. We compare the pairs with at least ten floor answers that take either reading. Over all modes, models answer without asking in 51\% of conversations on the pairs where at least 90\% of these answers take one reading, against 22\% on the other pairs. Answers without a clarification thus follow a default reading that models share, not a random choice.}

{\textbf{Asking on answerable and unanswerable questions.} We examine how often models ask on answerable and unanswerable questions, and what asking costs there. With a label, models ask on 3\% to 4\% of answerable questions. Without one, the four large systems ask on 4\% to 7\% and the other five on 22\% to 27\%. In the \textit{Taxonomy-Free} mode, the ask rate of individual models ranges from 2\% for Qwen3.8-Flash to 42\% for Qwen2.5-7B. These clarifications cost no measurable EX against the floor. On unanswerable questions, by contrast, models often guess after the simulated user's refusal. Where they asked, they end with SQL in 51\% of conversations, against 24\% for the same questions in the floor. Qwen2.5-7B, which keeps asking after the refusal, is the only model that ends in SQL less often after asking than in the floor. For most models, asking lowers the share of accepted feedback.}

{\textbf{Pairs resolved for both users.} EX scores each interpretation of an ambiguous question on its own. A model can therefore score on a pair while serving only one of its two simulated users, one for each interpretation. We ask how often a model serves both users of a pair. We call a pair \textit{resolved} when both of its conversations end in SQL that matches their own Ground Truth SQL and not the other interpretation's. A hedge, a query that returns the columns of both interpretations, thus does not count. In the floor's \textit{Ground Truth} mode, models resolve 0.1\% of the ABISS-BIRD pairs and 0.3\% of the ABISS-Spider pairs. Their ambiguous EX in the floor is nevertheless far higher (Section~\ref{sec:interaction_analysis}), so almost every correct answer that is not a hedge serves one user of a pair and fails the other. In the \textit{Predicted} mode of the dialogue, averaged over the two datasets, models resolve 8.7\% of the Attachment Ambiguity pairs and 10.5\% of the Scope Ambiguity pairs. On all pairs of these two subcategories, EX is 40.3\% and 47.7\%. We compare these resolved shares with the share that two independently answered interpretations would reach at that accuracy. The four large systems reach about half of that share, and the other five a quarter or less. Answers after a Relevant reply rarely take the other interpretation. On average, at most 0.9\% of them do, against 31\% to 37\% of the answers given without a clarification.}

\FloatBarrier

\section{{Checks of the Scores}}
\label{app:score_checks}

\subsection{{The Council's Roles}}

{\textbf{Council members judging their own systems.} A council member could favor its own system when it labels that system's clarifications or grades its feedback (Table~\ref{tab:roles}). Because a tie gives Irrelevant, a member's vote decides whether a label is Relevant when one of the two other members votes Relevant and the second does not. In these turns, only Qwen3.5 gives Relevant more often to its own system than to the other nine systems, at 61\% against 45\%. GPT-OSS and Nemotron give it less often to their own system, at 60\% against 63\% and 22\% against 32\%. At Qwen3.5's rate on the other systems, only about 19 of the 3{,}143 labeled turns of its own system would have received another label. However, each system asks different clarification questions, so this comparison is not controlled. As graders, the members vote to accept feedback in 97\% to 100\% of cases, for their own system and for the others alike.}

{\textbf{Questions written by a council member.} A council system might score higher on the questions that its own model wrote. For each council system, we take its score on these questions minus its score on the other writers' questions. The score is EX on answerable and ambiguous questions and FB on unanswerable ones, pooled over the four modes and both datasets. To correct for how hard each writer's questions are, we then subtract the same difference for the seven systems outside the council, whose models wrote no question. The resulting advantage is +1.8 points for GPT-OSS, \textminus{}0.2 for Qwen3.5, and \textminus{}1.1 for Nemotron, and all three 95\% intervals include zero. On average, the advantage is +0.2 points, with an interval from \textminus{}0.8 to +1.1. With the other three large systems as the baseline (the two other council systems and Qwen3.8-Flash), the average is again +0.2, with an interval from \textminus{}1.0 to +1.5. The held-out Qwen3.8-Flash, from the same family as Qwen3.5, does not score higher on Qwen3.5's questions either. Its advantage on them is \textminus{}2.0 points, with an interval from \textminus{}4.5 to +0.6. We therefore detect no advantage for a council member on the questions that its own model wrote, and the average interval excludes an advantage above about one point.}

{\textbf{Questions kept by one witness.} Stage~6 keeps a solvable question when at least one \textit{witness}, a council member other than the writer, reaches its Ground Truth result (Appendix~\ref{app:val_stages}). A single witness keeps 24.8\% of the answerable questions and ambiguous interpretations of ABISS-BIRD and 17.9\% of those of ABISS-Spider. A council system might therefore score higher on the questions whose Ground Truth result it reached itself. We checked this on ABISS-BIRD in the \textit{Ground Truth} mode. Questions whose result all three council models reached are the easiest for every system. On ambiguous interpretations, the council systems score 81.1\%, 56.5\%, and 41.2\% EX when three, two, and one council models reached the result. The seven systems outside the council score 71.9\%, 39.8\%, and 36.1\% in the same three cases. We therefore correct for difficulty as in the previous check. For each council system, we leave out its own model's questions and take its EX on the questions whose result it reached in Stage~6 minus its EX on those that it missed. We then subtract the same difference for the seven systems outside the council. The result is +4.5 points for GPT-OSS, +1.9 for Qwen3.5, and +0.0 for Nemotron, and all three 95\% intervals include zero. On average, the advantage is +2.1 points, with an interval from \textminus{}4.5 to +9.0. Over both datasets and all four modes, it is +5.2 points, with an interval from +1.9 to +8.4. A model tends to solve again a question that it solved once, so keeping the questions that a council model solved favors it slightly. Its effect on the scores is small: without the questions that a council system alone let in, its EX lead over the systems outside the council changes by 0.2 to 1.0 points, against leads of 4.7 to 17.0 points. In the \textit{Ground Truth} mode, the change is 0.4 to 0.9 points, far less than the 8.4 and 11.2 points that Qwen3.5 gains over GPT-OSS and Nemotron when it builds an ABISS-BIRD set alone (Appendix~\ref{app:qwen_only}). Validation by a council of three models thus greatly reduces the advantage that a single-model pipeline gives its model. The number of council models that reach a result mainly reflects how hard the question is.}

\subsection{{The Metrics}}

{\textbf{Feedback grades.} FB counts feedback as correct when a majority of the council accepts it, so lenient graders could inflate it. The three graders accept between 97.1\% and 98.6\% of the feedback that systems produce on unanswerable questions, in every mode and on both datasets. By contrast, the share of unanswerable questions that receive feedback changes with the mode. On ABISS-BIRD, it rises from 80.9\% in the \textit{Taxonomy-Free} mode to 94.0\% in the \textit{Ground Truth} mode, and on ABISS-Spider from 83.9\% to 96.0\%. FB therefore mostly reflects whether a system explains instead of answering.}

{Two annotators jointly checked the graders' verdicts on 200 graded feedback answers, 25 for each dataset and number of accepting graders, weighted to represent all 19{,}751 graded feedback answers of the nine open systems. They applied the graders' own rule (Appendix~\ref{app:prompts}) and agree with the graders' majority on 95.0\% of the answers. Most disagreements are answers that the graders accept and the annotators reject. When we run the graders again, they repeat 99.5\% of their verdicts.}

{Finally, two controls grade feedback that is not specific to the question. The first grades a fixed generic sentence, \textit{``The database does not contain the information needed to answer this question''}, against every unanswerable question. The sentence passes on 97.4\% of the Improper Questions of ABISS-BIRD and on 96.3\% of those of ABISS-Spider. On this subcategory, FB therefore measures whether a system declines rather than how it explains. By contrast, the sentence passes on 34.2\% and 19.5\% of the Missing External Knowledge questions, and on at most 9.2\% of the Missing Schema Elements questions. Models do not write such generic sentences, so this control bounds what a vague explanation could gain rather than describing real feedback. The second control grades the feedback that a model gave on one question against a random other question. On ABISS-BIRD and ABISS-Spider, such feedback passes on 1.6\% and 2.7\% of the nine open systems' pairs and on 0.7\% and 1.7\% of Claude Haiku~5.5's. When the other question is on the same database, such feedback passes on 6.8\% and 7.2\% (5.5\% and 7.8\% for Claude Haiku~5.5). Feedback written for another question thus passes more often when both questions concern the same database, but still rarely.}

{\textbf{Hedged answers.} EX accepts additional columns (Appendix~\ref{app:equivalence}). Therefore, when the two Ground Truth SQL of an ambiguous question return the same number of rows, one query that returns the columns of both interpretations passes both. Such a query is correct whichever interpretation the user holds, and we call it a \textit{hedge}. Over both datasets, a hedge is possible in 69 of the 79 Lexical Overlap pairs of interpretations. In each of Missing User Knowledge, Entity Ambiguity, and Conflicting Knowledge, it is possible in 34 to 47 of the 79 pairs. In the other subcategories, it is possible in at most 3 pairs. These counts are a ceiling. In Missing User Knowledge, for instance, the two interpretations differ in a fact that only the user knows, so no model can compute both. Over all modes, hedges are 4.7\% of the correct ambiguous answers. However, 1{,}295 of the 4{,}333 correct Lexical Overlap answers, or 29.9\%, are hedges, against at most 0.8\% in every other subcategory. Moreover, 97\% of the hedges come in conversations without a Relevant reply, that is, without the simulated user's answer to a Relevant clarification. A hedge therefore replaces the clarification instead of following it.}

{Table~\ref{tab:hedges} shows that, in the \textit{Ground Truth} mode, hedges reorder the systems on Lexical Overlap. Qwen3.5 hedges most and falls from second to eighth place when hedges count as wrong, whereas Qwen2.5-32B hedges least and rises to second. On this subcategory, EX thus partly rewards hedging. Hedges also raise the mean ambiguous EX of ABISS-Spider, by 5.8 to 6.6 points outside the \textit{Ground Truth} mode and by 2.2 points in it. On ABISS-BIRD, by contrast, the increase is at most 0.9 points. On ABISS-Spider, hedges therefore hide part of the cost of not asking. They also hide part of the gain from the correct label. Without hedges, ambiguous EX on ABISS-Spider would rise by 15.2 points from the \textit{Predicted} to the \textit{Ground Truth} mode, instead of 10.9 (Table~\ref{tab:ablation}).}

\begin{table}[!htb]
\centering
\small
\setlength{\tabcolsep}{5pt}
\begin{tabular}{@{}lrrr@{}}
\toprule
\textbf{Model} & \textbf{EX} & \textbf{w/o hedges} & \textbf{Hedges} \\
\midrule
Qwen3.8-Flash$^\dagger$ & 83.5 & 75.3 & 13 \\
Qwen3.5 & 82.9 & 64.6 & 29 \\
Claude Haiku 5.5 & 80.4 & 66.5 & 22 \\
GPT-OSS & 77.2 & 68.4 & 14 \\
Nemotron & 75.3 & 65.2 & 16 \\
Qwen2.5-32B & 75.3 & 73.4 & 3 \\
Gemma-3-27B & 73.4 & 68.4 & 8 \\
Qwen2.5-Coder-32B & 72.8 & 66.5 & 10 \\
Mistral-Small-24B & 70.9 & 59.5 & 18 \\
Qwen2.5-7B & 51.3 & 44.3 & 11 \\
\bottomrule
\end{tabular}
\caption{{Lexical Overlap EX (\%) in the \textit{Ground Truth} mode on both datasets, over 158 conversations per system. \textit{EX} counts hedged answers as correct, \textit{w/o hedges} counts them as wrong, and \textit{Hedges} gives their number. $^\dagger$Held-out system.}}
\label{tab:hedges}
\end{table}

\subsection{{Run-to-Run Variation}}

{\textbf{Repeated prompts.} Each system runs once in each mode, so its score could change on a rerun. When a system classifies a question correctly, its \textit{Predicted} and \textit{Ground Truth} prompts are the same text. The runs of the nine open systems therefore contain 11{,}286 pairs of conversations in which the same prompt ran twice, once in each mode. Their scores differ in 4.1\% of the pairs, and the differences go equally in both directions. Pooled over all its pairs of one dataset, no system's score moves by more than 2.4 points between the two runs. Both runs use the same seed, so 4.1\% is a lower bound on the variation of an independent rerun. Claude Haiku~5.5 takes no seed, so its 1{,}542 such pairs measure the variation of an unseeded rerun. Its score on them moves by +0.6 points on ABISS-BIRD and by \textminus{}0.4 points on ABISS-Spider.}

\FloatBarrier

\section{{Qwen-Only Comparison}}
\label{app:qwen_only}

{\textbf{Setup.} ABISS uses a council of three models so that no single model shapes the data or the simulated user. To measure what one model would change, we built a second ABISS-BIRD set with the same pipeline, but with Qwen3.5 as the only writer and the only council member. Qwen3.5 thus wrote every question, cast every validation vote, and simulated the user. For each generation request, it produced three samples, as many as the three writers of the main pipeline. We drew 988 questions, as many per subcategory as in the balanced subset of the main set. We ran GPT-OSS, Qwen3.5, and Nemotron on them in all four modes. All other benchmark settings stayed the same, including the three-model council that grades the feedback. We call this setup the \textit{Qwen-only setting}. If this setting favors Qwen3.5, that model should gain more than the two other systems between the main benchmark and this setting. We therefore compare the systems' changes between the two sets for EX and FB in the \textit{Ground Truth} and \textit{Predicted} modes. We call Qwen3.5 favored when its change exceeds the changes of both other systems and both differences have a 95\% bootstrap interval above zero. We set this criterion before the run. The intervals resample the questions as in Appendix~\ref{app:metrics}, independently in the two sets.}

\begin{table*}[!htb]
\centering
\small
\setlength{\tabcolsep}{5pt}
\begin{tabular}{@{}llrrrccl@{}}
\toprule
& & \multicolumn{3}{c}{\textbf{Change}} & \multicolumn{2}{c}{\textbf{Qwen3.5's change minus}} & \\
\cmidrule(lr){3-5} \cmidrule(lr){6-7}
\textbf{Metric} & \textbf{Mode} & \textbf{GPT-OSS} & \textbf{Qwen3.5} & \textbf{Nemotron} & \textbf{GPT-OSS's} & \textbf{Nemotron's} & \textbf{Favored} \\
\midrule
EX & GT & \textminus{}1.6 & +6.8 & \textminus{}4.3 & +8.4 [+3.5, +13.1] & +11.2 [+6.2, +16.1] & yes \\
EX & Pred. & \textminus{}2.0 & +0.3 & \textminus{}3.3 & +2.3 [\textminus{}2.5, +7.1] & +3.6 [\textminus{}1.2, +8.6] & undecided \\
FB & GT & \textminus{}3.0 & \textminus{}2.2 & \textminus{}7.0 & +0.8 [\textminus{}3.7, +5.2] & +4.8 [\textminus{}0.5, +10.0] & undecided \\
FB & Pred. & \textminus{}5.3 & \textminus{}3.6 & \textminus{}8.5 & +1.7 [\textminus{}3.1, +6.5] & +4.8 [\textminus{}0.5, +10.1] & undecided \\
\bottomrule
\end{tabular}
\caption{{Qwen-only comparison on ABISS-BIRD, with 988 questions in each set. \textit{Change} is the score in the Qwen-only setting minus the score in the main benchmark, in points. The last columns give Qwen3.5's change minus another system's change, with 95\% intervals. Qwen3.5 is favored when both intervals lie above zero.}}
\label{tab:qwen_only}
\end{table*}

{\textbf{Result.} Qwen3.5 is favored in EX in the \textit{Ground Truth} mode (Table~\ref{tab:qwen_only}), and this result holds after Holm's correction over the eight differences of the table (Appendix~\ref{app:metrics}). In the three other rows, Qwen3.5's change also exceeds both others', so all eight differences are positive. In these rows, however, both intervals include zero, so the criterion leaves them undecided. A pipeline run by one model can thus favor that model, which supports a council of several models.}

{\textbf{Where the gain comes from.} We split Qwen3.5's gain in \textit{Ground Truth} EX by question type. Per question, the gain is largest on answerable questions, where the simulated user plays almost no role. On them, Qwen3.5's \textit{Ground Truth} EX rises from 73.7\% to 87.5\%, while GPT-OSS's falls from 86.2\% to 80.3\% and Nemotron's from 84.9\% to 77.0\%. In total, however, ambiguous questions still contribute more than half of the gain. Part of the answerable gain may be chance, because Qwen3.5's starting value is low. In the \textit{Predicted} mode of the main benchmark, its answerable EX is 79.6\%. Yet on 145 of the 152 answerable questions, both modes gave it the same prompt (Appendix~\ref{app:score_checks}). On these questions alone, the two runs still reached 73.8\% and 80.7\%, so the low \textit{Ground Truth} value is probably in part a chance draw.}

{\textbf{Data or simulated user.} The Qwen-only setting changes the questions and the simulated user together. To tell the two apart, we ran one more setting, which pairs the Qwen-only questions with the three-model simulated user of the main benchmark. It splits the change into a \textit{data step}, from the main benchmark to this setting, and a \textit{user step}, from this setting to the Qwen-only setting. Only the questions change in the data step, and only the simulated user in the user step. Table~\ref{tab:qwen_only_steps} applies the same criterion to each step. For the data step, the intervals draw the two sets of questions independently, and for the user step, they draw the shared questions once. The data step favors Qwen3.5 in three of the four cases, and the user step in none. After Holm's correction over the 16 differences of Table~\ref{tab:qwen_only_steps}, only the two differences of the data step for EX in the \textit{Ground Truth} mode remain above zero. The gain therefore comes from the questions that Qwen3.5 alone writes and validates, and we detect no gain from Qwen3.5 as the only simulated user. In line with this, Qwen3.5 alone labels Relevant 92.6\% of the three systems' clarifications on ambiguous questions in the \textit{Ground Truth} mode, as the three-model council does on the main set. On FB, by contrast, the two steps point in opposite directions, which is why Table~\ref{tab:qwen_only} leaves FB undecided. Both simulated users answer a clarification on an unanswerable question with the same fixed refusal, so the user step's FB differences come mainly from run-to-run variation of the tested systems.}

\begin{table*}[!htb]
\centering
\small
\setlength{\tabcolsep}{4.5pt}
\begin{tabular}{@{}lllrrrccl@{}}
\toprule
& & & \multicolumn{3}{c}{\textbf{Change}} & \multicolumn{2}{c}{\textbf{Qwen3.5's change minus}} & \\
\cmidrule(lr){4-6} \cmidrule(lr){7-8}
\textbf{Metric} & \textbf{Mode} & \textbf{Step} & \textbf{GPT-OSS} & \textbf{Qwen3.5} & \textbf{Nemotron} & \textbf{GPT-OSS's} & \textbf{Nemotron's} & \textbf{Favored} \\
\midrule
EX & GT & data & \textminus{}3.7 & +5.4 & \textminus{}2.6 & +9.0 [+4.1, +14.0] & +8.0 [+2.8, +13.2] & yes \\
EX & GT & user & +2.1 & +1.5 & \textminus{}1.7 & \textminus{}0.7 [\textminus{}5.0, +3.6] & +3.2 [\textminus{}1.3, +7.8] & undecided \\
EX & Pred. & data & \textminus{}3.9 & +0.7 & \textminus{}0.6 & +4.6 [\textminus{}0.2, +9.5] & +1.3 [\textminus{}3.5, +6.1] & undecided \\
EX & Pred. & user & +1.9 & \textminus{}0.4 & \textminus{}2.7 & \textminus{}2.3 [\textminus{}5.9, +1.4] & +2.3 [\textminus{}1.6, +6.3] & undecided \\
\midrule
FB & GT & data & \textminus{}5.5 & +0.3 & \textminus{}5.2 & +5.8 [+1.1, +10.5] & +5.6 [+0.7, +10.4] & yes \\
FB & GT & user & +2.5 & \textminus{}2.5 & \textminus{}1.7 & \textminus{}5.0 [\textminus{}8.9, \textminus{}1.5] & \textminus{}0.8 [\textminus{}5.2, +3.4] & undecided \\
FB & Pred. & data & \textminus{}6.7 & \textminus{}1.0 & \textminus{}8.6 & +5.6 [+0.9, +10.4] & +7.5 [+2.1, +12.7] & yes \\
FB & Pred. & user & +1.4 & \textminus{}2.6 & +0.1 & \textminus{}3.9 [\textminus{}8.1, +0.3] & \textminus{}2.7 [\textminus{}7.0, +1.7] & undecided \\
\bottomrule
\end{tabular}
\caption{{Qwen-only change of Table~\ref{tab:qwen_only} in two steps, on ABISS-BIRD. The \textit{data} step goes from the main benchmark to the Qwen-only questions with the three-model simulated user, and the \textit{user} step from there to Qwen3.5 as the only simulated user. The two steps add up to the changes of Table~\ref{tab:qwen_only}. A step favors Qwen3.5 when both of its intervals lie above zero.}}
\label{tab:qwen_only_steps}
\end{table*}

{\textbf{Limits.} The two steps separate data and user in one order, changing the questions first and the simulated user second. We did not run the fourth possible setting, which pairs the three-model questions with Qwen3.5 as the only simulated user. We therefore do not estimate an interaction between data and user. The one-model pipeline also removes several safeguards of Sections~\ref{sec:generation_pipeline} and~\ref{sec:simulated_user} (Appendices~\ref{app:validation_detail} and~\ref{app:user_sim}). In the generation, Qwen3.5 judges its own questions, and Stage~6 accepts its Ground Truth result when Qwen3.5 itself reproduces it with another seed. The main pipeline leaves both decisions to the council members other than the writer. Moreover, Stage~7 tests unanswerability only against Qwen3.5, so GPT-OSS or Nemotron may answer some \textit{unanswerable} questions and lose FB on them. These missing safeguards of the generation are candidate causes of the data effect, but we did not test them separately. The one-member simulated user, finally, lacks the council's tie rule, which labels a tied vote Irrelevant, and the pooling of the SQL fragments that the Technical voters select.}

\FloatBarrier
\section{Detailed Dialogue Statistics}
\label{app:interaction_stats}
Tables~\ref{tab:interaction_stats_bird} and~\ref{tab:interaction_stats_spider} give the
per-model breakdown of the dialogue statistics {summarized per dataset in Table~\ref{tab:ablation}},
for ABISS-BIRD and ABISS-Spider respectively.

\begin{table*}[p]
\centering
\fontsize{7.6}{10.5}\selectfont
\setlength{\tabcolsep}{5pt}

\begin{tabular}{@{}llrrrrrrrr@{}}
\toprule
& & & & \multicolumn{3}{c}{\textbf{Relevancy (\%)}} & & & \\
\cmidrule(lr){5-7}
\textbf{Model} & \textbf{Mode} & \textbf{Turns} & \textbf{Clar.} & \textbf{Rel.} & \textbf{Tech.} & \textbf{Irrel.} & \textbf{AIR} & \textbf{Reach} & \textbf{TTR} \\
\midrule
\multirow{4}{*}{Qwen3.5} & GT & 1.91\ci{0.04} & 88.2\ci{3.2} & 94.6\ci{2.3} & 4.1\ci{2.1} & 1.2\ci{1.1} & 97.5\ci{2.0} & 84.2\ci{3.6} & 1.00\ci{0.00} \\
 & Pred. & 1.62\ci{0.04} & 59.8\ci{3.8} & 94.6\ci{3.0} & 3.6\ci{2.3} & 1.8\ci{1.8} & 96.4\ci{3.0} & 57.5\ci{3.6} & 1.00\ci{0.01} \\
 & Tax.-Only & 1.62\ci{0.04} & 59.4\ci{3.8} & 94.2\ci{3.4} & 4.6\ci{3.1} & 1.2\ci{1.3} & 97.1\ci{2.8} & 57.5\ci{3.6} & 1.00\ci{0.01} \\
 & Tax.-Free & 1.61\ci{0.05} & 58.5\ci{3.9} & 91.1\ci{3.6} & 6.4\ci{3.0} & 2.5\ci{1.7} & 95.5\ci{3.1} & 54.5\ci{4.0} & 1.00\ci{0.00} \\
\midrule
\multirow{4}{*}{GPT-OSS} & GT & 1.81\ci{0.05} & 77.3\ci{4.0} & 92.8\ci{3.2} & 4.6\ci{2.5} & 2.5\ci{2.1} & 96.2\ci{2.5} & 74.2\ci{4.1} & 1.00\ci{0.00} \\
 & Pred. & 1.65\ci{0.05} & 61.3\ci{4.3} & 92.2\ci{4.1} & 5.5\ci{3.1} & 2.3\ci{2.4} & 96.1\ci{2.8} & 58.1\ci{4.2} & 1.00\ci{0.00} \\
 & Tax.-Only & 1.60\ci{0.04} & 57.3\ci{3.8} & 94.7\ci{3.6} & 3.8\ci{2.4} & 1.6\ci{1.7} & 97.0\ci{2.8} & 55.8\ci{3.8} & 1.00\ci{0.01} \\
 & Tax.-Free & 1.61\ci{0.05} & 58.5\ci{4.5} & 91.4\ci{3.8} & 7.1\ci{3.6} & 1.5\ci{1.7} & 96.2\ci{2.9} & 54.7\ci{4.5} & 1.00\ci{0.00} \\
\midrule
\multirow{4}{*}{Nemotron} & GT & 1.89\ci{0.05} & 81.6\ci{4.0} & 90.3\ci{3.3} & 6.8\ci{2.6} & 3.0\ci{1.7} & 92.4\ci{3.2} & 78.8\ci{4.0} & 1.00\ci{0.01} \\
 & Pred. & 1.74\ci{0.06} & 64.5\ci{3.8} & 84.5\ci{4.3} & 8.1\ci{3.2} & 7.4\ci{3.2} & 88.8\ci{4.3} & 60.5\ci{4.0} & 1.01\ci{0.02} \\
 & Tax.-Only & 1.68\ci{0.05} & 61.5\ci{3.7} & 89.5\ci{3.9} & 6.3\ci{3.1} & 4.1\ci{3.0} & 90.9\ci{3.9} & 60.0\ci{3.6} & 1.00\ci{0.01} \\
 & Tax.-Free & 1.69\ci{0.05} & 63.9\ci{4.0} & 86.9\ci{4.1} & 9.0\ci{3.3} & 4.1\ci{2.2} & 93.3\ci{3.6} & 58.8\ci{4.2} & 1.00\ci{0.01} \\
\midrule
\multirow{4}{*}{Qwen3.8-Flash$^\dagger$} & GT & 1.99\ci{0.03} & 95.9\ci{2.3} & 94.7\ci{2.4} & 2.7\ci{1.7} & 2.7\ci{1.5} & 97.8\ci{1.8} & 92.7\ci{2.9} & 1.01\ci{0.01} \\
 & Pred. & 1.68\ci{0.05} & 63.9\ci{4.1} & 93.9\ci{3.4} & 0.8\ci{1.4} & 5.3\ci{3.1} & 96.7\ci{2.8} & 61.8\ci{4.2} & 1.02\ci{0.03} \\
 & Tax.-Only & 1.60\ci{0.04} & 57.9\ci{3.9} & 95.0\ci{3.0} & 1.6\ci{1.4} & 3.5\ci{2.5} & 97.6\ci{2.3} & 55.8\ci{4.0} & 1.01\ci{0.01} \\
 & Tax.-Free & 1.53\ci{0.05} & 51.7\ci{3.8} & 94.0\ci{3.3} & 1.1\ci{1.3} & 4.9\ci{3.2} & 97.7\ci{2.6} & 49.4\ci{4.0} & 1.01\ci{0.01} \\
\midrule
\multirow{4}{*}{Qwen2.5-Coder-32B} & GT & 2.13\ci{0.05} & 98.1\ci{1.9} & 88.7\ci{3.1} & 9.1\ci{2.8} & 2.2\ci{1.5} & 88.1\ci{3.2} & 93.4\ci{2.4} & 1.00\ci{0.01} \\
 & Pred. & 1.64\ci{0.06} & 55.6\ci{4.2} & 87.1\ci{4.4} & 10.6\ci{4.0} & 2.3\ci{2.2} & 88.2\ci{4.7} & 52.4\ci{4.0} & 1.01\ci{0.01} \\
 & Tax.-Only & 1.80\ci{0.06} & 70.3\ci{4.7} & 80.5\ci{5.0} & 13.9\ci{3.9} & 5.6\ci{3.5} & 88.6\ci{3.9} & 60.9\ci{4.5} & 1.01\ci{0.01} \\
 & Tax.-Free & 1.87\ci{0.07} & 75.0\ci{4.7} & 76.5\ci{5.0} & 15.9\ci{4.0} & 7.5\ci{3.1} & 90.5\ci{3.7} & 63.5\ci{4.5} & 1.02\ci{0.02} \\
\midrule
\multirow{4}{*}{Qwen2.5-32B} & GT & 1.94\ci{0.04} & 92.7\ci{3.0} & 95.2\ci{2.7} & 2.8\ci{1.9} & 2.0\ci{1.7} & 99.4\ci{1.1} & 88.9\ci{3.4} & 1.00\ci{0.01} \\
 & Pred. & 1.46\ci{0.04} & 45.1\ci{3.8} & 97.1\ci{3.3} & 1.6\ci{2.2} & 1.2\ci{1.5} & 98.7\ci{1.6} & 44.2\ci{3.6} & 1.00\ci{0.00} \\
 & Tax.-Only & 1.57\ci{0.04} & 56.4\ci{4.1} & 94.7\ci{3.5} & 4.0\ci{2.9} & 1.3\ci{1.8} & 99.3\ci{1.1} & 53.6\ci{3.7} & 1.00\ci{0.00} \\
 & Tax.-Free & 1.58\ci{0.05} & 57.9\ci{4.5} & 92.3\ci{3.7} & 4.2\ci{3.0} & 3.5\ci{2.5} & 99.7\ci{0.8} & 53.8\ci{4.2} & 1.00\ci{0.01} \\
\midrule
\multirow{4}{*}{Gemma-3-27B} & GT & 2.04\ci{0.06} & 92.1\ci{3.0} & 88.0\ci{3.6} & 8.5\ci{3.4} & 3.4\ci{1.7} & 90.8\ci{3.5} & 87.8\ci{3.4} & 1.01\ci{0.01} \\
 & Pred. & 1.54\ci{0.05} & 47.4\ci{3.8} & 88.5\ci{4.6} & 8.4\ci{4.1} & 3.1\ci{2.7} & 89.7\ci{4.9} & 45.7\ci{3.6} & 1.01\ci{0.02} \\
 & Tax.-Only & 1.90\ci{0.06} & 80.8\ci{4.3} & 81.5\ci{4.4} & 14.4\ci{3.9} & 4.2\ci{1.9} & 90.7\ci{3.7} & 70.9\ci{4.4} & 1.01\ci{0.01} \\
 & Tax.-Free & 1.72\ci{0.06} & 64.7\ci{4.7} & 83.5\ci{4.6} & 13.1\ci{4.2} & 3.4\ci{2.2} & 90.4\ci{4.2} & 58.6\ci{4.6} & 1.01\ci{0.01} \\
\midrule
\multirow{4}{*}{Mistral-Small-24B} & GT & 1.99\ci{0.07} & 83.8\ci{4.1} & 85.4\ci{4.3} & 12.3\ci{3.9} & 2.3\ci{2.0} & 88.2\ci{3.8} & 81.2\ci{4.1} & 1.01\ci{0.02} \\
 & Pred. & 1.67\ci{0.07} & 56.4\ci{4.9} & 82.9\ci{5.6} & 15.2\ci{5.3} & 2.0\ci{2.2} & 87.7\ci{4.7} & 53.4\ci{4.3} & 1.01\ci{0.02} \\
 & Tax.-Only & 1.65\ci{0.07} & 54.9\ci{4.5} & 82.6\ci{5.8} & 10.7\ci{4.5} & 6.7\ci{4.0} & 87.9\ci{4.8} & 51.3\ci{4.3} & 1.01\ci{0.02} \\
 & Tax.-Free & 1.85\ci{0.08} & 70.3\ci{4.9} & 76.2\ci{5.1} & 17.9\ci{4.7} & 6.0\ci{2.7} & 84.5\ci{4.8} & 62.0\ci{4.9} & 1.02\ci{0.02} \\
\midrule
\multirow{4}{*}{Qwen2.5-7B} & GT & 2.11\ci{0.09} & 79.9\ci{4.3} & 79.2\ci{4.7} & 14.7\ci{3.6} & 6.1\ci{2.9} & 79.3\ci{4.2} & 72.7\ci{4.5} & 1.02\ci{0.02} \\
 & Pred. & 1.52\ci{0.08} & 39.1\ci{4.7} & 81.7\ci{7.4} & 9.7\ci{5.0} & 8.6\ci{5.5} & 82.4\ci{5.8} & 36.3\ci{4.3} & 1.03\ci{0.04} \\
 & Tax.-Only & 1.73\ci{0.10} & 53.9\ci{5.1} & 77.0\ci{5.8} & 18.4\ci{5.1} & 4.6\ci{2.6} & 77.7\ci{6.2} & 48.9\ci{4.9} & 1.02\ci{0.03} \\
 & Tax.-Free & 1.58\ci{0.09} & 44.4\ci{5.7} & 71.2\ci{7.0} & 18.6\ci{5.7} & 10.1\ci{4.8} & 84.4\ci{5.3} & 36.1\ci{5.3} & 1.06\ci{0.05} \\
\midrule
\multirow{4}{*}{Claude Haiku 5.5} & GT & 1.89\ci{0.04} & 87.0\ci{3.7} & 97.7\ci{1.8} & 0.8\ci{0.9} & 1.5\ci{1.1} & 97.8\ci{1.9} & 85.5\ci{3.7} & 1.00\ci{0.00} \\
 & Pred. & 1.71\ci{0.05} & 69.4\ci{4.6} & 93.9\ci{3.0} & 3.2\ci{2.0} & 2.9\ci{2.1} & 97.1\ci{2.3} & 65.8\ci{4.5} & 1.00\ci{0.00} \\
 & Tax.-Only & 1.70\ci{0.05} & 68.2\ci{4.5} & 94.4\ci{3.1} & 2.1\ci{2.2} & 3.5\ci{2.4} & 96.8\ci{2.6} & 65.4\ci{4.5} & 1.00\ci{0.00} \\
 & Tax.-Free & 1.56\ci{0.05} & 53.9\ci{4.2} & 94.6\ci{2.8} & 2.3\ci{2.0} & 3.0\ci{2.3} & 95.7\ci{3.4} & 51.9\ci{4.3} & 1.00\ci{0.00} \\
\bottomrule
\end{tabular}
\caption{{Dialogue statistics for ambiguous questions on ABISS-BIRD, per model and mode. \textit{Turns} is the average number of interaction steps per conversation; \textit{Clar.}\ is the percentage of conversations with at least one clarification; \textit{Rel.}, \textit{Tech.}, and \textit{Irrel.}\ report the relevancy-label distribution; \textit{AIR} is the percentage of first Relevant replies after which the next system turn is final; \textit{Reach} is the percentage of conversations that obtain a Relevant reply; and \textit{TTR} is the number of clarifications up to the first Relevant one. Superscripts give the larger distance to the bounds of the 95\% interval. $^\dagger$Held-out system.}}
\label{tab:interaction_stats_bird}
\end{table*}

\begin{table*}[p]
\centering
\fontsize{7.6}{10.5}\selectfont
\setlength{\tabcolsep}{5pt}

\begin{tabular}{@{}llrrrrrrrr@{}}
\toprule
& & & & \multicolumn{3}{c}{\textbf{Relevancy (\%)}} & & & \\
\cmidrule(lr){5-7}
\textbf{Model} & \textbf{Mode} & \textbf{Turns} & \textbf{Clar.} & \textbf{Rel.} & \textbf{Tech.} & \textbf{Irrel.} & \textbf{AIR} & \textbf{Reach} & \textbf{TTR} \\
\midrule
\multirow{4}{*}{Qwen3.5} & GT & 1.90\ci{0.04} & 87.5\ci{3.4} & 95.4\ci{2.3} & 1.9\ci{1.5} & 2.7\ci{1.6} & 98.2\ci{1.4} & 85.0\ci{3.5} & 1.00\ci{0.01} \\
 & Pred. & 1.62\ci{0.04} & 61.0\ci{3.5} & 95.3\ci{2.8} & 1.4\ci{1.6} & 3.4\ci{2.2} & 98.8\ci{1.2} & 59.1\ci{3.5} & 1.01\ci{0.02} \\
 & Tax.-Only & 1.65\ci{0.04} & 62.4\ci{3.6} & 95.4\ci{2.6} & 2.9\ci{2.4} & 1.6\ci{1.3} & 96.5\ci{2.4} & 60.5\ci{3.5} & 1.00\ci{0.01} \\
 & Tax.-Free & 1.65\ci{0.04} & 63.6\ci{3.6} & 93.8\ci{2.7} & 2.9\ci{2.0} & 3.2\ci{1.9} & 98.8\ci{1.2} & 60.5\ci{3.5} & 1.01\ci{0.01} \\
\midrule
\multirow{4}{*}{GPT-OSS} & GT & 1.85\ci{0.04} & 82.9\ci{3.8} & 95.7\ci{2.4} & 2.0\ci{1.3} & 2.2\ci{1.7} & 97.8\ci{1.7} & 80.7\ci{3.9} & 1.00\ci{0.01} \\
 & Pred. & 1.64\ci{0.04} & 62.0\ci{3.9} & 96.2\ci{2.5} & 1.9\ci{1.7} & 1.9\ci{1.9} & 98.0\ci{1.8} & 60.6\ci{3.7} & 1.00\ci{0.01} \\
 & Tax.-Only & 1.62\ci{0.04} & 58.7\ci{3.5} & 93.2\ci{3.6} & 3.7\ci{2.3} & 3.1\ci{2.3} & 97.6\ci{1.8} & 57.0\ci{3.3} & 1.02\ci{0.02} \\
 & Tax.-Free & 1.65\ci{0.05} & 62.7\ci{4.0} & 91.9\ci{3.5} & 5.1\ci{2.7} & 3.0\ci{2.2} & 97.6\ci{2.1} & 58.9\ci{3.8} & 1.01\ci{0.01} \\
\midrule
\multirow{4}{*}{Nemotron} & GT & 1.91\ci{0.04} & 87.6\ci{3.1} & 96.9\ci{1.6} & 1.0\ci{0.9} & 2.1\ci{1.4} & 96.9\ci{1.7} & 85.5\ci{3.2} & 1.00\ci{0.01} \\
 & Pred. & 1.67\ci{0.05} & 62.9\ci{3.7} & 93.2\ci{2.9} & 3.7\ci{1.9} & 3.1\ci{2.0} & 94.8\ci{2.7} & 60.1\ci{3.5} & 1.00\ci{0.00} \\
 & Tax.-Only & 1.69\ci{0.05} & 65.0\ci{3.9} & 92.9\ci{2.7} & 4.5\ci{2.3} & 2.5\ci{1.7} & 95.0\ci{2.4} & 62.2\ci{3.7} & 1.01\ci{0.01} \\
 & Tax.-Free & 1.70\ci{0.05} & 65.9\ci{3.8} & 90.2\ci{3.7} & 5.8\ci{2.6} & 4.0\ci{2.2} & 95.8\ci{2.3} & 61.5\ci{3.7} & 1.01\ci{0.01} \\
\midrule
\multirow{4}{*}{Qwen3.8-Flash$^\dagger$} & GT & 1.98\ci{0.02} & 96.2\ci{2.1} & 96.8\ci{1.9} & 1.1\ci{1.0} & 2.1\ci{1.4} & 99.1\ci{0.9} & 94.4\ci{2.2} & 1.01\ci{0.01} \\
 & Pred. & 1.68\ci{0.05} & 65.0\ci{3.9} & 93.3\ci{3.1} & 1.5\ci{1.6} & 5.2\ci{2.6} & 97.8\ci{2.4} & 62.4\ci{3.9} & 1.01\ci{0.02} \\
 & Tax.-Only & 1.61\ci{0.04} & 58.9\ci{3.7} & 95.4\ci{2.9} & 1.1\ci{1.6} & 3.4\ci{2.4} & 98.5\ci{1.4} & 57.1\ci{3.7} & 1.01\ci{0.02} \\
 & Tax.-Free & 1.58\ci{0.04} & 55.7\ci{3.4} & 93.7\ci{3.0} & 2.4\ci{1.7} & 3.9\ci{2.6} & 97.7\ci{2.0} & 53.8\ci{3.3} & 1.02\ci{0.02} \\
\midrule
\multirow{4}{*}{Qwen2.5-Coder-32B} & GT & 2.11\ci{0.04} & 97.9\ci{1.7} & 88.4\ci{3.8} & 6.3\ci{2.5} & 5.3\ci{2.7} & 91.8\ci{2.6} & 93.6\ci{2.7} & 1.01\ci{0.01} \\
 & Pred. & 1.71\ci{0.06} & 61.8\ci{4.1} & 84.7\ci{5.6} & 8.1\ci{3.9} & 7.2\ci{4.1} & 92.7\ci{3.3} & 57.1\ci{4.0} & 1.01\ci{0.01} \\
 & Tax.-Only & 1.85\ci{0.07} & 73.2\ci{4.6} & 80.2\ci{4.8} & 13.6\ci{3.9} & 6.2\ci{3.1} & 91.3\ci{3.0} & 66.2\ci{4.5} & 1.02\ci{0.01} \\
 & Tax.-Free & 1.98\ci{0.06} & 81.4\ci{4.0} & 79.0\ci{4.4} & 12.5\ci{3.6} & 8.5\ci{3.2} & 86.5\ci{3.5} & 70.9\ci{4.3} & 1.02\ci{0.02} \\
\midrule
\multirow{4}{*}{Qwen2.5-32B} & GT & 1.97\ci{0.03} & 96.5\ci{2.1} & 97.8\ci{1.6} & 0.7\ci{1.1} & 1.4\ci{1.3} & 100.0\ci{0.0} & 94.8\ci{2.6} & 1.00\ci{0.01} \\
 & Pred. & 1.48\ci{0.04} & 48.3\ci{3.7} & 98.2\ci{2.0} & 0.7\ci{1.5} & 1.1\ci{1.3} & 99.6\ci{0.7} & 47.4\ci{3.5} & 1.00\ci{0.00} \\
 & Tax.-Only & 1.60\ci{0.05} & 59.4\ci{4.2} & 93.9\ci{3.3} & 3.8\ci{2.5} & 2.3\ci{1.9} & 100.0\ci{0.0} & 55.9\ci{4.0} & 1.00\ci{0.00} \\
 & Tax.-Free & 1.58\ci{0.05} & 57.3\ci{4.3} & 92.7\ci{3.8} & 4.2\ci{3.1} & 3.0\ci{2.4} & 100.0\ci{0.0} & 53.5\ci{4.2} & 1.00\ci{0.00} \\
\midrule
\multirow{4}{*}{Gemma-3-27B} & GT & 2.04\ci{0.04} & 96.3\ci{2.2} & 90.4\ci{2.8} & 5.9\ci{2.3} & 3.7\ci{1.7} & 94.5\ci{2.5} & 92.3\ci{2.8} & 1.02\ci{0.02} \\
 & Pred. & 1.56\ci{0.05} & 53.5\ci{3.8} & 90.4\ci{3.7} & 3.1\ci{2.6} & 6.5\ci{2.8} & 97.6\ci{2.4} & 49.8\ci{3.5} & 1.02\ci{0.03} \\
 & Tax.-Only & 2.00\ci{0.05} & 89.0\ci{3.3} & 80.7\ci{3.7} & 12.4\ci{3.2} & 7.0\ci{2.2} & 89.5\ci{3.5} & 77.9\ci{3.9} & 1.00\ci{0.01} \\
 & Tax.-Free & 1.78\ci{0.05} & 73.3\ci{4.1} & 90.6\ci{3.3} & 6.3\ci{2.6} & 3.1\ci{2.4} & 95.6\ci{2.1} & 67.9\ci{3.9} & 1.01\ci{0.01} \\
\midrule
\multirow{4}{*}{Mistral-Small-24B} & GT & 2.10\ci{0.06} & 94.8\ci{2.5} & 86.9\ci{3.1} & 9.6\ci{3.0} & 3.5\ci{1.7} & 87.8\ci{3.2} & 91.6\ci{2.9} & 1.01\ci{0.01} \\
 & Pred. & 1.67\ci{0.06} & 60.3\ci{4.4} & 88.1\ci{4.3} & 7.5\ci{2.9} & 4.4\ci{2.8} & 91.4\ci{3.6} & 56.8\ci{4.3} & 1.00\ci{0.01} \\
 & Tax.-Only & 1.65\ci{0.06} & 57.7\ci{4.7} & 84.6\ci{5.4} & 8.4\ci{3.4} & 7.0\ci{4.6} & 92.2\ci{3.5} & 53.5\ci{4.4} & 1.01\ci{0.01} \\
 & Tax.-Free & 1.81\ci{0.06} & 72.8\ci{4.2} & 78.9\ci{4.5} & 13.8\ci{3.7} & 7.3\ci{3.0} & 90.9\ci{3.6} & 61.3\ci{4.2} & 1.01\ci{0.01} \\
\midrule
\multirow{4}{*}{Qwen2.5-7B} & GT & 2.37\ci{0.09} & 90.4\ci{3.3} & 76.2\ci{4.3} & 14.0\ci{2.7} & 9.8\ci{3.2} & 73.7\ci{4.4} & 84.1\ci{3.6} & 1.04\ci{0.02} \\
 & Pred. & 1.63\ci{0.08} & 46.9\ci{4.1} & 84.5\ci{4.6} & 10.2\ci{3.8} & 5.3\ci{2.6} & 81.3\ci{5.3} & 45.6\ci{4.2} & 1.03\ci{0.04} \\
 & Tax.-Only & 1.98\ci{0.08} & 70.2\ci{4.9} & 78.1\ci{4.8} & 14.1\ci{3.6} & 7.8\ci{3.5} & 74.1\ci{5.1} & 62.5\ci{4.7} & 1.02\ci{0.02} \\
 & Tax.-Free & 1.62\ci{0.09} & 44.6\ci{5.4} & 74.6\ci{6.3} & 16.9\ci{5.1} & 8.5\ci{4.0} & 78.9\ci{6.1} & 38.9\ci{5.2} & 1.04\ci{0.03} \\
\midrule
\multirow{4}{*}{Claude Haiku 5.5} & GT & 1.88\ci{0.04} & 87.6\ci{3.3} & 97.4\ci{1.6} & 0.6\ci{1.0} & 2.0\ci{1.3} & 99.2\ci{1.1} & 85.7\ci{3.5} & 1.00\ci{0.00} \\
 & Pred. & 1.75\ci{0.05} & 73.3\ci{4.1} & 97.0\ci{2.0} & 0.7\ci{1.2} & 2.3\ci{1.7} & 98.8\ci{1.4} & 71.8\ci{4.2} & 1.01\ci{0.01} \\
 & Tax.-Only & 1.68\ci{0.04} & 67.6\ci{4.0} & 97.7\ci{1.9} & 0.8\ci{1.3} & 1.5\ci{1.6} & 99.5\ci{0.8} & 66.2\ci{4.0} & 1.00\ci{0.00} \\
 & Tax.-Free & 1.58\ci{0.04} & 57.3\ci{3.8} & 96.1\ci{2.5} & 0.9\ci{1.1} & 3.0\ci{2.3} & 99.4\ci{1.0} & 55.2\ci{3.9} & 1.00\ci{0.00} \\
\bottomrule
\end{tabular}
\caption{{Dialogue statistics for ambiguous questions on ABISS-Spider, per model and mode. \textit{Turns} is the average number of interaction steps per conversation; \textit{Clar.}\ is the percentage of conversations with at least one clarification; \textit{Rel.}, \textit{Tech.}, and \textit{Irrel.}\ report the relevancy-label distribution; \textit{AIR} is the percentage of first Relevant replies after which the next system turn is final; \textit{Reach} is the percentage of conversations that obtain a Relevant reply; and \textit{TTR} is the number of clarifications up to the first Relevant one. Superscripts give the larger distance to the bounds of the 95\% interval. $^\dagger$Held-out system.}}
\label{tab:interaction_stats_spider}
\end{table*}

\FloatBarrier
\input{prompts_appendix}

\end{document}

%% file: prompts_appendix.tex
{
\section{Prompts}\label{app:prompts}

This appendix gives the prompt templates of all model roles in the generation pipeline (Section~\ref{sec:generation_pipeline}), the validation stages (Appendix~\ref{app:validation_detail}) and ABISS (Section~\ref{sec:abiss}), as our code renders them. Placeholders in capitals between angle brackets, such as \texttt{<SCHEMA>}, stand for the content of an instance, whereas angle brackets inside a JSON template belong to the prompt. A line that holds an optional field, such as the evidence or the hidden knowledge, appears only when the instance has that field. We break long lines and indent each continuation by two spaces. Each prompt is the single user message of a chat rendered with the model's own template, and the code adds no system message. Every role reasons before it ends its answer with the JSON object of its template.

\paragraph{Copy into the format.} No role has a separate prompt for the copy step. After the free answer (Appendix~\ref{app:setup}), the engine continues the model's own output with a line break, after closing any reasoning that is still open, and decodes the JSON object greedily under the role's JSON schema or, for the tested system, under a grammar that writes the same JSON. A free answer that already ends with a JSON object fitting the schema is used as it is. The schema can be stricter than the template in the prompt: every verdict field admits only its listed values, the classifier's field only the 13 subcategory labels, the tested system's answer exactly one non-null action with at least one visible character, and the simulated user's \texttt{node\_ids} only the listed nodes and, for an answerable question, no \texttt{FROM}, \texttt{JOIN}, \texttt{WHERE}, \texttt{HAVING} or non-aggregate \texttt{SELECT} node.

\subsection*{Shared Blocks}

\paragraph{Database schema.} \texttt{<SCHEMA>} stands for every table of the database in the format below, separated by blank lines; only BIRD tables have column descriptions. All roles that see the schema receive the same block; the simulated user does not see it.

{\scriptsize
\begin{verbatim}
<CREATE TABLE STATEMENT>
 /*
 Column descriptions:
 <COLUMN>: <EXPANDED NAME>. <DESCRIPTION>. Values:
   <VALUES>
 ...
 */
 /*
 5 example rows:
 SELECT * FROM "<TABLE>" LIMIT 5;
 <COLUMN NAMES>
 <ROW 1>
 ...
 <ROW 5>
 */
\end{verbatim}
}
\paragraph{Category block.} \texttt{<CATEGORY BLOCK>} stands for a subcategory's label, definition and examples in the layout below. In the tested system's prompts the block also has the type line (\textit{Answerable}, \textit{Ambiguous} or \textit{Unanswerable}) and is written \texttt{<CATEGORY BLOCK WITH TYPE>}. The texts of the 13 subcategories are in the category classes of the released code (folder \texttt{categories}); the question writer below shows those of Attachment Ambiguity.

{\scriptsize
\begin{verbatim}
**Category:** <LABEL>
**Type:** <TYPE>
**Definition:** <DEFINITION>
**Examples:**
- <EXAMPLE 1>
- <EXAMPLE 2>
\end{verbatim}
}
\paragraph{Conversation history.} \texttt{<CONVERSATION HISTORY>} stands for the completed clarification turns, here two. Before the first answer it is empty, and the simulated user's prompts then also leave out the blank line after it.

{\scriptsize
\begin{verbatim}
The conversation so far, after the question above:
System: <SYSTEM QUESTION 1>
User: <USER ANSWER 1>
System: <SYSTEM QUESTION 2>
User: <USER ANSWER 2>
\end{verbatim}
}
\paragraph{Result preview.} In Stages~3, 4, 7 and~9, \texttt{<RESULT PREVIEW>} stands for the result of the SQL query under judgment, as below. With five rows or fewer, the first line lists all rows after a colon; with none, it ends after the column names. A text value longer than 100 characters is cut and followed by its full length. If the query does not run, Stages~4 and~7 show \texttt{**Query Execution Results:** Query execution returned no results.}\ instead, and Stages~3 and~9 show nothing.

{\scriptsize
\begin{verbatim}
**Query Execution Results:** <N> rows with the columns
  <COLUMNS>. The first 5 rows: <FIRST 5 ROWS>
**Sizes of the tables that the query reads:**
  <TABLE 1> <N 1> rows, <TABLE 2> <N 2> rows
\end{verbatim}
}
\paragraph{Question styles.} Stage~10 receives the required style's description up to its first blank line (\texttt{<STYLE DESCRIPTION>}). The question writer receives it with the example question and distinguishing features (\texttt{<STYLE DESCRIPTION WITH EXAMPLE QUESTION>}), and the simulated user with the example answer (\texttt{<STYLE DESCRIPTION WITH EXAMPLE ANSWER>}), except in the interrogative style, whose answer description is the second block.

{\scriptsize
\begin{verbatim}
**Formal Style:**
- Uses standard grammar and professional vocabulary
- Complete, well-structured sentences with proper
  punctuation
- Neutral tone without emotional language
- First-person wording such as "I would like to know"
  is acceptable, because formality is a matter of
  register, not of grammatical person
- No contractions, slang, or colloquialisms
- Clear and precise expression following formal
  writing conventions

**Example Question:** "Find all students older than 18
  years and return their home addresses."
**Distinguishing Features:** Standard grammar, neutral
  tone, formal vocabulary.

**Example Answer:** "Provide all records from the
  engineering department pertaining to senior-level
  positions."

**Colloquial Style:**
- Employs informal vocabulary and conversational
  expressions
- Includes interjections ("Hey!", "Oh"), filler words,
  and casual phrases
- Uses contractions and relaxed grammar as in everyday
  speech
- May include personal pronouns (I, me, you) and
  emotional expressions
- Friendly, approachable tone with phrases like "I'd
  love to", "Could you help me"

**Example Question:** "Hey! Could you help me find all
  the students who are over 18? I'd love to know their
  names and home addresses."
**Distinguishing Features:** Interjection "Hey!",
  "Could you help me", "I'd love to", conversational
  tone.

**Example Answer:** "Hey! I mean the engineering
  folks, and I'd really love to see just the senior
  people there."

**Interrogative Style:**
- Uses question forms that explicitly seek information
- Begins with phrases like: "Could you tell me", "Can
  you show me", "Would you explain"
- Or starts with interrogative words: "What", "Which",
  "Who", "Where", "How"
- Always ends with a question mark
- Polite inquiry tone without urgency or commands

**Example Question:** "Could you tell me which
  students are older than 18 and what their home
  addresses are?"
**Distinguishing Features:** "Could you tell me",
  question mark, inquiry-focused without urgency.
\end{verbatim}
}
{\scriptsize
\begin{verbatim}
**Interrogative Style:**
- Polite, measured tone as in the original question,
  without urgency or commands
- Gives the answer as a declarative sentence, not as a
  new question
- Never ends with a question mark

**Example Answer:** "I mean the employees in the
  engineering department who hold senior positions."
\end{verbatim}
}
\paragraph{SQL difficulty levels.} The question writer receives one of these criteria as \texttt{<DIFFICULTY CRITERIA>}.

{\scriptsize
\begin{verbatim}
**Simple SQL Criteria:**
Simple SQL queries satisfy one or more of the
  following:
- Select data from a single table only
- Use basic filtering with WHERE clauses
- Apply simple sorting with ORDER BY
- Include basic aggregate functions (COUNT, SUM, AVG,
  MIN, MAX) without GROUP BY
- No joins, subqueries, or advanced SQL features

**Example:**
```sql
SELECT name, department_name
FROM employees
WHERE level > 5
ORDER BY age DESC;
```

**Moderate SQL Criteria:**
Moderate SQL queries satisfy one or more of the
  following:
- Use table joins (INNER JOIN, LEFT JOIN, RIGHT JOIN,
  CROSS JOIN)
- Include subqueries in SELECT or WHERE clauses
- Combine aggregate functions with GROUP BY
- Use complex WHERE conditions (IN, BETWEEN, LIKE,
  multiple AND/OR)
- Apply HAVING clauses to filter grouped results
- Involve multiple aggregate functions

**Example:**
```sql
SELECT e.name, d.department_name, AVG(s.salary) AS
  average_salary
FROM employees e
INNER JOIN departments d ON e.department_id =
  d.department_id
LEFT JOIN salaries s ON e.employee_id = s.employee_id
WHERE e.age > 30 AND e.status = 'active'
GROUP BY e.name, d.department_name
HAVING AVG(s.salary) > 50000;
```

**Complex SQL Criteria:**
Complex SQL queries satisfy one or more of the
  following:
- Use nested subqueries with multiple levels
- Apply multiple types of joins, including self-joins
- Implement window functions (ROW_NUMBER, RANK,
  DENSE_RANK, LAG, LEAD)
- Use Common Table Expressions (CTEs) for query
  organization
- Combine multiple aggregate functions with complex
  grouping
- Include complex WHERE and HAVING clauses with nested
  conditions
- Apply advanced SQL functions and operators

**Example:**
```sql
WITH EmployeeCTE AS (
    SELECT employee_id, name, department_id,
           ROW_NUMBER() OVER (PARTITION BY
             department_id ORDER BY salary DESC) AS
             rank
    FROM employees
)
SELECT e.name, d.department_name
FROM EmployeeCTE e
INNER JOIN departments d ON e.department_id =
  d.department_id
WHERE e.rank <= 3;
```

**Highly Complex SQL Criteria:**
Highly complex SQL queries satisfy one or more of the
  following:
- Use multiple CTEs with interdependencies
- Implement recursive CTEs for hierarchical data
  traversal
- Combine nested subqueries with various join types
- Apply extensive window functions with complex
  partitioning
- Use set operations (UNION, UNION ALL, INTERSECT or
  EXCEPT) to combine multiple result sets
- Implement advanced analytical functions and complex
  aggregations
- Employ a wide range of SQL clauses with intricate
  logic
- Utilize advanced SQL features like PIVOT, lateral
  joins, or array operations

**Example:**
```sql
WITH RECURSIVE EmployeeHierarchy AS (
    SELECT employee_id, name, manager_id,
      department_id, 1 as level
    FROM employees
    WHERE manager_id IS NULL
    UNION ALL
    SELECT e.employee_id, e.name, e.manager_id,
      e.department_id, eh.level + 1
    FROM employees e
    JOIN EmployeeHierarchy eh ON e.manager_id =
      eh.employee_id
),
DepartmentSalaries AS (
    SELECT eh.employee_id, eh.name, eh.level,
      d.department_name, s.salary, d.department_id
    FROM EmployeeHierarchy eh
    INNER JOIN departments d ON eh.department_id =
      d.department_id
    INNER JOIN salaries s ON eh.employee_id =
      s.employee_id
),
DepartmentStats AS (
    SELECT
        d.department_id,
        COUNT(e.employee_id) AS employee_count,
        AVG(s.salary) AS average_salary
    FROM employees e
    INNER JOIN salaries s ON e.employee_id =
      s.employee_id
    INNER JOIN departments d ON e.department_id =
      d.department_id
    GROUP BY d.department_id
)
SELECT ds.name, ds.level,
       SUM(ds.salary) OVER (PARTITION BY
         ds.department_id ORDER BY ds.level, ds.name)
         AS cumulative_salary
FROM DepartmentSalaries ds
INNER JOIN DepartmentStats dstat ON ds.department_id =
  dstat.department_id
ORDER BY ds.level, ds.name;
```
\end{verbatim}
}
\subsection*{Generation}

\paragraph{Question writer.} The writer receives the schema, a subcategory with its output format, a style and a difficulty level, and returns the question and the fields of the subcategory's output format. Below is the template of ambiguous subcategories, with the subcategory-specific parts (Category Details, Output Format and the second item of the Generation Process) of Attachment Ambiguity, whose fields are its two hidden-knowledge texts.

{\scriptsize
\begin{verbatim}
You are an expert in creating text-to-SQL benchmarks
  for natural language questions. Your task is to
  generate realistic, high-quality questions that test
  the boundaries of text-to-SQL systems.

## Database Schema
Generate questions based on the following database
  schema:

<SCHEMA>

## Task Overview
Generate one natural language question that:
1. Is genuinely ambiguous and has multiple valid
  interpretations
2. Cannot be answered without additional clarification
  from the user
3. Is a realistic question a user might actually ask
  on the given database
4. Can be resolved by using the hidden knowledge which
  represents the disambiguation information
5. Fits the specific category defined below

## Category Details
**Category:** Structure Ambiguity (Attachment
  Ambiguity)
**Definition:** Attachment Ambiguity occurs when a
  modifier follows a conjunction of two nouns and can
  attach either to the nearest noun only (low
  attachment) or to both (high attachment): in
  'professors and students in engineering', either
  only the students or both groups must be in
  engineering. The modifier is part of the noun
  phrase, such as a prepositional phrase, a relative
  clause ('suppliers and manufacturers who are
  certified') or a participle ('managers and employees
  hired after 2020'). The two nouns must name separate
  sets of entities, and the modifier must fit both.
  Two words that describe one noun ('male and female
  clients', 'home and away teams') and a condition
  that is the predicate of the sentence ('which
  clients and accounts are located in Prague') do not
  create this ambiguity. The two readings produce SQL
  queries with different filtering scopes. This is not
  about whether a quantified or plural noun phrase,
  such as 'every department' or 'all stores', applies
  to the entities together or to each one separately.
**Examples:**
- List the professors and students in engineering.
  (Attachment ambiguity: does 'in engineering' modify
  only 'students' or both 'professors and students'?)
- Find the suppliers and manufacturers who are
  certified organic. (Attachment ambiguity: does the
  relative clause 'who are certified organic' apply
  only to 'manufacturers' or to both 'suppliers and
  manufacturers'?)
- Display the managers and employees hired after 2020.
  (Attachment ambiguity: does the participial phrase
  'hired after 2020' filter only 'employees' or both
  'managers and employees'?)
- Show the songs and albums released in 2023.
  (Attachment ambiguity: does 'released in 2023' apply
  only to 'albums' or to both 'songs and albums'?)

## Output Format
Provide one JSON object with the following fields:
{
    "question": "<A natural language question in which
      a modifier follows a conjunction of two separate
      nouns ('A and B [modifier]'), belongs to their
      noun phrase and fits both nouns, so that it can
      attach to the nearest noun only (low attachment)
      or to both (high attachment). The ambiguity must
      arise purely from sentence structure.>",
    "hidden_knowledge_last_only": "<A statement
      clarifying that the modifier attaches only to
      the nearest element (low attachment). It should
      clearly identify which element(s) the modifier
      applies to and which it does not. For example:
      'The condition in engineering applies only to
      students, not to professors.'>",
    "hidden_knowledge_all_elements": "<A statement
      clarifying that the modifier attaches to the
      entire conjunction (high attachment). It should
      clearly state that all elements are affected by
      the modifier. For example: 'The condition in
      engineering applies to both professors and
      students.'>"
}

## Quality Guidelines
Ensure that:
- The ambiguity is **genuine and non-trivial**:
  Multiple interpretations should be reasonable
- The question MUST **strictly isolate the target
  category**: It should not exhibit traits or
  confusing overlaps with any other category in the
  taxonomy
- The question sounds **natural and realistic**:
  Something a real user would ask
- The hidden knowledge clearly **resolves the
  ambiguity**
- The question complexity **matches the difficulty
  level specified below**
- The question style **matches the style requirements
  specified below**

## Generation Process
Develop your question through an iterative refinement
  process, moving back and forth between the output
  fields as needed to ensure coherence and quality.
  Consider the output schema fields in the order they
  are defined, but feel free to revise earlier fields
  based on insights from later ones:

- **question**: Craft a natural language question that
  fits the category and style
- **hidden_knowledge_last_only**,
  **hidden_knowledge_all_elements**: Fill each field
  as its description in the Output Format asks
- Iterate between these fields to ensure the question
  genuinely requires the hidden knowledge, and that
  each piece of hidden knowledge resolves the
  ambiguity toward one interpretation

## Question Style Requirements
The question MUST be written in the following style:
<STYLE DESCRIPTION WITH EXAMPLE QUESTION>

## SQL Difficulty Requirements
The SQL query (or queries for ambiguous questions)
  MUST match the following difficulty level:
<DIFFICULTY CRITERIA>

Generate the question sample now: think step by step,
  then end your answer with the JSON object described
  in the Output Format section.
\end{verbatim}
}
For answerable subcategories, items 1--4 of the Task Overview, the Quality Guidelines before the two shared ones, and the last item of the Generation Process read:

{\scriptsize
\begin{verbatim}
1. Can be directly answered using the database without
  ambiguity or missing information
2. Clearly maps to the database schema with all
  necessary information available
3. Is a realistic question a user might actually ask
  on the given database and evidence (if available)
4. Has a single, unambiguous interpretation

- The question is **unambiguous and clear**: There is
  only one reasonable interpretation
- The question sounds **natural and realistic**:
  Something a real user would ask
- All information needed to answer the question is
  **available in the schema or evidence (if
  available)**

- Ensure the question clearly and unambiguously maps
  to the database schema
\end{verbatim}
}
For unanswerable subcategories they read as follows, and the SQL Difficulty Requirements as shown last:

{\scriptsize
\begin{verbatim}
1. Appears to be a valid question a user might ask
2. Cannot be answered with either the current database
  schema, evidence, or both
3. Is a realistic and well-formed question
4. Is not trivially solvable or easily answerable

- The question is **well-formed and natural**:
  Something a real user would reasonably ask
- The question **clearly cannot be answered**, for the
  reason that the category definition gives
- The question MUST **strictly isolate the target
  category**: It should not exhibit traits or
  confusing overlaps with any other category in the
  taxonomy
- Avoid questions that the database can answer after
  all, for example through joins or aggregations

- Iterate between these fields to ensure the question
  clearly cannot be answered and the feedback
  precisely explains why

If this question were answerable, the expected SQL
  query should match the following difficulty level:
<DIFFICULTY CRITERIA>
Consider this when crafting the question's complexity
  and scope.
\end{verbatim}
}
\paragraph{SQL writer.} The writer receives the schema, the question, its evidence and, for an ambiguous interpretation, its hidden knowledge, together with the last rule of Step~2, and returns \texttt{sql}.

{\scriptsize
\begin{verbatim}
You are an expert SQL developer specializing in
  text-to-SQL systems.

## Database Schema
<SCHEMA>

## Chain-of-Thought Instructions
Follow these steps to generate the SQL query:

**Step 1 - Schema Linking:**
Identify the relevant tables, columns, and foreign key
  relationships for the question below. List them
  explicitly.

**Step 2 - SQL Generation:**
Write the final SQL query following these rules:
- Use valid SQLite syntax
- Use double quotes for identifiers with spaces or
  special characters
- Use explicit aliases for aggregated or calculated
  columns (e.g., `SELECT COUNT(*) AS count ...`)
- The query must be executable and return the correct
  results
- Return only the values that the question explicitly
  asks for. Do not add identifier, key, name or
  grouping columns, or intermediate values, unless the
  question asks for them.
- For a result per group or per entity, return only
  the requested values. For example, for "What is the
  average salary in each department?" return the
  average salary, one row per department, without the
  department column.
- When the question asks for entities without naming
  an attribute, as in "List the schools that ...",
  return their name column if the table has one,
  otherwise their primary key.
- The SQL must faithfully represent the interpretation
  described in the hidden knowledge

## Output Format
Provide your SQL query as a JSON object with:
{
    "sql": "<The SQL query that answers the given
      question based on the database schema.>"
}

## Question
**Natural Language Question:** <QUESTION>

**Evidence:** <EVIDENCE>

**Hidden Knowledge:** <HIDDEN KNOWLEDGE>

Generate the SQL query now: think step by step, then
  end your answer with the JSON object described in
  the Output Format section.
\end{verbatim}
}
\subsection*{Validation}

\paragraph{SQL attempts of Stages~5 to~7.} The council's attempts use the SQL writer's prompt with the information of each stage: Stage~5 gives the question without its evidence, Stage~6 gives what the writer received, and Stage~7 gives the question and its evidence.

\paragraph{Stage~3.} Each judge receives the schema, the subcategory and the question section, and returns \texttt{belongs\_to\_category} (Yes or No).

{\scriptsize
\begin{verbatim}
You are an expert in text-to-SQL ambiguity and
  unanswerability classification. Your task is to
  determine whether a question genuinely belongs to a
  specific category based on the category's
  definition, characteristics, and examples.

## Classification Task
Determine whether the question genuinely belongs to
  the specified category. A question belongs to a
  category if it clearly exhibits the core
  characteristics described in the category's
  definition.

**Evaluation Criteria (in order of importance):**
1. **Core Definition Alignment**: Does the question
  exhibit the fundamental characteristics described in
  the category's definition?
2. **Distinctive Features**: Does the question display
  the specific, distinguishing features of this
  category?
3. **Problem Type Match**: Does the underlying issue
  in the question match the type of problem this
  category describes?
4. **Disambiguation or Feedback Alignment**: If the
  question has two interpretations, each with its
  disambiguation information, do both follow naturally
  from the question as written, and does the
  difference between them match the nature of the
  ambiguity described by the category? If the question
  has unsolvability feedback, does the feedback match
  the specific limitation described by the category?
  Skip this criterion if the question has neither.
5. **Example Similarity**: Is the question similar to
  the category's examples?

**Answer 'Yes' if:** The question clearly and
  genuinely belongs to this category based on the
  criteria above.
**Answer 'No' if:** The question does not fit this
  category, or fits it only superficially without
  exhibiting its core characteristics.

## Response Format
Provide a brief analysis addressing:
1. **Question Analysis**: Identify the key
  characteristics of the question
2. **Category Alignment**: How well does the question
  match the category's definition and examples?
3. **Verdict**: Does it belong to this category?

Keep your reasoning concise (approximately 100-200
  words). Be decisive and objective.

Then provide your final verdict as a JSON object with:
{
    "belongs_to_category": "<'Yes' if the question
      genuinely belongs to the specified category,
      'No' otherwise. Put only 'Yes' or 'No'.>"
}

## Database Schema
<SCHEMA>

## Category
<CATEGORY BLOCK>

<QUESTION SECTION>

Think step by step, then end your answer with the JSON
  object described in the Response Format section.
\end{verbatim}
}
\paragraph{Question section of Stages~3 and~9.} \texttt{<QUESTION SECTION>} shows an ambiguous question with both interpretations, an answerable question with its Ground Truth SQL, or an unanswerable question with its feedback:

{\scriptsize
\begin{verbatim}
## Question Under Analysis
**Natural Language Question:** <QUESTION>
**Evidence:** <EVIDENCE>
The question has 2 interpretations. Each comes with
  the disambiguation information that selects it and
  its ground truth SQL query.
### Interpretation 1
**Disambiguation Information** (clarifies the user's
  intended interpretation among the possible ones):
<HIDDEN KNOWLEDGE 1>
**Ground Truth SQL Query:** <SQL 1>
<RESULT PREVIEW>
### Interpretation 2
**Disambiguation Information** (clarifies the user's
  intended interpretation among the possible ones):
<HIDDEN KNOWLEDGE 2>
**Ground Truth SQL Query:** <SQL 2>
<RESULT PREVIEW>

## Question Under Analysis
**Natural Language Question:** <QUESTION>
**Evidence:** <EVIDENCE>
**Ground Truth SQL Query:** <SQL>
<RESULT PREVIEW>

## Question Under Analysis
**Natural Language Question:** <QUESTION>
**Evidence:** <EVIDENCE>
**Unsolvability Feedback** (explains why the question
  cannot be answered with the current
  schema/knowledge):
<FEEDBACK>
\end{verbatim}
}
\paragraph{Stage~4.} Each judge receives the schema, the question with its evidence, its Ground Truth SQL with the result preview and, for an ambiguous interpretation, its hidden knowledge, and returns \texttt{answer} (Yes or No). For an answerable question the prompt is:

{\scriptsize
\begin{verbatim}
You are an expert SQL validator for text-to-SQL
  benchmarks. Your task is to verify whether a ground
  truth SQL query correctly answers a natural language
  question.

## Context
This is a straightforward question that can be
  answered directly using the database schema. Your
  job is to evaluate whether the provided SQL query
  correctly answers the question.

## Validation Task
Answer **Yes** if:
- The SQL correctly uses tables and columns from the
  schema
- The query logic (joins, WHERE clauses, aggregations,
  etc.) matches the question's intent
- The results align with what the question asks for
- The query returns the values that the question asks
  for. A query that leaves out identifier or label
  columns that the question does not request is
  complete.

Answer **No** if:
- The SQL uses incorrect or non-existent schema
  elements
- The query logic doesn't match the question's
  requirements
- The query would return incorrect or incomplete
  results

## Response Format
Before answering, consider whether the SQL query
  correctly implements the question's intent: (1)
  whether the query uses the correct tables and
  columns from the schema, (2) whether the joins,
  filters, and aggregations match the question's
  requirements, and (3) whether the query results
  align with what the question is asking for.

Then provide your evaluation as a JSON object with:
{
    "answer": "<Final verdict: 'Yes' if the SQL query
      correctly answers the question, or 'No' if it
      fails to capture the intended meaning or
      contains errors. Put only 'Yes' or 'No'.>"
}

## Database Schema
<SCHEMA>

## Question Information
**Natural Language Question:** <QUESTION>
**Evidence:** <EVIDENCE>
**Ground Truth SQL Query:** <SQL>
<RESULT PREVIEW>

Think step by step, then end your answer with the JSON
  object described in the Response Format section.
\end{verbatim}
}
For an ambiguous interpretation, the text before the JSON template reads as follows, and the hidden-knowledge line shown last precedes the SQL line.

{\scriptsize
\begin{verbatim}
You are an expert SQL validator for text-to-SQL
  benchmarks. Your task is to verify whether a ground
  truth SQL query correctly answers a natural language
  question after disambiguation through hidden
  knowledge.

## Context
The original question was ambiguous or underspecified.
  The user has provided hidden knowledge that
  clarifies their intent. Your job is to evaluate
  whether the provided SQL query correctly implements
  this disambiguated intent and reflects the intended
  meaning of the question.

## Validation Task
Answer **Yes** if:
- The SQL correctly uses tables and columns from the
  schema
- The query logic (joins, WHERE clauses, aggregations,
  etc.) matches the disambiguated intent
- The results align with what the question asks for
- The query returns the values that the question asks
  for. A query that leaves out identifier or label
  columns that the question does not request is
  complete.
- The hidden knowledge is properly reflected in the
  SQL structure

Answer **No** if:
- The SQL uses incorrect or non-existent schema
  elements
- The query logic doesn't match the question's
  requirements
- The hidden knowledge is ignored or incorrectly
  applied
- The query includes unnecessary columns, JOINs, or
  table references that belong to an alternative
  interpretation of the ambiguity rather than the one
  specified by the hidden knowledge
- The query would return incorrect or incomplete
  results

## Response Format
Before answering, consider whether the SQL query
  correctly implements the disambiguated intent: (1)
  whether the query uses the correct tables and
  columns from the schema, (2) whether the joins,
  filters, and aggregations match the question's
  requirements given the hidden knowledge, (3) whether
  the query results align with what the question is
  asking for, and (4) whether the query includes any
  unnecessary elements (columns, JOINs, or tables)
  that belong to an alternative interpretation rather
  than the one specified by the hidden knowledge.

**Hidden Knowledge (Disambiguating Information):**
  <HIDDEN KNOWLEDGE>
\end{verbatim}
}
\paragraph{Stage~7 judges.} Each judge receives the schema, the question with its evidence and one council model's SQL attempt as a candidate, without the question's feedback, and returns \texttt{answer} (Yes or No). The prompt is that of Stage~4 for an answerable question, except that the first two paragraphs read as below, the assumed-value rule shown next precedes the last No criterion, and the SQL line reads as shown last.

{\scriptsize
\begin{verbatim}
You are an expert SQL validator for text-to-SQL
  benchmarks. Your task is to verify whether a
  candidate SQL query correctly answers a natural
  language question.

## Context
Your job is to evaluate whether the candidate SQL
  query answers the question as it is asked. When the
  database does not hold what the question asks for, a
  query that approximates the request, for example
  through a related stored value, an assumed threshold
  or definition, or a simplified formula, does not
  answer it. Rejecting such a query is better than
  accepting an answer that does not match what the
  user asked for, even when it is the best query that
  the schema allows.

- The query depends on a value or definition that is
  not in the database, the question or its evidence,
  such as an answer written into the query as a
  constant, an assumed threshold or definition, or a
  statistic computed from the database in place of an
  outside one

**Candidate SQL Query:** <CANDIDATE SQL>
\end{verbatim}
}
\paragraph{Stage~8.} Each judge receives the schema, the subcategory, the question with its evidence and its feedback, and returns \texttt{is\_valid} (Yes or No).

{\scriptsize
\begin{verbatim}
You are an expert validator for text-to-SQL question
  generation. Your task is to assess whether the
  provided feedback correctly explains why a question
  is fundamentally unsolvable.

## Validation Task
Determine whether the feedback correctly and clearly
  explains why the question cannot be answered.

**The feedback should:**
- Correctly identify the type of unsolvability that
  the category definition describes
- Clearly specify what information, schema elements,
  or knowledge is missing or problematic
- Be specific enough to be useful (not vague or
  generic)
- Align with the category definition and the question
  content
- Explain the fundamental problem that makes the
  question unanswerable

**Consider it VALID if:**
- The feedback identifies the correct type of problem
  for this category
- The explanation is clear, specific, and actionable
- It explains what is missing or wrong without being
  too vague
- The reasoning aligns with both the question and the
  category definition

**Consider it INVALID if:**
- The feedback identifies the wrong type of
  unsolvability
- The explanation is too vague or generic (e.g.,
  "missing information")
- It doesn't specify what exactly is missing or
  problematic
- The feedback contradicts the category definition
- The feedback suggests the question could be answered
  with simple clarification
- The feedback relies only on a missing declared
  constraint, such as a foreign key, while a join on
  matching columns could still connect the data, or it
  admits an indirect path that would answer the
  question

## Response Format
Provide an analysis addressing:
1. **Category Alignment:** Does the feedback match the
  unsolvability category type?
2. **Specificity:** Is the feedback specific enough?
  What exactly is missing or wrong?
3. **Clarity:** Is the explanation clear and
  understandable?
4. **Correctness:** Does the feedback correctly
  identify why this question is fundamentally
  unsolvable?

Your analysis should be concise but thorough
  (approximately 512 characters).

Then provide your final verdict as a JSON object with:
{
    "is_valid": "<'Yes' if the feedback correctly and
      clearly explains why the question is unsolvable,
      'No' otherwise. Put only 'Yes' or 'No'.>"
}

## Database Schema
<SCHEMA>

## Question Category
<CATEGORY BLOCK>

## Question Information
**Natural Language Question:** <QUESTION>
**Evidence:** <EVIDENCE>

## Provided Feedback
This is the feedback that explains why the question is
  unsolvable:
**Feedback:** <FEEDBACK>

Remember: Focus on whether the feedback correctly and
  clearly explains the unsolvability.

Think step by step, then end your answer with the JSON
  object described in the Response Format section.
\end{verbatim}
}
\paragraph{Stage~9.} Each judge receives the schema, the question section of Stage~3, and the question's subcategory and one competitor as Categories~A and~B, in an order drawn from the comparison's content, and returns \texttt{answer} (A or B). A vote for the competitor counts only if the judge repeats it when the two categories are swapped.

{\scriptsize
\begin{verbatim}
You are an expert in text-to-SQL ambiguity and
  unanswerability classification. Your task is to
  perform a comparative analysis between two
  categories and determine which one a question more
  closely aligns with based on their definitions,
  characteristics, and examples.

## Comparative Evaluation Task
You must select which category the question fits
  BETTER. This is a forced choice between Category A
  and Category B. Even if the question partially fits
  both or neither perfectly, you must determine which
  category's core characteristics and defining
  features the question more strongly exhibits.

**Evaluation Criteria (in order of importance):**
1. **Core Definition Alignment**: Which category's
  fundamental definition the question best
  exemplifies. Pay close attention to subtle
  distinctions.
2. **Distinctive Characteristics**: Which category's
  unique/specific features are present in the question
3. **Problem Type Match**: Whether the underlying
  issue matches one category's problem type better
4. **Disambiguation or Feedback Alignment**: If the
  question has two interpretations, each with its
  disambiguation information, which category's type of
  ambiguity better describes the difference between
  the two readings (i.e., the difference should match
  the nature of the ambiguity described by the
  category). If the question has unsolvability
  feedback, which category's unsolvability cause the
  feedback better describes (i.e., the feedback should
  match the specific limitation described by the
  category)
5. **Example Similarity**: Which category's examples
  are most similar to the question

**Select 'A' if:** Category A is the better fit based
  on the criteria above.
**Select 'B' if:** Category B is the better fit based
  on the criteria above.

## Response Format
Provide a structured comparative analysis with the
  following sections:
1. **Question Analysis**: Briefly identify the key
  characteristics of the question
2. **Category A Match**: Evaluate how well the
  question aligns with Category A (consider
  definition, characteristics, examples)
3. **Category B Match**: Evaluate how well the
  question aligns with Category B (consider
  definition, characteristics, examples)
4. **Final Determination**: Clearly state which
  category is the better fit and why

Keep your reasoning concise but thorough
  (approximately 200-400 words total). Be decisive and
  objective in your comparison.

Then provide your final selection as a JSON object
  with:
{
    "answer": "<Final selection: 'A' if the question
      better fits Category A, 'B' if it better fits
      Category B. Put only 'A' or 'B'.>"
}

## Database Schema
<SCHEMA>

<QUESTION SECTION>

## Category A
<CATEGORY A BLOCK>

## Category B
<CATEGORY B BLOCK>

Important: Choose the category that most closely
  matches the question's primary characteristics. If
  the question exhibits features of both categories,
  select the one with the STRONGEST alignment.

Think step by step, then end your answer with the JSON
  object described in the Response Format section.
\end{verbatim}
}
\paragraph{Stage~10.} Each judge receives the schema, the required style (\texttt{<STYLE NAME>} is \texttt{formal}, \texttt{colloquial} or \texttt{interrogative}) and the question with its evidence, and returns \texttt{style\_matches} (Yes or No). The two interpretations of an ambiguous question share this prompt.

{\scriptsize
\begin{verbatim}
You are an expert in evaluating natural language
  question styles. Your task is to verify whether a
  generated question's style matches its specified
  requirements.

## Validation Task
Verify that the natural language question matches the
  specified style:
- Check the vocabulary, tone, and sentence structure
- Compare against the style description in the Style
  Requirement section
- Ensure the question exhibits the key characteristics
  of the specified style

**Answer 'Yes' for style_matches if:** The question
  clearly follows the specified style requirements.
**Answer 'No' for style_matches if:** The question
  uses a different style or doesn't match the
  requirements.

## Response Format
Provide a detailed analysis addressing:
**Style Analysis:** How well does the question match
  the specified style? What characteristics support or
  contradict this?

Then provide your final verdict as a JSON object with:
{
    "style_matches": "<'Yes' if the question's style
      matches the specified style requirements, 'No'
      otherwise. Put only 'Yes' or 'No'.>"
}

## Database Schema
<SCHEMA>

## Style Requirement
**Required Style:** <STYLE NAME>
<STYLE DESCRIPTION>

## Question Information
**Natural Language Question:** <QUESTION>
**Evidence:** <EVIDENCE>

Think step by step, then end your answer with the JSON
  object described in the Response Format section.
\end{verbatim}
}
\subsection*{Benchmark}

\paragraph{Classification phase.} The tested system receives the 13 category blocks, the schema and the question with its evidence, and returns \texttt{category}.

{\scriptsize
\begin{verbatim}
You are an expert in analyzing natural language
  questions for text-to-SQL systems. Your task is to
  classify a question into one of several categories
  based on whether it can be directly answered with a
  SQL query or requires clarification/additional
  information.

## Available Categories
Analyze the question and classify it into ONE of the
  following categories. The label after **Category:**
  is the exact value to answer with:

<CATEGORY BLOCK WITH TYPE 1>

...

<CATEGORY BLOCK WITH TYPE 13>

## Classification Task
Carefully analyze the question and determine which
  category it belongs to. Consider:
- Can the question be directly answered with the given
  schema?
- Is there any ambiguity or missing information?
- If ambiguous, can it be resolved through user
  interaction?
- If information is missing, is it schema-related or
  external knowledge?

## Output Format
Provide your classification as a JSON object with:
{
    "category": "<The category that best describes
      this question. Must be one of: 'Answerable
      (Without Evidence)', 'Answerable (With
      Evidence)', 'Conflicting Knowledge', 'Improper
      Question', 'Lexical Vagueness', 'Missing
      External Knowledge', 'Missing Schema Elements
      (Missing Entities or Attributes)', 'Missing
      Schema Elements (Missing Relationships)',
      'Missing User Knowledge', 'Semantic Mapping
      Ambiguity (Entity Ambiguity)', 'Semantic Mapping
      Ambiguity (Lexical Overlap)', 'Structure
      Ambiguity (Attachment Ambiguity)', 'Structure
      Ambiguity (Scope Ambiguity)'.>"
}

## Database Schema
<SCHEMA>

## Question to Classify
**Natural Language Question:** <QUESTION>
**Evidence:** <EVIDENCE>

Classify the question now: think step by step, then
  end your answer with the JSON object described in
  the Output Format section.
\end{verbatim}
}
\paragraph{Interactive phase.} At each turn, the tested system receives the guide to the question types, the response options, the schema, the subcategory information of the mode, the question with its evidence and the completed turns, and the turn status. It returns \texttt{system\_question}, \texttt{system\_sql} and \texttt{system\_feedback}, of which exactly one is non-null. Below is the first turn in the \textit{Ground Truth} and \textit{Predicted} modes, with our limit of three clarification questions; \texttt{<CATEGORY BLOCK WITH TYPE>} is the question's subcategory or the classifier's prediction. Neither mode adds an instruction tied to the type, and the guide to the question types is the same in all four modes.

{\scriptsize
\begin{verbatim}
You are an expert text-to-SQL assistant. Your job is
  to help a user get the correct answer from a
  database by generating SQL, asking clarification
  questions when the question is ambiguous, or
  explaining why a question cannot be answered.

## Question Types and Expected Behavior
Every user question falls into one of three types.
  Your response must match the type:

1. **Answerable** — The question can be directly
  translated into SQL with the given schema and
  context. You should generate a SQL query without
  asking clarification questions.
2. **Ambiguous** — The question has a genuine
  natural-language ambiguity (e.g., a vague term, an
  unclear entity reference, a structural reading that
  could go two ways). You should ask a focused
  clarification question that targets the specific
  ambiguity, then generate SQL once the user
  clarifies. Do NOT ask about anything that the
  schema, the evidence or the conversation already
  settles.
3. **Unanswerable** — The question requires data,
  relationships, or external knowledge that the
  database does not contain. No amount of
  clarification can fix this. When the database does
  not hold what the question asks for, a query that
  approximates the request, for example through a
  related stored value, an assumed threshold or
  definition, or a simplified formula, does not answer
  it. You should provide feedback explaining exactly
  what is missing and why the question cannot be
  answered.

## Response Options
Provide exactly ONE of the following:

**Option 1: Clarification Question** (system_question)
Use when the question is ambiguous and you need the
  user to disambiguate. Your question should be
  specific, non-technical and directly target the
  ambiguity. It must be atomic: ask exactly one thing,
  not multiple subquestions, not a list, and not a
  semantic plus technical question together. If
  several clarifications seem useful, ask only the
  single most blocking one.

**Option 2: SQL Query** (system_sql)
Use when you have enough information to generate a
  correct query. Write valid SQLite using only tables
  and columns from the database schema below.

**Option 3: Feedback** (system_feedback)
Use when the question cannot be answered with the
  current database. Explain what is missing and why.

## Output Format
{
    "system_question": "<A clear, concise
      clarification question to ask the user that will
      help resolve ambiguity or obtain missing
      information needed to generate a SQL query.
      Should be null if you're providing SQL or
      feedback instead.>",
    "system_sql": "<A valid SQLite query that
      correctly answers the user's question based on
      the database schema. Should be null if you're
      asking a question or providing feedback
      instead.>",
    "system_feedback": "<A clear, helpful explanation
      of why the question cannot be answered from the
      database and the information given, and what
      would be needed to answer it. Should be null if
      you're asking a question or providing SQL
      instead.>"
}

Provide exactly ONE non-null field. The others MUST be
  null.

## Database Schema
<SCHEMA>

## Diagnosed Category
Treat the category below as strong contextual
  evidence, not as an infallible instruction.
<CATEGORY BLOCK WITH TYPE>

## User Question
**Natural Language Question:** <QUESTION>
**Evidence:** <EVIDENCE>

## Conversation Status
**Turn:** 1 of 4
You may ask up to 3 more clarification question(s).

Think step by step, then end your answer with the JSON
  object described in the Output Format section: fill
  exactly one field, and set the others to null.
\end{verbatim}
}
In the \textit{Taxonomy-Only} mode, the Diagnosed Category section is replaced by the following one, which lists the 13 subcategories; in the \textit{Taxonomy-Free} mode it is left out.

{\scriptsize
\begin{verbatim}
## Taxonomy Reference
No category has been diagnosed for this question. Use
  the taxonomy below to determine the question type
  and respond accordingly.

<CATEGORY BLOCK WITH TYPE 1>

...

<CATEGORY BLOCK WITH TYPE 13>
\end{verbatim}
}
From the second turn on, the completed turns follow the question, and the Conversation Status reads as below (second turn).

{\scriptsize
\begin{verbatim}
### Conversation History
<CONVERSATION HISTORY>

## Conversation Status
**Turn:** 2 of 4
You may ask up to 2 more clarification question(s).

**Important:** Do not repeat questions already asked
  or request information already provided. If the user
  did not answer your question, decide again with the
  question types above. You may assume what the user
  most plausibly means. You may not assume information
  that the database does not contain.
\end{verbatim}
}
On the final turn, the Response Options, the Output Format, the Conversation Status and the closing line read as follows, and the copy admits only \texttt{system\_sql} or \texttt{system\_feedback}.

{\scriptsize
\begin{verbatim}
## Response Options
Provide exactly ONE of the following:

**Option 1: SQL Query** (system_sql)
Use when you have enough information to generate a
  correct query. Write valid SQLite using only tables
  and columns from the database schema below.

**Option 2: Feedback** (system_feedback)
Use when the question cannot be answered with the
  current database. Explain what is missing and why.

## Output Format
{
    "system_sql": "<A valid SQLite query that
      correctly answers the user's question based on
      the database schema. Should be null if you're
      providing feedback instead.>",
    "system_feedback": "<A clear, helpful explanation
      of why the question cannot be answered from the
      database and the information given, and what
      would be needed to answer it. Should be null if
      you're providing SQL instead.>"
}

Provide exactly ONE non-null field. The other MUST be
  null. system_question is not available on the final
  turn.

## Conversation Status
**Turn:** 4 of 4
**This is the final turn.** You MUST provide either a
  SQL query or feedback. You cannot ask another
  clarification question.

**Important:** If the user did not answer your
  question, decide with the question types above. You
  may assume what the user most plausibly means. You
  may not assume information that the database does
  not contain.

Think step by step, then end your answer with the JSON
  object described in the Output Format section: fill
  exactly one of system_sql and system_feedback, and
  set the other to null.
\end{verbatim}
}
In the floor, the first turn is the final one: the prompt is the first turn above with these final-turn parts, its Conversation Status reads \texttt{**Turn:** 1 of 1}, and it has no \texttt{**Important:**} paragraph. In the oracle, each conversation on an ambiguous question starts with one completed turn, in which the system asks the fixed question ``Could you clarify what you mean?'' and the user answers with the interpretation's hidden knowledge. The system then answers on its final turn (\texttt{**Turn:** 2 of 2}), and the simulated user is not called.

\paragraph{Simulated user: labels.} For a clarification question on an ambiguous question, each council member receives the label definitions, the question with its evidence and hidden knowledge, the Ground Truth SQL split into numbered nodes (the fragments of Appendix~\ref{app:user_sim}), and the conversation, and returns \texttt{relevancy} and \texttt{node\_ids}. \texttt{<SQL REFERENCE NODES>} has one node per line, such as \texttt{[2] WHERE (subquery 1): <SQL FRAGMENT>}.

{\scriptsize
\begin{verbatim}
You are a classification engine for a text-to-SQL user
  simulator. A text-to-SQL system has asked a
  clarification question. The system cannot see the
  hidden knowledge: only the user knows it. Classify
  the question and identify which SQL nodes it relates
  to.

## Relevancy Definitions
**1. RELEVANT** (Asks what the user means or knows):
- Directly addresses what is unclear or missing in the
  question, using hidden knowledge
- Asks about a point that the hidden knowledge
  settles: a reply that states the hidden knowledge
  would answer the question, or would correct an
  option or an assumption in it. A question only about
  points that the hidden knowledge does not settle is
  NOT Relevant, even when those points are unclear in
  the query
- Helps clarify which interpretation the user intends
- Focuses on natural language disambiguation (not SQL
  implementation)
- Includes questions about which columns or tables to
  use when that choice resolves the ambiguity -- what
  counts is WHICH MEANING is intended

**2. TECHNICAL** (Asks about SQL implementation
  details):
- Focuses on output preferences and implementation:
  ordering, limits, grouping, aggregation
- Asks about which columns, tables, or fields to use,
  when that choice does not resolve the ambiguity
- Can be answered using the SQL reference nodes listed
  below
- NOT about semantic meaning -- about HOW to implement
  or present results

**3. IRRELEVANT** (Doesn't help resolve the query):
- Doesn't address the ambiguity or provide useful
  technical details
- Tries to extract the SQL solution directly
- Completely off-topic or tangential to the query

## Classification Rules
- RELEVANT = asks what the user means or knows, on a
  point that the hidden knowledge settles
- A question is still RELEVANT when the hidden
  knowledge contradicts an option or an assumption in
  it: the user's reply then corrects it
- TECHNICAL = asks implementation details (ordering,
  limits, grouping, aggregation, or columns and tables
  when that choice does not resolve the ambiguity)
- IRRELEVANT = doesn't help or tries to extract SQL
- Questions asking about columns/tables to use are
  RELEVANT if the choice resolves the ambiguity, and
  TECHNICAL otherwise
- A question only about points that the hidden
  knowledge does not settle is never RELEVANT:
  classify it as TECHNICAL or IRRELEVANT by the
  definitions above
- If TECHNICAL, list ALL node IDs the question relates
  to, using only the IDs listed in the SQL Reference
  Nodes. If none fits, leave node_ids as []
- If RELEVANT or IRRELEVANT, leave node_ids as []

## How to Read the SQL Reference Nodes
The SQL reference nodes split the reference SQL query
  of the question into numbered clauses, one per line:
  [ID] CLAUSE: SQL fragment. If the query has several
  parts, the clause type is followed by its part in
  parentheses: "main query", "part k of UNION" (also
  UNION ALL, INTERSECT or EXCEPT), "subquery N", or
  "subquery <name>" for a named subquery. A part
  inside another part names the outer part first,
  separated by a comma. In an EXCEPT, "rows kept"
  marks the part whose rows the result keeps, and
  "rows removed" marks the part whose rows the result
  removes: these are rows that the user does not want.

## Response Format
Provide brief reasoning (approximately 128
  characters), then a JSON object with:
{
    "relevancy": "<Classification of the clarification
      question: 'Relevant' if it asks what the user
      means or knows, on a point that the hidden
      knowledge settles, also when it asks which
      columns or tables to use and that choice
      resolves the ambiguity, 'Technical' if it asks
      about SQL implementation details, including
      which columns or tables to use when that choice
      does not resolve the ambiguity, or 'Irrelevant'
      if it doesn't help or tries to extract SQL. Put
      only 'Relevant', 'Technical', or
      'Irrelevant'.>",
    "node_ids": [<If Technical, list ALL SQL node IDs
      that the clarification question relates to,
      using only the IDs listed in the SQL Reference
      Nodes. Leave the list empty if none fits. If
      Relevant or Irrelevant, leave the list empty.>]
}

## Context
**Question Type:** Solvable -- has ambiguity that can
  be resolved through clarification
**Natural Language Question:** <QUESTION>
**Evidence:** <EVIDENCE>
**Hidden Knowledge (Disambiguating Information):**
  <HIDDEN KNOWLEDGE>

## SQL Reference Nodes
<SQL REFERENCE NODES>

## Conversation
<CONVERSATION HISTORY>

**Clarification Question:** <CLARIFICATION QUESTION>

Classify the clarification question: give brief
  reasoning, then end your answer with the JSON object
  described in the Response Format section.
\end{verbatim}
}
For an answerable question, the prompt shows no hidden knowledge and admits only Technical and Irrelevant; the text before the node guide, the JSON template and the question type read:

{\scriptsize
\begin{verbatim}
You are a classification engine for a text-to-SQL user
  simulator. A text-to-SQL system has asked a
  clarification question about an already-clear query.
  Classify it and identify which SQL nodes it relates
  to.

## Relevancy Definitions
**1. TECHNICAL** (Asks about SQL implementation
  details):
- Focuses on output preferences and implementation:
  ordering, limits, grouping, aggregation
- Can be answered using the SQL reference nodes listed
  below
- NOT about semantic meaning -- about HOW to implement
  or present results
- NOT about which columns, tables, or fields to use or
  to return

**2. IRRELEVANT** (Doesn't help resolve the query):
- Asks only which columns, tables, or fields to use or
  to return: the question already says what it asks
  for
- Doesn't address the ambiguity or provide useful
  technical details
- Tries to extract the SQL solution directly
- Completely off-topic or tangential to the query

## Classification Rules
- TECHNICAL = asks implementation details (ordering,
  limits, grouping, aggregation)
- IRRELEVANT = doesn't help or tries to extract SQL
- Questions about what the user means are IRRELEVANT
  (question is already clear)
- Questions asking which columns/tables/fields to use
  or to return are IRRELEVANT (the question already
  says what it asks for). If such a question also asks
  about ordering, limits, grouping, or aggregation,
  classify it as TECHNICAL and list only the nodes of
  that part
- If TECHNICAL, list ALL node IDs the question relates
  to, using only the IDs listed in the SQL Reference
  Nodes. Leave out the FROM, JOIN, WHERE and HAVING
  nodes and every SELECT node without an aggregation.
  If none fits, leave node_ids as []
- If IRRELEVANT, leave node_ids as []

{
    "relevancy": "<Classification of the clarification
      question: 'Technical' if it asks about SQL
      implementation details such as ordering, limits,
      grouping, or aggregation, but not which columns,
      tables, or fields to use or to return, or
      'Irrelevant' if it doesn't help, tries to
      extract SQL, or asks only which columns, tables,
      or fields to use or to return. Put only
      'Technical' or 'Irrelevant'.>",
    "node_ids": [<If Technical, list ALL SQL node IDs
      that the clarification question relates to,
      using only the IDs listed in the SQL Reference
      Nodes. Leave out the FROM, JOIN, WHERE and
      HAVING nodes and every SELECT node without an
      aggregation. Leave the list empty if none fits.
      If Irrelevant, leave the list empty.>]
}

**Question Type:** Answerable -- question is already
  clear, nothing is unclear or missing
\end{verbatim}
}
\paragraph{Simulated user: answers.} Under a Relevant label, the drawn council member receives the question with its evidence and hidden knowledge, the conversation and the answer style, and returns \texttt{answer}.

{\scriptsize
\begin{verbatim}
You are a user in a text-to-SQL scenario. A system
  asked you a clarification question about what is
  unclear or missing in your query. Answer it using
  your knowledge of what you actually want.

**CRITICAL CONSTRAINT:** You are a non-technical user.
  Respond in natural language ONLY. NEVER use SQL
  syntax, column names, table names, table aliases,
  CTE names, or query structure. Express the intent
  behind the information, not its SQL representation.
  The source material is provided as reference for
  what you want; your answer must sound like a real
  non-technical user speaking in their own voice.
Bad examples: 'Use the rcp_ing table', 'Join on the
  recipe_id column', 'Filter by the cuisine_cd field'.
Good examples: 'Use the list of ingredients of each
  recipe', 'Match each ingredient to its recipe', 'I
  want the type of cuisine'.

## Instructions
Use Your Hidden Knowledge to disambiguate. Be direct
  about which interpretation you mean, or state the
  fact about yourself that settles it. If the
  clarification question offers options that do not
  include what you mean, or assumes something that
  Your Hidden Knowledge contradicts, say so and state
  what you mean. State what you know as a fact, not as
  a guess. When you give a value, give it exactly and
  say in plain words what it is. Do NOT mention SQL,
  databases, or technical implementation.

**GROUNDING CONSTRAINT:** ONLY state what Your Hidden
  Knowledge, your Natural Language Question or its
  Evidence settles. Do NOT invent details that they do
  not settle, such as a date, a name, or a value. For
  any point of the clarification question that they do
  not settle, say you have no preference for that
  point.

## Response Format
Provide your answer as a JSON object with:
{
    "answer": "<Your natural language answer to the
      clarification question.>"
}

## Expected Answer Style
<STYLE DESCRIPTION WITH EXAMPLE ANSWER>
State your answer in this style. The style changes
  only the wording, so do not answer with a question.

## Context
**Natural Language Question:** <QUESTION>
**Evidence:** <EVIDENCE>

## Your Hidden Knowledge
What you mean, or a fact about yourself, that the
  system cannot see:
<HIDDEN KNOWLEDGE>

## Conversation
<CONVERSATION HISTORY>

**Clarification Question:** <CLARIFICATION QUESTION>

Write your reply in plain language, without SQL or
  table and column names, in the Expected Answer
  Style.

Think step by step, then end your answer with the JSON
  object described in the Response Format section.
\end{verbatim}
}
Under a Technical label, the first paragraph and the Instructions change, and the selected nodes, each rendered in plain language, replace the hidden knowledge; on BIRD, a line that names a column with its table ends with the dataset's description of that column:

{\scriptsize
\begin{verbatim}
You are a user in a text-to-SQL scenario. A system
  asked you a technical question about implementation
  details. Answer it based on what you want from the
  results.

## Instructions
Express the intent behind the reference information in
  natural language. For example, 'ORDER BY count DESC'
  should become something like 'I want the most
  popular ones first' (colloquial) or 'Sort by
  frequency in descending order' (formal). Match the
  Expected Answer Style.

A reference line may name, in parentheses after its
  label, the part of your request that it belongs to,
  for example "main query", "part 2 of UNION" or
  "subquery 1". Nested parts are separated by commas,
  outermost first. A part marked "rows removed"
  describes rows that you do NOT want in the results,
  and a part marked "rows kept" describes the rows
  that you want before those are removed.

**GROUNDING CONSTRAINT:** ONLY describe what the
  reference fragments express. Do NOT add new
  requirements, filters, aggregations, time
  constraints, or conditions that are not reflected in
  the fragments. Do NOT 'improve' on the fragments
  (e.g., do not add tie-breaking, deduplication, or
  date filtering unless a fragment says so). If the
  fragments don't cover what the system is asking
  about, say you have no preference for that aspect.
  This also holds for a choice between options: if the
  fragments name none of them, do not choose one.

## What You Want (Reference)
The following describes what you want from the
  results. Use this to inform your answer, but express
  it in your own words:
- <FRAGMENT DESCRIPTION 1>
- <FRAGMENT DESCRIPTION 2>
\end{verbatim}
}
\paragraph{Simulated user: fixed refusals.} A clarification question on an unanswerable question, one labeled Irrelevant, or one labeled Technical whose selected nodes give no reference line receives a fixed sentence in the question's style without a model call: ``I have nothing to add to my question.'' (formal), ``Hmm, I don't really have anything to add to what I asked.'' (colloquial) or ``I am afraid I have nothing to add to my question.'' (interrogative).

\paragraph{Feedback graders.} Each council member receives the schema, the question with its evidence, the expected feedback (the question's hidden knowledge) and the system's feedback, but not the subcategory, and returns \texttt{matches} (Yes or No).

{\scriptsize
\begin{verbatim}
You are an expert evaluator for text-to-SQL systems.
  Your task is to assess whether the system correctly
  identified why a question is fundamentally
  unsolvable.

## Evaluation Task
Determine whether the system's feedback correctly
  identifies why the question is unsolvable.

**The feedback should:**
- Explain what specific information, schema elements,
  or knowledge is missing or problematic
- Name at least one problem that the expected feedback
  gives and that by itself makes the question
  unanswerable (exact wording doesn't need to match)

**Consider it a match if:**
- The system identifies the same fundamental problem
  as the expected feedback, or one of the problems
  that the expected feedback gives when that problem
  by itself makes the question unanswerable
- The system does not need to name every problem of
  the expected feedback, and other reasons that it
  adds do not make the feedback wrong unless a
  criterion below fails them

**Consider it NOT a match if:**
- None of the problems that the system names is a
  problem of the expected feedback that by itself
  makes the question unanswerable
- The feedback is too vague or generic to be useful
- The system's explanation contradicts the expected
  feedback
- The system's feedback states something about the
  database that the schema below shows to be false,
  for example that a column does not exist when it
  does

## Response Format
Provide an analysis addressing:
1. **Expected Feedback Analysis:** What specific
  problem does the expected feedback identify?
2. **System Feedback Analysis:** What does the
  system's feedback claim is the problem?
3. **Comparison:** Does the system name a problem of
  the expected feedback that by itself makes the
  question unanswerable? Does anything in the system's
  feedback contradict the expected feedback or the
  schema?

Your analysis should be concise but thorough
  (approximately 512 characters).

Then provide your final verdict as a JSON object with:
{
    "matches": "<'Yes' if the system's feedback names
      a problem that the expected feedback gives and
      that by itself makes the question unanswerable,
      and no criterion above fails it, 'No' otherwise.
      Put only 'Yes' or 'No'.>"
}

## Database Schema
<SCHEMA>

## Context
**Natural Language Question:** <QUESTION>
**Evidence:** <EVIDENCE>

## Expected Feedback
This is the expected explanation of why the question
  is unsolvable:
**Expected:** <FEEDBACK>

## System's Feedback
This is what the system provided as feedback:
**System's Response:** <SYSTEM FEEDBACK>

Remember: Focus on semantic equivalence, not exact
  wording. The system should identify the same
  fundamental problem.

Think step by step, then end your answer with the JSON
  object described in the Response Format section.
\end{verbatim}
}
}

%% file: acl_latex.bbl
\begin{thebibliography}{39}
\providecommand{\natexlab}[1]{#1}

\bibitem[{Bhaskar et~al.(2023)Bhaskar, Tomar, Sathe, and Sarawagi}]{benchmarking-ambiguity}
Adithya Bhaskar, Tushar Tomar, Ashutosh Sathe, and Sunita Sarawagi. 2023.
\newblock \href {https://doi.org/10.18653/v1/2023.emnlp-main.436} {Benchmarking and improving text-to-{SQL} generation under ambiguity}.
\newblock In \emph{Proceedings of the 2023 Conference on Empirical Methods in Natural Language Processing}, pages 7053--7074, Singapore. Association for Computational Linguistics.

\bibitem[{Chang and Fosler-Lussier(2023)}]{chang2023prompt}
Shuaichen Chang and Eric Fosler-Lussier. 2023.
\newblock \href {https://arxiv.org/abs/2305.11853} {How to prompt {LLMs} for text-to-{SQL}: A study in zero-shot, single-domain, and cross-domain settings}.
\newblock In \emph{NeurIPS 2023 Table Representation Learning Workshop}.

\bibitem[{Ding et~al.(2026)Ding, Lin, Zeng, Zhu, Ding, and Zhou}]{ding2025ambisql}
Zhongjun Ding, Yin Lin, Tianjing Zeng, Rong Zhu, Bolin Ding, and Jingren Zhou. 2026.
\newblock \href {https://doi.org/10.1145/3788853.3801581} {{AmbiSQL}: Interactive ambiguity detection and resolution for text-to-{SQL}}.
\newblock In \emph{Companion of the 2026 International Conference on Management of Data}, pages 26--29, Bengaluru, India. ACM.

\bibitem[{Dong et~al.(2025)Dong, Ashok~Kumar, Hu, Chauhan, Hang, Chang, Pan, Lan, Zhu, Jiang, Ng, and Wang}]{practiq}
Mingwen Dong, Nischal Ashok~Kumar, Yiqun Hu, Anuj Chauhan, Chung-Wei Hang, Shuaichen Chang, Lin Pan, Wuwei Lan, Henghui Zhu, Jiarong Jiang, Patrick Ng, and Zhiguo Wang. 2025.
\newblock \href {https://doi.org/10.18653/v1/2025.naacl-long.13} {{PRACTIQ}: A practical conversational text-to-{SQL} dataset with ambiguous and unanswerable queries}.
\newblock In \emph{Proceedings of the 2025 Conference of the Nations of the Americas Chapter of the Association for Computational Linguistics: Human Language Technologies (Volume 1: Long Papers)}, pages 255--273, Albuquerque, New Mexico. Association for Computational Linguistics.

\bibitem[{Dror et~al.(2018)Dror, Baumer, Shlomov, and Reichart}]{dror-etal-2018-hitchhikers}
Rotem Dror, Gili Baumer, Segev Shlomov, and Roi Reichart. 2018.
\newblock \href {https://doi.org/10.18653/v1/P18-1128} {The hitchhiker's guide to testing statistical significance in natural language processing}.
\newblock In \emph{Proceedings of the 56th Annual Meeting of the Association for Computational Linguistics (Volume 1: Long Papers)}, pages 1383--1392, Melbourne, Australia. Association for Computational Linguistics.

\bibitem[{Efron and Tibshirani(1993)}]{efron1993introduction}
Bradley Efron and Robert~J. Tibshirani. 1993.
\newblock \emph{An Introduction to the Bootstrap}, volume~57 of \emph{Monographs on Statistics and Applied Probability}.
\newblock Chapman \& Hall, New York.

\bibitem[{Floratou et~al.(2024)Floratou, Psallidas, Zhao, Deep, Hagleither, Tan, Cahoon, Alotaibi, Henkel, Singla, Grootel, Chow, Deng, Lin, Campos, Emani, Pandit, Shnayder, Wang, and Curino}]{Floratou2024NL2SQLIA}
Avrilia Floratou, Fotis Psallidas, Fuheng Zhao, Shaleen Deep, Gunther Hagleither, Wangda Tan, Joyce Cahoon, Rana Alotaibi, Jordan Henkel, Abhik Singla, Alex~Van Grootel, Brandon Chow, Kai Deng, Katherine Lin, Marcos Campos, K.~Venkatesh Emani, Vivek Pandit, Victor Shnayder, Wenjing Wang, and Carlo Curino. 2024.
\newblock \href {https://www.cidrdb.org/cidr2024/papers/p74-floratou.pdf} {{NL2SQL} is a solved problem... {Not}!}
\newblock In \emph{Conference on Innovative Data Systems Research}.

\bibitem[{F{\"{u}}rst et~al.(2025)F{\"{u}}rst, Kosten, Nooralahzadeh, Zhang, and Stockinger}]{EvaluatingDataModel}
Jonathan F{\"{u}}rst, Catherine Kosten, Farhad Nooralahzadeh, Yi~Zhang, and Kurt Stockinger. 2025.
\newblock \href {https://doi.org/10.48786/EDBT.2025.13} {Evaluating the data model robustness of text-to-{SQL} systems based on real user queries}.
\newblock In \emph{Proceedings 28th International Conference on Extending Database Technology, {EDBT} 2025, Barcelona, Spain, March 25-28, 2025}, pages 158--170. OpenProceedings.org.

\bibitem[{Gao et~al.(2024)Gao, Wang, Li, Sun, Qian, Ding, and Zhou}]{gao2024text}
Dawei Gao, Haibin Wang, Yaliang Li, Xiuyu Sun, Yichen Qian, Bolin Ding, and Jingren Zhou. 2024.
\newblock \href {https://doi.org/10.14778/3641204.3641221} {Text-to-{SQL} empowered by large language models: A benchmark evaluation}.
\newblock \emph{Proc. VLDB Endow.}, 17(5):1132–1145.

\bibitem[{{Gemma Team} et~al.(2025){Gemma Team}, Kamath, Ferret, Pathak, Vieillard, Merhej, Perrin, Matejovicova, Ramé, Rivière, Rouillard, Mesnard, Cideron, Grill, Ramos, Yvinec, Casbon, Pot, Penchev, Liu, Visin, Kenealy, Beyer, Zhai, Tsitsulin, Busa-Fekete, Feng, Sachdeva, Coleman, Gao, Mustafa, Barr, Parisotto, Tian, Eyal, Cherry, Peter, Sinopalnikov, Bhupatiraju, Agarwal, Kazemi, Malkin, Kumar, Vilar, Brusilovsky, Luo, Steiner, Friesen, Sharma, Sharma, Gilady, Goedeckemeyer, Saade, Feng, Kolesnikov, Bendebury, Abdagic, Vadi, György, Pinto, Das, Bapna, Miech, Yang, Paterson, Shenoy, Chakrabarti, Piot, Wu, Shahriari, Petrini, Chen, Lan, Choquette-Choo, Carey, Brick, Deutsch, Eisenbud, Cattle, Cheng, Paparas, Sreepathihalli, Reid, Tran, Zelle, Noland, Huizenga, Kharitonov, Liu, Amirkhanyan, Cameron, Hashemi, Klimczak-Plucińska, Singh, Mehta, Lehri, Hazimeh, Ballantyne, Szpektor, Nardini, Pouget-Abadie, Chan, Stanton, Wieting, Lai, Orbay, Fernandez, Newlan, yeong Ji, Singh, Black, Yu, Hui, Vodrahalli,
  Greff, Qiu, Valentine, Coelho, Ritter, Hoffman, Watson, Chaturvedi, Moynihan, Ma, Babar, Noy, Byrd, Roy, Momchev, Chauhan, Sachdeva, Bunyan, Botarda, Caron, Rubenstein, Culliton, Schmid, Sessa, Xu, Stanczyk, Tafti, Shivanna, Wu, Pan, Rokni, Willoughby, Vallu, Mullins, Jerome, Smoot, Girgin, Iqbal, Reddy, Sheth, Põder, Bhatnagar, Panyam, Eiger, Zhang, Liu, Yacovone, Liechty, Kalra, Evci, Misra, Roseberry, Feinberg, Kolesnikov, Han, Kwon, Chen, Chow, Zhu, Wei, Egyed, Cotruta, Giang, Kirk, Rao, Black, Babar, Lo, Moreira, Martins, Sanseviero, Gonzalez, Gleicher, Warkentin, Mirrokni, Senter, Collins, Barral, Ghahramani, Hadsell, Matias, Sculley, Petrov, Fiedel, Shazeer, Vinyals, Dean, Hassabis, Kavukcuoglu, Farabet, Buchatskaya, Alayrac, Anil, Lepikhin, Borgeaud, Bachem, Joulin, Andreev, Hardin, Dadashi, and Hussenot}]{gemmateam2025gemma3technicalreport}
{Gemma Team}, Aishwarya Kamath, Johan Ferret, Shreya Pathak, Nino Vieillard, Ramona Merhej, Sarah Perrin, Tatiana Matejovicova, Alexandre Ramé, Morgane Rivière, Louis Rouillard, Thomas Mesnard, Geoffrey Cideron, {Jean-bastien} Grill, Sabela Ramos, Edouard Yvinec, Michelle Casbon, Etienne Pot, Ivo Penchev, and 196 others. 2025.
\newblock \href {https://arxiv.org/abs/2503.19786} {{Gemma} 3 technical report}.
\newblock \emph{Preprint}, arXiv:2503.19786.

\bibitem[{G{\"u}r et~al.(2018)G{\"u}r, Yavuz, Su, and Yan}]{gur2018dialsql}
Izzeddin G{\"u}r, Semih Yavuz, Yu~Su, and Xifeng Yan. 2018.
\newblock \href {https://doi.org/10.18653/v1/P18-1124} {{DialSQL}: Dialogue based structured query generation}.
\newblock In \emph{Proceedings of the 56th Annual Meeting of the Association for Computational Linguistics (Volume 1: Long Papers)}, pages 1339--1349.

\bibitem[{Holm(1979)}]{holm1979simple}
Sture Holm. 1979.
\newblock \href {https://www.jstor.org/stable/4615733} {A simple sequentially rejective multiple test procedure}.
\newblock \emph{Scandinavian Journal of Statistics}, 6(2):65--70.

\bibitem[{Hui et~al.(2024)Hui, Yang, Cui, Yang, Liu, Zhang, Liu, Zhang, Yu, Lu, Dang, Fan, Zhang, Yang, Men, Huang, Zheng, Miao, Quan, Feng, Ren, Ren, Zhou, and Lin}]{hui2024qwen25coder}
Binyuan Hui, Jian Yang, Zeyu Cui, Jiaxi Yang, Dayiheng Liu, Lei Zhang, Tianyu Liu, Jiajun Zhang, Bowen Yu, Keming Lu, Kai Dang, Yang Fan, Yichang Zhang, An~Yang, Rui Men, Fei Huang, Bo~Zheng, Yibo Miao, Shanghaoran Quan, and 5 others. 2024.
\newblock \href {https://arxiv.org/abs/2409.12186} {{Qwen2.5-Coder} technical report}.
\newblock \emph{Preprint}, arXiv:2409.12186.

\bibitem[{Huo et~al.(2026)Huo, Xu, Li, Jacobsson, Lin, Qin, Hui, Li, Qu, Si, Han, Alexander, Zhu, Qin, Yu, Jin, Zhou, Zhong, Chen, Liu, Ma, Ozcan, Papakonstantinou, and Cheng}]{huo2025birdinteractreimaginingtexttosqlevaluation}
Nan Huo, Xiaohan Xu, Jinyang Li, Per Jacobsson, Shipei Lin, Bowen Qin, Binyuan Hui, Xiaolong Li, Ge~Qu, Shuzheng Si, Linheng Han, Edward Alexander, Xintong Zhu, Rui Qin, Ruihan Yu, Yiyao Jin, Feige Zhou, Weihao Zhong, Yun Chen, and 5 others. 2026.
\newblock \href {https://openreview.net/forum?id=nHrYBGujps} {{BIRD}-{INTERACT}: Re-imagining text-to-{SQL} evaluation via lens of dynamic interactions}.
\newblock In \emph{The Fourteenth International Conference on Learning Representations}.

\bibitem[{Koehn(2004)}]{koehn-2004-statistical}
Philipp Koehn. 2004.
\newblock \href {https://aclanthology.org/W04-3250/} {Statistical significance tests for machine translation evaluation}.
\newblock In \emph{Proceedings of the 2004 Conference on Empirical Methods in Natural Language Processing}, pages 388--395, Barcelona, Spain. Association for Computational Linguistics.

\bibitem[{Li and Jagadish(2014)}]{li2014constructing}
Fei Li and H.~V. Jagadish. 2014.
\newblock \href {https://doi.org/10.14778/2735461.2735468} {Constructing an interactive natural language interface for relational databases}.
\newblock \emph{Proceedings of the VLDB Endowment}, 8(1):73--84.

\bibitem[{Li et~al.(2025)Li, Wu, Zhang, Huang, Zhang, Jiang, Wang, Zhang, Chen, Shi, Chen, and Li}]{OMNI-SQL}
Haoyang Li, Shang Wu, Xiaokang Zhang, Xinmei Huang, Jing Zhang, Fuxin Jiang, Shuai Wang, Tieying Zhang, Jianjun Chen, Rui Shi, Hong Chen, and Cuiping Li. 2025.
\newblock \href {https://doi.org/10.14778/3749646.3749723} {{OmniSQL}: Synthesizing high-quality text-to-{SQL} data at scale}.
\newblock \emph{Proc. VLDB Endow.}, 18(11):4695–4709.

\bibitem[{Li et~al.(2023{\natexlab{a}})Li, Zhang, Li, and Chen}]{li2023resdsql}
Haoyang Li, Jing Zhang, Cuiping Li, and Hong Chen. 2023{\natexlab{a}}.
\newblock \href {https://doi.org/10.1609/aaai.v37i11.26535} {{RESDSQL}: Decoupling schema linking and skeleton parsing for text-to-{SQL}}.
\newblock In \emph{Proceedings of the Thirty-Seventh AAAI Conference on Artificial Intelligence and Thirty-Fifth Conference on Innovative Applications of Artificial Intelligence and Thirteenth Symposium on Educational Advances in Artificial Intelligence}, AAAI'23/IAAI'23/EAAI'23. AAAI Press.

\bibitem[{Li et~al.(2024)Li, Zhang, Liu, Fan, Zhang, Zhu, Wei, Pan, Li, and Chen}]{li2024codes}
Haoyang Li, Jing Zhang, Hanbing Liu, Ju~Fan, Xiaokang Zhang, Jun Zhu, Renjie Wei, Hongyan Pan, Cuiping Li, and Hong Chen. 2024.
\newblock \href {https://doi.org/10.1145/3654930} {{CodeS}: Towards building open-source language models for text-to-{SQL}}.
\newblock \emph{Proc. ACM Manag. Data}, 2(3).

\bibitem[{Li et~al.(2023{\natexlab{b}})Li, Hui, Qu, Yang, Li, Li, Wang, Qin, Geng, Huo, Zhou et~al.}]{Li2023bird}
Jinyang Li, Binyuan Hui, Ge~Qu, Jiaxi Yang, Binhua Li, Bowen Li, Bailin Wang, Bowen Qin, Ruiying Geng, Nan Huo, Xuanhe Zhou, et~al. 2023{\natexlab{b}}.
\newblock \href {https://proceedings.neurips.cc/paper_files/paper/2023/hash/83fc8fab1710363050bbd1d4b8cc0021-Abstract-Datasets_and_Benchmarks.html} {Can {LLM} already serve as {A} database interface? {A} {BIg} bench for large-scale database grounded text-to-{SQLs}}.
\newblock In \emph{Advances in Neural Information Processing Systems 36: Annual Conference on Neural Information Processing Systems 2023, NeurIPS 2023, New Orleans, LA, USA, December 10 - 16, 2023}.

\bibitem[{Li et~al.(2020)Li, Chen, Liu, Gao, Lou, Zhang, and Zhang}]{li-etal-2020-mean}
Yuntao Li, Bei Chen, Qian Liu, Yan Gao, Jian-Guang Lou, Yan Zhang, and Dongmei Zhang. 2020.
\newblock \href {https://doi.org/10.18653/v1/2020.emnlp-main.561} {``{W}hat do you mean by that?'' {A} parser-independent interactive approach for enhancing text-to-{SQL}}.
\newblock In \emph{Proceedings of the 2020 Conference on Empirical Methods in Natural Language Processing (EMNLP)}, pages 6913--6922, Online. Association for Computational Linguistics.

\bibitem[{{Mistral AI}(2025)}]{Mistral}
{Mistral AI}. 2025.
\newblock \href {https://huggingface.co/mistralai/Mistral-Small-3.2-24B-Instruct-2506} {{Mistral-Small-3.2-24B-Instruct-2506}}.
\newblock Hugging Face model card.

\bibitem[{{NVIDIA}(2025)}]{nvidia_nemotron_3_2025}
{NVIDIA}. 2025.
\newblock \href {https://arxiv.org/abs/2512.20856} {{NVIDIA} {Nemotron} 3: Efficient and open intelligence}.
\newblock White Paper.

\bibitem[{{OpenAI}(2023)}]{openai2023evals}
{OpenAI}. 2023.
\newblock \href {https://github.com/openai/evals} {{Evals}}.
\newblock GitHub repository.

\bibitem[{OpenAI et~al.(2025)OpenAI, Agarwal, Ahmad, Ai, Altman, Applebaum, Arbus, Arora, Bai, Baker, Bao, Barak, Bennett, Bertao, Brett, Brevdo, Brockman, Bubeck, Chang, Chen, Chen, Cheung, Clark, Cook, Dukhan, Dvorak, Fives, Fomenko, Garipov, Georgiev, Glaese, Gogineni, Goucher, Gross, Guzman, Hallman, Hehir, Heidecke, Helyar, Hu, Huet, Huh, Jain, Johnson, Koch, Kofman, Kundel, Kwon, Kyrylov, Le, Leclerc, Lennon, Lessans, Lezcano-Casado, Li, Li, Lin, Liss, Lily, Liu, Liu, Lu, Lu, Martinovic, McCallum, McGrath, McKinney, McLaughlin, Mei, Mostovoy, Mu, Myles, Neitz, Nichol, Pachocki, Paino, Palmie, Pantuliano, Parascandolo, Park, Pathak, Paz, Peran, Pimenov, Pokrass, Proehl, Qiu, Raila, Raso, Ren, Richardson, Robinson, Rotsted, Salman, Sanjeev, Schwarzer, Sculley, Sikchi, Simon, Singhal, Song, Stuckey, Sun, Tillet, Toizer, Tsimpourlas, Vyas, Wallace, Wang, Wang, Watkins, Weil, Wendling, Whinnery, Whitney, Wong, Yang, Yang, Yasunaga, Ying, Zaremba, Zhan, Zhang, Zhang, Zhang, and
  Zhao}]{openai2025gptoss120bgptoss20bmodel}
OpenAI, Sandhini Agarwal, Lama Ahmad, Jason Ai, Sam Altman, Andy Applebaum, Edwin Arbus, Rahul~K. Arora, Yu~Bai, Bowen Baker, Haiming Bao, Boaz Barak, Ally Bennett, Tyler Bertao, Nivedita Brett, Eugene Brevdo, Greg Brockman, Sebastien Bubeck, Che Chang, and 107 others. 2025.
\newblock \href {https://arxiv.org/abs/2508.10925} {gpt-oss-120b \& gpt-oss-20b model card}.
\newblock \emph{Preprint}, arXiv:2508.10925.

\bibitem[{Panickssery et~al.(2024)Panickssery, Bowman, and Feng}]{LLMFavor}
Arjun Panickssery, Samuel~R. Bowman, and Shi Feng. 2024.
\newblock \href {https://proceedings.neurips.cc/paper_files/paper/2024/hash/7f1f0218e45f5414c79c0679633e47bc-Abstract-Conference.html} {{LLM} evaluators recognize and favor their own generations}.
\newblock In \emph{Proceedings of the 38th International Conference on Neural Information Processing Systems}, NIPS '24, Red Hook, NY, USA. Curran Associates Inc.

\bibitem[{Pourreza and Rafiei(2023)}]{DINSQL}
Mohammadreza Pourreza and Davood Rafiei. 2023.
\newblock \href {https://proceedings.neurips.cc/paper_files/paper/2023/hash/72223cc66f63ca1aa59edaec1b3670e6-Abstract-Conference.html} {{DIN}-{SQL}: Decomposed in-context learning of text-to-{SQL} with self-correction}.
\newblock In \emph{Proceedings of the 37th International Conference on Neural Information Processing Systems}, NIPS '23, Red Hook, NY, USA. Curran Associates Inc.

\bibitem[{Qiu et~al.(2025)Qiu, Li, Su, and Chen}]{InteractiveT2SQLViaExpected}
Luyu Qiu, Jianing Li, Chi Su, and Lei Chen. 2025.
\newblock \href {https://arxiv.org/abs/2507.06467} {Interactive text-to-{SQL} via expected information gain for disambiguation}.
\newblock \emph{ArXiv}, abs/2507.06467.

\bibitem[{Qwen et~al.(2025)Qwen, Yang, Yang, Zhang, Hui, Zheng, Yu, Li, Liu, Huang, Wei, Lin, Yang, Tu, Zhang, Yang, Yang, Zhou, Lin, Dang, Lu, Bao, Yang, Yu, Li, Xue, Zhang, Zhu, Men, Lin, Li, Tang, Xia, Ren, Ren, Fan, Su, Zhang, Wan, Liu, Cui, Zhang, and Qiu}]{qwen2025qwen25technicalreport}
Qwen, An~Yang, Baosong Yang, Beichen Zhang, Binyuan Hui, Bo~Zheng, Bowen Yu, Chengyuan Li, Dayiheng Liu, Fei Huang, Haoran Wei, Huan Lin, Jian Yang, Jianhong Tu, Jianwei Zhang, Jianxin Yang, Jiaxi Yang, Jingren Zhou, Junyang Lin, and 24 others. 2025.
\newblock \href {https://arxiv.org/abs/2412.15115} {Qwen2.5 technical report}.
\newblock \emph{Preprint}, arXiv:2412.15115.

\bibitem[{{Qwen Team}(2026{\natexlab{a}})}]{qwen3.5}
{Qwen Team}. 2026{\natexlab{a}}.
\newblock \href {https://qwen.ai/blog?id=qwen3.5} {{Qwen3.5}: Towards native multimodal agents}.

\bibitem[{{Qwen Team}(2026{\natexlab{b}})}]{qwen38flash}
{Qwen Team}. 2026{\natexlab{b}}.
\newblock \href {https://huggingface.co/Qwen/Qwen3.8-Flash-Next-FP8} {{Qwen3.8-Flash-Next}}.

\bibitem[{Saparina and Lapata(2024)}]{Ambrosia}
Irina Saparina and Mirella Lapata. 2024.
\newblock \href {https://proceedings.neurips.cc/paper_files/paper/2024/file/a4c942a8405cc910f0a833d28d2573cc-Paper-Datasets_and_Benchmarks_Track.pdf} {{AMBROSIA}: A benchmark for parsing ambiguous questions into database queries}.
\newblock In \emph{Advances in Neural Information Processing Systems}, volume~37, pages 90600--90628. Curran Associates, Inc.

\bibitem[{Tam et~al.(2024)Tam, Wu, Tsai, Lin, Lee, and Chen}]{tam-etal-2024-speak}
Zhi~Rui Tam, Cheng-Kuang Wu, Yi-Lin Tsai, Chieh-Yen Lin, Hung-yi Lee, and Yun-Nung Chen. 2024.
\newblock \href {https://doi.org/10.18653/v1/2024.emnlp-industry.91} {Let me speak freely? {A} study on the impact of format restrictions on large language model performance.}
\newblock In \emph{Proceedings of the 2024 Conference on Empirical Methods in Natural Language Processing: Industry Track}, pages 1218--1236, Miami, Florida, US. Association for Computational Linguistics.

\bibitem[{Wang et~al.(2023)Wang, Gao, Li, and Lou}]{KnowWhatIDontKnow}
Bing Wang, Yan Gao, Zhoujun Li, and Jian-Guang Lou. 2023.
\newblock \href {https://doi.org/10.18653/v1/2023.findings-acl.352} {Know what {I} don{'}t know: Handling ambiguous and unknown questions for text-to-{SQL}}.
\newblock In \emph{Findings of the Association for Computational Linguistics: ACL 2023}, pages 5701--5714, Toronto, Canada. Association for Computational Linguistics.

\bibitem[{Wang et~al.(2025)Wang, Ren, Yang, Liang, Bai, Chai, Yan, Zhang, Yin, Sun, and Li}]{wang2025macsql}
Bing Wang, Changyu Ren, Jian Yang, Xinnian Liang, Jiaqi Bai, Linzheng Chai, Zhao Yan, Qian-Wen Zhang, Di~Yin, Xing Sun, and Zhoujun Li. 2025.
\newblock \href {https://aclanthology.org/2025.coling-main.36/} {{MAC}-{SQL}: A multi-agent collaborative framework for text-to-{SQL}}.
\newblock In \emph{Proceedings of the 31st International Conference on Computational Linguistics (COLING 2025)}, pages 540--557.

\bibitem[{Yao et~al.(2019)Yao, Su, Sun, and Yih}]{yao2019model}
Ziyu Yao, Yu~Su, Huan Sun, and Wen-tau Yih. 2019.
\newblock \href {https://doi.org/10.18653/v1/D19-1547} {Model-based interactive semantic parsing: A unified framework and a text-to-{SQL} case study}.
\newblock In \emph{Proceedings of the 2019 Conference on Empirical Methods in Natural Language Processing and the 9th International Joint Conference on Natural Language Processing (EMNLP-IJCNLP)}, pages 5447--5458.

\bibitem[{Yu et~al.(2019)Yu, Zhang, Er, Li, Xue, Pang, Lin, Tan, Shi, Li, Jiang, Yasunaga, Shim, Chen, Fabbri, Li, Chen, Zhang, Dixit, Zhang, Xiong, Socher, Lasecki, and Radev}]{cosql}
Tao Yu, Rui Zhang, Heyang Er, Suyi Li, Eric Xue, Bo~Pang, Xi~Victoria Lin, Yi~Chern Tan, Tianze Shi, Zihan Li, Youxuan Jiang, Michihiro Yasunaga, Sungrok Shim, Tao Chen, Alexander Fabbri, Zifan Li, Luyao Chen, Yuwen Zhang, Shreya Dixit, and 5 others. 2019.
\newblock \href {https://doi.org/10.18653/v1/D19-1204} {{C}o{SQL}: A conversational text-to-{SQL} challenge towards cross-domain natural language interfaces to databases}.
\newblock In \emph{Proceedings of the 2019 Conference on Empirical Methods in Natural Language Processing and the 9th International Joint Conference on Natural Language Processing (EMNLP-IJCNLP)}, pages 1962--1979, Hong Kong, China. Association for Computational Linguistics.

\bibitem[{Yu et~al.(2018)Yu, Zhang, Yang, Yasunaga, Wang, Li, Ma, Li, Yao, Roman, Zhang, and Radev}]{yu2018spider}
Tao Yu, Rui Zhang, Kai Yang, Michihiro Yasunaga, Dongxu Wang, Zifan Li, James Ma, Irene Li, Qingning Yao, Shanelle Roman, Zilin Zhang, and Dragomir Radev. 2018.
\newblock \href {https://doi.org/10.18653/v1/D18-1425} {{S}pider: A large-scale human-labeled dataset for complex and cross-domain semantic parsing and text-to-{SQL} task}.
\newblock In \emph{Proceedings of the 2018 Conference on Empirical Methods in Natural Language Processing}, pages 3911--3921, Brussels, Belgium. Association for Computational Linguistics.

\bibitem[{Zhang et~al.(2020)Zhang, Dong, Chang, Yu, Shi, and Zhang}]{didaskgoodquestion}
Yusen Zhang, Xiangyu Dong, Shuaichen Chang, Tao Yu, Peng Shi, and Rui Zhang. 2020.
\newblock \href {https://arxiv.org/abs/2010.12634} {Did you ask a good question? {A} cross-domain question intention classification benchmark for text-to-{SQL}}.
\newblock \emph{Preprint}, arXiv:2010.12634.

\end{thebibliography}
